\documentclass{aa}

\usepackage{natbib}
\usepackage{graphicx}
\usepackage{txfonts}
\usepackage{placeins}

\begin{document}

\title{Extended correlations between diffuse interstellar bands. 
} 
\titlerunning{Extended DIB correlations}

\author{Alain Omont \inst{1}}
\institute{Sorbonne Universit\'e, UPMC Universit\'e Paris 6 and CNRS, UMR 7095, Institut d'Astrophysique de Paris, France\\
}

\abstract{The systematic analysis of the correlations between diffuse interstellar bands (DIBs) is extended to weak DIBs through the comprehensive catalogue of the Apache Peak Observatory (APO) of 559 DIBs in 25 lines of sight with diverse interstellar properties. The main results are the following: 1) An extension of the number of DIBs identified to be related to C$_2$, that is, those  that need very shielded interstellar regions for their carriers to survive UV photo-dissociation. Based on the correlations with the reference C$_2$ and $\zeta$ DIBs, anticorrelations with UV-favoured ($\sigma$) DIBs, and the strength ratios in shielded and unshielded sight lines, we propose 12 new  C$_2$ candidates and 34 possible "C$_2$-related" DIBs (mostly at $\lambda$\,$<$\,5950\,\AA ) in addition to the $\sim$20 known confirmed C$_2$ DIBs.
With these additions, the census of C$_2$ DIBs might approach completion.
2) We discovered that the intensities of a large set of poorly studied DIBs are strongly enhanced in one  or two of the sight lines of HD\,175156 and HD\,148579. This tentative class, denoted $\chi$ for the time being, extends over the whole wavelength range of visible DIBs. It might include up to 50-100 members, half at $\lambda$\,$>$\,6000\,\AA , and a number of C$_2$ DIBs. These possible enhancements might reflect specific formation processes of their carriers that are yet to be identified in the interstellar medium of these two sight lines. 
The possible matches of the wavelength of five very broad DIBs, including three $\chi$ DIBs, with the strong bands that were recently measured by action spectroscopy might favour some long carbon chains and rings as carriers of some DIBs. 
These correlations and findings justify further theoretical and laboratory efforts for improving our understanding of the complex physics, spectroscopy, and chemistry of the various  carbon chains and rings, and their possible formation and destruction in the diffuse interstellar medium.
}


\maketitle 

\section{Introduction}	\label{Intro}

Diffuse interstellar bands (DIBs), which are hundreds of unidentified broad absorption lines that are observed in interstellar lines of sight, have been discovered a century ago by  Mary Lea Heger (Heger 1922a,b). After  confirmation that they universally form in the diffuse interstellar medium (ISM), the search for their origin is almost as old (Merrill 1934, 1936; Merrill \& Wilson 1938). During this long time (McCall \& Griffin 2013; McCabe 2019), a vast diversity of carriers have been proposed. Many of them were discarded, such as impurities in interstellar dust grains (see e.g. Herbig (1975, 1995); Smith et al.\ (1977); Tielens \& Snow (1995); Cami et al.\ (1997); Sarre (2006); Snow \& McCall (2006); Cami \& Cox (2014) for reviews about DIBs. See also e.g. Fan et al.\ (2017, 2019); Cami et al.\ (2018); Cox et al.\ (2017); York (2024) for more recent references).

A very small part of this mystery has been solved with the confirmation that five near-infrared DIBs are carried by the fullerene cation C$_{60}^+$ (Foing \& Ehrenfreund 1994, 1997; Fulara et al\ 1993; Campbell et al.\ 2015; Campbell \& Maier 2017; Cordiner et al.\ 2019; Linnartz et al.\ 2020). However, the nature of the carriers of the immense majority of the more than 600 known visible and 
near-infrared DIBs (Fan et al.\ 2019; Ebenblicher et al.\ 2022; Hamano et al.\ 2022) is still unknown. 

In addition to fullerenes, it is generally agreed (see e.g. Cami \& Cox 2014) that DIB carriers are probably other large carbon-based molecules (with
$\sim$10-100 atoms) in the gas phase, such as polycyclic aromatic hydrocarbons (PAHs), long carbon chains, or rings. However, other carriers, such as smaller molecules, are possible (see e.g.\ Tielens 2014, for the complexity of the problem).  

Since their convincing identification (L\'eger \& Puget, 1984; Allamandola et al.\ 1985), relatively large PAHs have remained good DIB carrier candidates (see e.g. van der Zwet \& Allamandola 1985; L\'eger \& D'Hendecourt 1985; Crawford et al.\ 1985; Salama et al.\ 1996, 2011 and references therein; Salama \& Ehrenfreund 2014). From their prominent mid-infrared emission bands in galaxies, it is well established that as a whole, PAHs contain several percent (up to $\ga$10\%) of the total interstellar carbon (e.g. Puget \& L\'eger 1989; Tielens 2008, 2013; Draine \& Li 2007; Shivaei et al.\ 2024). To date, however, all attempts to identify a single interstellar PAH as DIB carrier have failed, while a few small PAHs have been identified  in dense clouds (McGuire et al.\ 2021; McGuire 2022;  Burkhardt et al.\ 2021; Cernicharo et al.\ 2021; Sita et al.\ 2022).   
Nevertheless, PAHs may remain important DIB carriers if their distribution is dominated by a small number of species, especially in the form of cations. It is generally thought that the most abundant PAHs are the most compact (pericondensed), such as circumcoronene  C$_{54}$H$_{18}$. They are often called grand PAHs (e.g. Tielens 2013; Andrews et al.\ 2015). Their importance seems to have been confirmed by the first 
JWST
\footnote{John Webb Space Telescope}  
observations of the Orion Bar (Chown et al.\ 2023), for example.  However, these large PAHs remain difficult to produce and study in the laboratory.

After the confirmation of the five DIBs carried by C$_{60}^+$, a variety of DIBs carried by fullerene compounds may be possible (e.g. Omont 2016), including pure fullerene cages (Candian et al.\ 2019; Omont \& Bettinger 2021), hydrogenated fullerenes (fulleranes; e.g. Zhang et al.\ 2020), other exohedral or endohedral hetero-atom compounds, such as exohedral FeC$_{60}$, in various charge states. However, the overall abundance of fullerenes is relatively low, as found for C$_{60}^+$ (a few 10$^{-4}$ of the total interstellar carbon), pure fullerenes have no strong bands in the visible, and the abundance of possible hetero-fullerenes is unknown.

Carbon chains, with n\,$\la$\,10, a polyyne or cumulenic C$_n$ skeleton,  and possible hetero atoms, H, N, O, and so on, have been observed since the early 1970s in dense molecular clouds (see e.g. the review by Taniguchi et al.\ 2024).  They have repeatedly been proposed as DIB carriers (see e.g.  Douglas 1977; Thaddeus 1995; Snow 1995a,b; Allamandola et al.\ 1999; Maier et al.\ 2004; Rice \& Maier 2013; Zack \& Maier 2014a,b; Campbell \& Maier 2017). Long chains and rings remain good DIB carrier candidates, even though short chains (n$\la$10) are excluded because their known wavelengths do not correlate with those of DIBs (e.g. Zack \& Maier 2014a,b), probably because of their photo-destruction in the diffuse ISM. Their spectra (e.g. Jochnowitz \& Maier 2008a,b; Buntine et al.\ 2022; Marlton et al.\ 2022, 2023, 2024; Rademacher et al.\ 2022) generally  include a series of very strong broad bands throughout the visible range, for which  the wavelength can be roughly proportional to the number n of carbon atoms. This property  has made them especially appealing as DIB carrier candidates (e.g. Campbell \& Maier 2017, and Section \ref{sec:6chains}).

A key feature of DIBs that provides fundamental clues for the identification of their carriers, is the level of correlation between them in various sight lines. Strong correlations within some DIB subsets that are much higher than with other DIBs allow us to define DIB families. The three most important DIB families are related to the behaviour  and the abundance of their carriers in various intensities of the ultraviolet (UV) interstellar radiation. They are $\sigma$ DIBs, which are strongest in the normal diffuse ISM (with a  fraction of molecular hydrogen of f$_{\rm H2}$ $\sim$ 0.1-0.3; e.g.\ Vos et al.\ 2011, Fig.\ 20); $\zeta$ DIBs, which are strong in moderately shielded regions of the diffuse ISM (Krelowski and Walters 1987; see also Herbig 1995); and C$_2$ DIBs (Thorburn et al.\ 2003), which are stronger in more shielded regions, where the bands of the C$_2$ molecule are generally observed (see Section \ref{sec:4.1currentC2}). 

More detailed correlation studies (Fan et al.\ 2017; Ensor et al.\ 2017) have provided further insights into the properties of DIB carriers (see Section \ref{sec:2.1early}).   
Using the data of the DIB Atlas of Hobbs et al.\ (2008, 2009), Omont \& Bettinger (2020) analysed the correlation between the DIB wavelength and the apparent UV resilience (or boost) of their carriers. They  noted that this property fits  linear carriers well, such as carbon chains or polyacenes, whose series of very strong bands have wavelengths that linearly vary with the length of the molecule.

The availability of the Apache Point Observatory Catalog of Optical Diffuse Interstellar Bands (APO Catalog) of 559 DIBs in 25 sight lines with diverse interstellar properties (Fan et al.\ 2019, 2020; see details in Section \ref{sec:3.1catalog}) has given a new dimension to possible correlation studies. The first studies, such as Fan et al.\ (2022) and Smith et al.\ (2021), are briefly analysed in Section \ref{sec:2.2comprehensive}, together with other studies, notably with the ESO Diffuse Interstellar Bands Large Exploration Survey (EDIBLES; Cox et al.\ 2017; Cami et al.\ 2018).

The aim of the present paper is to further explore the richness of the APO Catalog for studying DIB correlations, with a particular focus on unexplored correlations involving weak DIBs. A first goal is to extend the census of C$_2$-like DIBs, and to further study their correlation (or anticorrelation) with longer-wavelength DIBs. Special attention is paid to all clues that may be of interest for confirming the conjecture that the carriers of C$_2$ DIBs or associated DIBs might be long carbon chains or rings. In addition to correlations, a key input for this purpose is the result of laboratory spectroscopy of long chains and rings. This mainly includes breakthrough data reported in the past from matrix spectroscopy by the Basel group (e.g. Jochnowitz \& Maier 2008a,b; Zack \& Maier 2014a,b) or recently from action spectroscopy by the Melbourne and Edinburgh groups (Buntine et al.\ 2022; Marlton et al.\ 2022, 2023, 2024; Rademacher et al.\ 2022).
 
The paper is organised as follows: Section \ref{sec:2previous} summarises the results of previous DIB correlation studies. Section \ref{sec:3APO} describes the capabilities of the APO Catalog and the method used here to explore weak-DIB correlations. These methods are applied in Section \ref{sec:4C2} to propose an extension of the C$_2$ family based on correlations with reference C$_2$ and $\zeta$ DIBs, strength ratios in shielded and unshielded sight lines, and anticorrelations with $\sigma$ DIBs, which confirms that the new members are also confined to a short wavelength, $\lambda$ $\la$ 5900\,\AA . By checking the correlations of the extended C$_2$ family, we found that a large fraction of its members are significantly correlated with a large set of longer-wavelength DIBs. Section \ref{sec:5peculiarsightlines} is devoted to the analysis of this set of poorly studied DIBs, which are enhanced in two peculiar lines of sight. We debate without a firm conclusion whether their correlation might allow us to tentatively define a new family of DIBs. 
 Section \ref{sec:6chains} summarises various possible indications that carbon chains and rings are DIB carriers, including five coincidences between wavelengths and widths of very broad strong DIBs and the strong bands of two carbon chains and one ring that were recently reported from action spectroscopy. Finally, Section \ref{sec:7conclusion} summarises the main findings of this study.

 Following Fan et al.\ (2019), the APO DIBs are generally denoted by their abbreviated wavelength in \AA\ (except when quoted otherwise). For instance, the DIB at $\lambda$ = 6613.70\,\AA\  is denoted $\lambda$6613.

\section{Previous correlation studies of diffuse interstellar bands}  \label{sec:2previous}

\subsection{Early studies of the main diffuse interstellar bands} \label{sec:2.1early}

The study of the various correlations implying strong DIBs among themselves or with reddening, atomic, or molecular column densities has a long history. It has  provided a good part of our current limited knowledge about DIB carriers. As early as 1938, Merrill and Wilson reported a correlation of the strongest DIBs with the sight-line reddening E(B-V). Much later, it was established mainly by Krelowski and coworkers (e.g. (Krelowski \& Walker 1987; Krelowski 1989; Krelowski et al.\ 1992; Fulara \& Krelowski 2000; see also Herbig 1995) that most of the strong DIBs may be classified into the  two broad $\sigma$ and $\zeta$ families. Still later, the C$_2$ family was identified (Thorburn et al.\ 2003; Ka{\'z}mierczak et al.\ 2010; Elyajouri et al.\ 2018;  Fan et al. 2024). The practical confinement of C$_2$ DIBs to short wavelengths, $\lambda$\,$<$\,5800\,\AA\,, is a striking property (see e.g.\ the list of the 16 C$_2$ DIBs given by Thorburn et al.\ 2003, reproduced in Table \ref{tab:B1thorburn}) .

More sophisticated correlations were explored among DIBs that were observed along lines of sight with diverse properties,  
including a study of the Orion sight lines with extremely strong UV radiation; diverse correlations and anti-correlations between pairs of DIBs and with E(B-V) and column densities of H or H$_2$ (Cami et al.\ 1997; Friedman et al.\ 2011; Vos et al.\ 2011; Kos \& Zwitter 2013; Baron et al.\ 2015; Xiang et al.\ 2017); a principal component analysis (Ensor et al.\ 2017); and a DIB ordering according to their sensitivity to UV destruction (Fan et  et al.\ 2017;  Elyajouri et al.\ 2018).

\subsection{Correlations of diffuse interstellar bands with comprehensive catalogues}
\label{sec:2.2comprehensive}
Correlation studies of DIBs have taken on a new dimension since the advent of comprehensive DIB catalogues that include all the DIBs that are detectable in multiple lines of sight. These catalogues have opened the possibility to extend correlation studies to hundreds of weak DIBs. The preliminary APO catalogue of Hobbs et al.\ (2008, 2009) was limited to two well-chosen lines of sight, HD\,204827, with moderate UV radiation, and HD\,183143, with relatively strong UV. In addition to the completeness of its list of $\sim$500 DIBs, which is fundamental for their identification, it provides a few parameters for each DIB, that is, the two equivalent widths (EW) and the DIB width (full width at half maximum; FWHM).  

The inferred equivalent-width ratio that compares the strength of a DIB along a low-UV ($\zeta$) and a high-UV ($\sigma$) line of sight, 
\begin{equation} \label{eq:1R21}
	{\rm R_{21} = EW(HD204827)/EW(HD183143)},
\end{equation}

gives a first view of the DIB sensitivity to UV and thus of the family ($\sigma$, $\zeta$, or C$_2$) to which the DIB belongs. It was shown by Omont \& Bettinger (2020) that R$_{21}$, and therefore, the UV resilience or boost of the DIB carrier, are significantly correlated with its wavelength. This may be a good argument for carbon chains as DIB carriers (Section \ref{sec:6chains}) because the wavelength of their strong bands is correlated with their length, and therefore, with their resilience against photo-destruction.   

The APO Catalog (Section \ref{sec:3.1catalog}) extends the survey of 559 DIBS to 25 sight lines (Table \ref{tab:A1sightlines}), providing hundreds of detections in each sight line. With sensitive data, whose quality is well characterised, it is a much more powerful tool for detailed studies of DIB correlations.  It was previously used for two discussions of the correlations of strong DIBs. Fan et al.\ (2022) used the correlation (and anticorrelation) of a sample of 54 strong DIBs for revisiting the UV-sensitivity sequence of DIB families and a possible hierarchy of DIB correlations. They introduced a new class of DIBs that are intermediate between the $\zeta$ and $\sigma$ families. This class is denoted $\zeta\sigma$ hereafter. The degree of correlation among all these $\zeta$, $\zeta\sigma$, and $\sigma$ DIBs is so high that they might belong to a single type of carriers, independently of the charge state. Smith et al.\ (2021) reviewed the case of DIB pairs presenting the highest degree of correlation. They identified a few groups of highly correlated DIBs, some of which might have the same carriers.

The new large DIB survey  EDIBLES (Cox et al.\ 2017; Cami et al.\ 2018) offers similar or better capabilities on a larger number of sight lines, including a few APO sight lines. However, its publication is at an earlier stage of availability. Its data were  used for a thorough study of the profiles of C$_2$ DIBs (Elyajouri et al.\ 2018), confirming the homogeneity of this DIB family. EDIBLES data have already provided  key extensions of DIB correlations: Smoker et al.\ (2023) have extended them to near-infrared DIBs (Joblin et al.\ 1990; Geballe et al.\ 2011; Cox et al.\ 2014; Hamano et al.\ 2022; Ebenbichler et al.\ 2022; Castellanos et al.\ 2024). Their conclusions showed that most prominent near-infrared DIBs are better correlated with visible $\sigma$ DIBs than with $\zeta$ DIBs. Ebenbichler et al.\ (2024) have extended the correlations to DIB profiles and showed tight correlations within eight DIB families.
Fan et al.\ (2024) reported a deeper insight into the correlations of C$_2$ DIBs with the C$_2$ and C$_3$ molecules. They showed that the relative enhancement of C$_2$ DIBs with respect to non-C$_2$ DIBs in denser regions of the ISM might result in good part from the weakening of non-C$_2$ DIBs there.  It is clear that much more is expected to come from EDIBLES data when complete correlation studies implying weak visible and infrared DIBs are performed.

\section{APO catalogue, and our method}
\label{sec:3APO}	

\subsection{APO catalogue}
\label{sec:3.1catalog}	

 The APO Catalog (Fan et al.\ 2019, 2020\footnote{ https://cdsarc.cds.unistra.fr/viz-bin/cat/J/ApJ/878/151}) is the main product of the DIB survey project that was  carried out since 1999 with the 3.5 telescope and the ARC echelle spectrograph (ARCES) at the Apache Point Observatory. 
 It provides the properties (wavelength, width, and equivalent width with the uncertainty) of 559 DIBs from 4259\,\AA\ to 8763\,\AA\ in  25 lines of sight of nearby O and B stars (Table \ref{tab:A1sightlines}). These sight lines were chosen with diverse interstellar properties from very low reddening and very strong UV to high-reddening regions that are well shielded from the UV. The catalogue is restricted to DIBs that were detected in at least 5 of the 25 sight lines. It is an extension of the DIB Atlas in the two sight lines of HD\,204827 and HD\,183143, which was published by Hobbs et al.\ (2008, 2009). The data of these two sight lines were examined again and are included in the new APO catalogue. The catalogue also includes 11 DIBs with a very broad width published by Sonnentrucker et al.\ (2018).  However, relatively few new DIBs were
found with respect to the 545 DIBs that were identified in the two previous papers (Hobbs et al.\ 2008, 2009).

The catalogue is available online at the Centre de Donn\'ees de Strasbourg (CDS)$^2$.  
Fan et al.\ (2019, 2020) 
provided all the details of the observations, the sight-line properties (summarised in Table \ref{tab:A1sightlines}), and data analysis. They listed the DIB equivalent widths, 
FWHM widths, 3$\sigma $ detection sensitivity (typical EW $\sim$ 1.5-3\,m\AA ), and so on. and statistics. 

The APO Catalog was recently used for the analysis of correlations among strong DIBs (Fan et al.\ 2022; Smith et al.\ 2021, 2022). Its data quality and the large number of sight lines in which these DIBs are detected allow the identification of correlations with large Pearson coefficients. In some cases, these correlation coefficients may even approach unity for relatively strong members of the $\sigma$ and $\zeta$ DIB families. As quoted by Smith et al.\ (2021), for example, this might open the search for DIBs that are carried by the same molecule. 

\subsection{Method for weak-DIB correlation studies}
\label{sec:3.2methodology}

It is obvious that extending the correlation studies to weak DIBs may bring much information about DIBs and their carriers, especially by extending the membership of classical DIB families or identifying new families or subfamilies. 
However, it is well known that an accurate estimate of weak DIBs
and even their detection 
are difficult because they are plagued by stellar lines, DIB blending and crowding, uncertain continuum placement, and so on.  Despite the care brought to DIB extraction by the APO group, the quality of the data cannot be homogeneous for the weakest DIBs. In addition, their correlations are weakened by the low signal-to-noise ratio, which is often close to the limit of S/N = 3, and mainly by the limited number of sight lines in which both weak DIBs are detected. Therefore, a rigorous extension of correlation studies to weak DIBs is difficult. Nevertheless, it is clear that these correlations including weak DIBs of the APO Catalog contain an enormous amount of information that is expected to help us make progress in the identification of DIB carriers. However, the huge number of possible correlations among 500 DIBs makes it imperative to adopt a simplified approach to explore the problem. 

Therefore, as a first step and despite the drawbacks, we adopted the most straightforward method for studying the correlations implying weak DIBs of the APO Catalog, as listed below.

- A systematic use of the plain Pearson correlation coefficients between the equivalent widths of DIBs, calculated without taking their estimated uncertainties into account.

- Following Fan et al.\ (2022), for example, two sets of Pearson coefficients were used for the correlation between the DIBs $i$ and $j$. The regular Pearson coefficients, between their  equivalent widths, EW$_i$ and EW$_j$, denoted q($ij$) here, are well adapted for highly correlated strong DIBs (e.g.\ Smith et al.\ 2021). The equivalent width normalised to reddening, 
\begin{equation}	\label{eq:2EN}
	{\rm EN_i = EW_i /E(B-V)},
\end{equation}
 gives greater weight to weak DIBs. The corresponding normalised Pearson coefficients, denoted kn($ij$), are better adapted to weak DIBs and to anticorrelations, as shown by Fan et al.\ (2022). However, the uncertainties on E(B-V) estimates for the sample sight lines also affect the accuracy of the correlation calculations. The large differences between kn($ij$) and q($ij$) are shown in Fig. \ref{fig:19C7-6010_R21chinew}. In some cases (Section 4 and Appendix \ref{app:BC2}), the mean of kn($ij$) and q($ij$),  
\begin{equation}	
	\label{eq:3knq}
	{\rm knq}(ij) =[{\rm kn}(ij)+{\rm q}(ij)]/2,
\end{equation}
is used for DIBs of intermediate strength, such as C$_2$ DIBs.

- In deriving q($ij$) or kn($ij$) values, if possible, all sight lines were used in which both DIBs $i$ and $j$ are detected, although these coefficients depend on the set of sight lines that is used. This maximised the limited number of sight lines n($ij$) where the two members of a pair of weak DIBs are both detected; nevertheless, n(ij) is often close to 7 (out of 25), or even lower.  Correlations are hardly meaningful when n($ij$) $\le$ 5. However, as described in Section \ref{sec:5peculiarsightlines}, it was discovered that the members of a peculiar set of DIBs have a peculiar behaviour in two  sight lines, HD\,175156 and HD\,148579. Keeping these lines of sight in the calculation of correlation coefficients may strongly distort their values (see examples in Appendix \ref{app:C.1tabsfigs}, Fig. \ref{fig:16C4-undeternew}).  Therefore, it may be better to discard these sight lines when calculating correlations implying these DIBs, that is, considering only those of the other 23 sight lines in which the two DIBs are detected. 
On the other hand,  HD\,175156 and HD\,148579 were kept to study special correlations that imply 
these DIBs that are enhanced there (Section \ref{sec:5peculiarsightlines} and Appendix \ref{app:Cchi}).

- No use was made of the provided APO upper limits for undetected DIBs. The caveat for contamination (DIBs quoted as ($c$) in the APO Catalog) was generally not taken into account in order to maximise the number of sight lines in which weak DIBs are detected.

-  The degree to which Pearson coefficients may be considered significant depends on the number of sight lines n($ij$) in which the two DIBs are both detected, on the nature of the coefficient, q or kn, and on the context. A value q $\sim$ 0.8 may mean a moderate correlation between strong DIBs with a high S/N ratio, especially for DIBs that are highly correlated with reddening (Friedman et al.\ 2011), while even kn $\sim$ 0.8 means a high correlation for strong DIBs and a fortiori for weak DIBs, because of the uncertainties on E(B-V). Roughly speaking, as confirmed, e.g. by the analysis of correlations implying reference C$_2$-DIBs (Section \ref{sec:4.1currentC2}), values of kn down to $\sim$0.6 and even $\sim$0.5, may reflect significant correlations, for instance for DIBs with a low correlation with reddening; values of q or kn in the range -0.2 to +0.4 mean a poor or no correlation, with little graduation for various values in this range; kn $\la$ -0.4, and even -0.3, generally reflect a significant anticorrelation (Fig. \ref{fig:3-R21_k6010new1}). 

A difficult case for quantitative DIB correlations was found for the two quoted  sight lines, HD\,175156 and HD\,148579, where the normalised equivalent widths of some DIBs are strongly enhanced compared with other sight lines (Section \ref{sec:5peculiarsightlines} and Appendix \ref{app:Cchi}). The values of coefficients kn for correlations between two of these enhanced DIBs may reach very high formal values, $\ga$ 0.9. These high values (Figs. \ref{fig:15C3-examplesnew} and \ref{fig:16C4-undeternew}) may seem meaningless because the correlation is dominated by a single line of sight and the number of other sight lines in which the two DIBs are detected is very low. However, the mere fact that the two DIBs are simultaneously enhanced may be enough to imply an important actual correlation (see Section \ref{sec:5peculiarsightlines} and Appendix \ref{app:C.2indeter}). 
 
 The Topcat tool\footnote{https://www.star.bris.ac.uk/~mbt/topcat/} (TOPCAT \& STIL: Starlink Table/VOTable Processing Software; Taylor 2005) is well fitted for studying correlations between DIBs of the APO Catalog. It is systematically used in this study.

\section{Extension of the C$_2$ diffuse interstellar band family}
\label{sec:4C2}

\subsection{Current C$_2$ diffuse interstellar bands}
\label{sec:4.1currentC2}	

The identification of a new class of DIBs whose carriers seem to survive better in regions that are very well shielded from UV radiation  
was a key point in the history of DIB studies (Thorburn et al.\ 2003; see also Ka{\'z}mierczak et al.\ 2010, 2014; Elyajouri et al.\ 2018; Ebenbichler et al.\ 2024). They  are generally seen in sight lines with substantial absorption by C$_2$ molecules in the interstellar gas, hence the name C$_2$ DIBs. However, the correlation between C$_2$ DIBs and the column density of C$_2$ molecules is loose, and a key property of C$_2$ DIBs is that their intensity does not seem to decrease in dense environments, in contrast to other regular DIBs  (Fan et al.\ 2024, 2017).

The reference sample of C$_2$ DIBs is the commonly used sample that was identified by Thorburn et al.\ (2003),   
which is broadly consistent with the samples of Elyajouri et al.\ (2018) and  Fan et al.\ (2022). It includes 16 DIBs, all at $\lambda$ $<$ 5800\,\AA , that are listed in Table \ref{tab:B1thorburn}. It excludes a few related DIBs that were added to the list by various authors, $\lambda\lambda$5245, 5547, 5769.9, 5793, 5828, 5849, 5910 and 6729 (Table \ref{tab:B2literature}). They often display a peculiar behaviour with respect to the other C$_2$ DIBs. 

The C$_2$ DIB family seems relatively homogeneous and well defined compared with the $\sigma$ and $\zeta$ families (Fan et al.\ 2017, 2022; Ensor et al.\ 2017; Elyajouri et al.\ 2018). Their band profiles are similar (Elyajouri et al.\ 2018; Ebenbichler et al.\ 2024). Their number does not seem to exceed a fraction of $\sim$10\% of the currently known DIBs (Fan et al.\ (2017). Importantly, they are confined to short wavelengths, mostly at $\lambda$ $\la$ 5900\,\AA\ (Tables \ref{tab:B1thorburn}-\ref{tab:B4related}). 
In addition, most of them are strongly anticorrelated to prominent $\sigma$ DIBs, and they display high values for equivalent width ratios in the $\zeta$ to $\sigma$ sight lines (Eq.\ \ref{eq:1R21}, see below, Figs. \ref{fig:1-kq_k6009new}-\ref{fig:3-R21_k6010new1}). In the following, the search for new C$_2$ DIBs was therefore performed first. We started by exploring the APO Catalog for all DIBs that are significantly correlated with well-established classical C$_2$ DIBs and share their key properties. 


\begin{figure}
	\begin{center}
		\includegraphics[scale=0.66, angle=0]{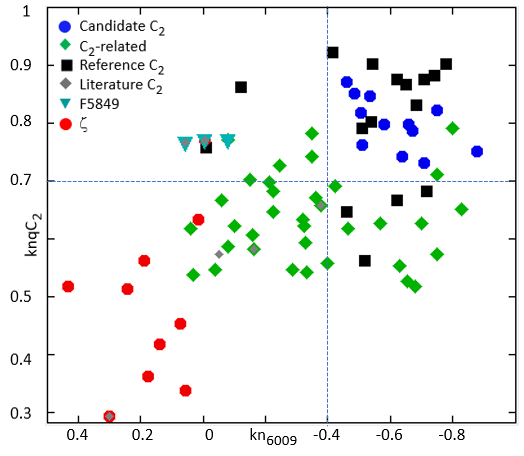}
		\caption{Correlation  diagram for candidate C$_2$ DIBs (blue dots, Table \ref{tab:B3robust}) and possible C$_2$-related DIBs (green diamonds, Table \ref{tab:B4related}) compared to reference C$_2$ DIBs (black squares, Table \ref{tab:B1thorburn}), candidate C$_2$ DIBs from the literature (grey diamonds, Table \ref{tab:B2literature}), possible DIBs similar to $\lambda$5849 (F5849, cyan triangles), and to a comparison sample of $\zeta$ DIBs (Fan et al.\ 2022, red dots). 
		The horizontal axis displays the normalised anticorrelation Pearson coefficient, kn$_{6009}$, with the $\sigma$ DIB $\lambda$6009 (excluding the sight lines of HD\,175156 and HD\,148579). The vertical axis displays the average correlation coefficient, knq$_{C2}$ (Eq.\ \ref{eq:3knq}), with a subsample of six reference C$_2$ DIBs ($\lambda\lambda$4726, 4963, 4984, 5418, 5512, and 5546). The selection criteria (Eqs.\ [\ref{eq:4knqC2},\ref{eq:5kn6009C2}]; dotted blue lines) distinguish C$_2$  DIBs well from $\zeta$ DIBs. Most  C$_2$  DIBs and related DIBs are significantly anticorrelated with $\sigma$ DIBs (kn$_{6009}$ $<$ 0).  See Fig. \ref{fig:10B1-kq_k6009longnew} for an extended version of this figure, including the DIB labels.}
		\label{fig:1-kq_k6009new}
	\end{center} 
\end{figure}

\begin{figure}
	\begin{center}
		\includegraphics[scale=0.63, angle=0]{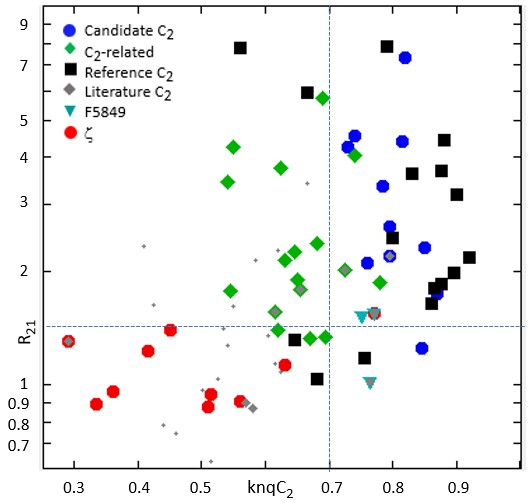}
		\caption{Same definitions as for the symbols in Fig. \ref{fig:1-kq_k6009new} for the correlation  diagram of candidate C$_2$ DIBs and possible C$_2$-related DIBs.  
		The horizontal axis displays the average Pearson correlation factor, knq$_{C2}$ (Eq. \ref{eq:3knq} and Fig. \ref{fig:1-kq_k6009new}). 
		The vertical axis displays the ratio of the DIB equivalent widths in the sight line of HD\,204827 to HD\,183143 (Eq.\ \ref{eq:1R21}; actual values or equivalent values, see Appendix \ref{app:B.3substitute}). The selection criteria (Eqs.\ [\ref{eq:4knqC2},\ref{eq:6R21C2}]; dotted blue lines) distinguish C$_2$  DIBs well from $\zeta$ DIBs. Most C$_2$ DIBs and related DIBs have R$_{21}$ $\ge$ 1.5. See Fig. \ref{fig:11B2-R21_knqlongnew} for an extended version of this figure, including DIB labels. }
		 \label{fig:2-R21_knqnew}		
	\end{center}
\end{figure}

\begin{figure}
	\begin{center} 
		\includegraphics[scale=0.65, angle=0]{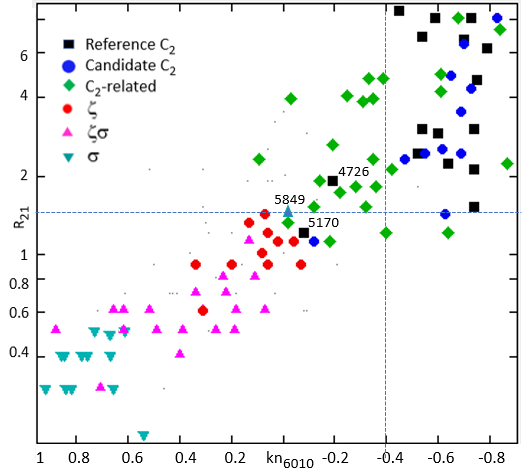}
		\caption{Correlation diagram for C$_2$ DIBs with the  same symbols and definitions as in Figs. \ref{fig:1-kq_k6009new}-\ref{fig:2-R21_knqnew}, where the horizontal axis displays the normalised correlation factor, kn$_{6010}$ (equivalent to kn$_{6009}$, Appendix \ref{app:B.3substitute}), and the vertical axis displays the ratio of the DIB equivalent widths in the  line of sight HD\,204827 to HD\,183143 (Eq.\ \ref{eq:1R21}). Reference  $\zeta\sigma$ and $\sigma$ DIBs of Fan et al.\ (2022) have been added to show the whole range of variation of kn$_{6010}$ and R$_{21}$.  The selection criteria (Eqs.\ (\ref{eq:5kn6009C2},\ref{eq:6R21C2}); dotted blue lines) distinguish C$_2$  DIBs well from $\zeta$ DIBs. The intermediate position of C$_2$-related DIBs (Table \ref{tab:B4related}, green diamonds) between C$_2$ and $\zeta$ DIBs, is clearly visible, as is the peculiar position of two reference DIBs, $\lambda\lambda$4726, 5170, and of $\lambda$5849 at the boundary of $\zeta$ DIBs in Fig. \ref{fig:2-R21_knqnew} of Fan et al.\ (2022).}
		 \label{fig:3-R21_k6010new1}
	\end{center}
\end{figure}

\begin{figure}
	\begin{center}
		\includegraphics[scale=0.88, angle=0]{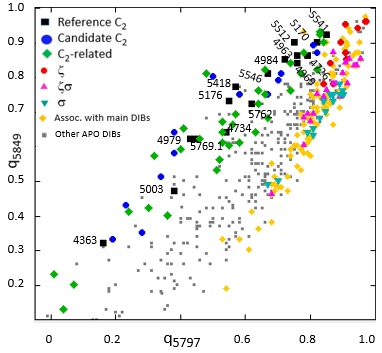}
		\caption{Regular correlation coefficients of all APO DIBs with $\lambda$5849 vs. $\lambda$5797 (excluding the sight lines of HD\,175156 and HD\,148579). Reference C$_2$, candidate C$_2$, and C$_2$-related DIBs span a sequence distinct from the sequence spanned by the 47 $\zeta$, $\zeta\sigma$, and $\sigma$ main DIBs of Fan et al.\ (2022), and 106 other DIBs (orange dots, Table \ref{tab:D1mainass}) that are strongly correlated with these main DIBs (Appendix \ref{app:D.1associated}).  Other APO DIBs are plotted as small grey squares}
		 \label{fig:4-q5797_5849new}
	\end{center} 
\end{figure}

\subsection{New C$_2$ diffuse interstellar bands}
\label{sec:4.2newC2}	

The main criterion for identifying new C$_2$ DIBs among weak DIBs is their high correlation with the classical C$_2$ DIBs in the reference sample of Thorburn et al.\ (2003). We assume below that this is well quantified by computing the average Pearson  coefficients  over the panel of the six C$_2$ DIBs with the best APO data used by Fan et al.\ (2022, see their Fig.\ B2), $\lambda$$\lambda$4726, 4963, 4984, 5418, 5512, and 5546, which we call the $C_2F22$ sample below. 
Because of the anomalous intensities of some DIBs in the sight lines of HD\,175156 and HD\,148579 (Section \ref{sec:5peculiarsightlines}), it seems more consistent to compute these correlation coefficients without taking these two lines of sight into account. 

These average normalised and regular correlation coefficients, kn$_{C2}$ and q$_{C2}$, are presented in Tables \ref{tab:B1thorburn}-\ref{tab:B4related} for the best C$_2$ DIB candidates, together with their mean, knq$_{C2}$ (Eq.\ \ref{eq:3knq}). The  latter seems best suited for the range of intensities of C$_2$-like DIBs, from strong reference C$_2$ DIBs to the weakest C$_2$ DIB candidates.
Most of the reference C$_2$ DIBs appear to be highly correlated with the $C_2F22$ sample, with knq$_{C2}$ $\ge$ 0.8 for 12 out of 16 of them, including the whole $C_{2}F22$ sample itself. 
 However, the four others, $\lambda\lambda$5003, 5541, 5170, and 4363, are moderately correlated with the reference $C_2F22$ sample, with knq$_{C2}\,\la$\,0.65.  
 
 A well-known property of classical C$_2$ DIBs is their anticorrelation with  $\sigma$ DIBs, which reflects the opposite behaviour of these two classes of DIBs with respect to the UV intensity (e.g.\ Fan et al.\ 2022). Tables \ref{tab:B1thorburn}-\ref{tab:B4related} give the normalised correlation coefficient  kn$_{6009}$ of the 16 reference DIBs and candidate C$_2$ DIBs  with the most extreme  $\sigma$ DIB, $\lambda$6009. Figs. \ref{fig:1-kq_k6009new} and \ref{fig:10B1-kq_k6009longnew} display the  correlation coefficient kn$_{6009}$ $vs$  knq$_{C2}$ for the 16 reference C$_2$ DIBs, the new candidate C$_2$ DIBs, and a comparison sample of $\zeta$ DIBs. The vast majority of the reference DIBs (14/16) are strongly anticorrelated with $\lambda$6009, with -0.8\,$<$\,kn$_{6009}\,<$\,-0.45.  
 Of the two exceptions, the case of $\lambda$4726 is known to be peculiar because it is much stronger and broader than all regular C$_2$ DIBs. The case of $\lambda$5170 is more surprising (Fig. \ref{fig:3-R21_k6010new1}, Appendix \ref{app:B.2groups}).
 
 Another traditional way of identifying C$_2$ DIBs is based on the DIB ratio, R$_{21}$, in a shielded sight line, such as HD\,204827, to an unshielded sight line, such as HD\,183143 (Eq.\ \ref{eq:1R21}). Fig. \ref{fig:2-R21_knqnew} displays  the APO value of R$_{21}$  
 or its equivalent (Appendix \ref{app:B.3substitute}) versus knq$_{C2}$. A comparison with the reference C$_2$ DIBs (Fig. \ref{fig:2-R21_knqnew} and Fig. \ref{fig:11B2-R21_knqlongnew}) shows that the combination of knq$_{C2}\,>$\,0.7 and a ratio R$_{21}$\,$>$\,1.5 clearly indicates a C$_2$ DIB. 

To summarise, 14 out of 16 reference C$_2$ DIBs meet the following three conditions:
\begin{equation}	
	\label{eq:4knqC2}  
	{\rm knq_{C2} \ga 0.70},
\end{equation}
\begin{equation}	
	\label{eq:5kn6009C2}
	{\rm kn_{6009}} \la-0.4, and
\end{equation}
\begin{equation}	
	\label{eq:6R21C2}	
	{\rm R_{21}} \ga 1.5	.
\end{equation}
Figures 1-3 show that these conditions distinguish  $\zeta$ DIBs well. It seems logical to exploit this specificity of C$_2$ DIBs to identify new C$_2$ DIB candidates. The three conditions of Eqs.\ (\ref{eq:4knqC2}-\ref{eq:6R21C2}) were used for this purpose in the following way: 
 
$\bullet$ Solid  C$_2$ DIB candidates must be significantly correlated with the basic panel of 6 prominent C$_2$ DIBs, $C_{2}F22$ (Table \ref{tab:B1thorburn}), as closely as the 16 classical C$_2$ DIBs. Therefore, they must strictly verify Eq.\ (\ref{eq:4knqC2}). The two other conditions, Eqs.\ (\ref{eq:5kn6009C2},\ref{eq:6R21C2}), are also required, at least approximately. In addition to the 16 reference C$_2$, 12  other DIBs meet these three conditions. They are listed in Table \ref{tab:B3robust}. All of them verify ${\rm knq_{C2}} > 0.65$, kn$_{6009} \le$ -0.46 and R$_{21} \ga$ 1.5 (Figs. \ref{fig:1-kq_k6009new}-\ref{fig:3-R21_k6010new1}).  
 \smallskip
 
$\bullet$ These conditions may be relaxed for additional  C$_2$-related DIB candidates. For instance, Table \ref{tab:B4related} lists 34 DIBs verifying ${\rm knq_{C2}} > 0.50$, kn$_{6009} <$ 0.05, and R$_{21} \ge$ 1.1.  Figs. \ref{fig:1-kq_k6009new}-\ref{fig:3-R21_k6010new1} show that these DIBs are located between the groups of reference C$_2$ DIBs and reference $\zeta$ DIBs from Fan et al.\ (2022). This intermediate position of the group of possible C$_2$-related DIBs is especially illustrated in the correlation diagram of Fig. \ref{fig:3-R21_k6010new1} that displays R$_{21}$ versus kn$_{6010}$ (equivalent to kn$_{6009}$, Appendix \ref{app:B.3substitute}). The extension of this diagram to the whole range, -0.9 $<$ kn$_{6010} <$ 1.0, shows the whole series of decaying or boosting UV-sensitivity of DIB families from C$_2$ to $\sigma$ through C$_2$-related, $\zeta$, and $\zeta\sigma$ DIBs.  
    \smallskip
    
However, the extension of the selection to such a large number of weak DIBs implies that correlations with limited statistical quality are included. An additional criterion was added to mitigate this risk and strengthen the selection of C$_2$-related DIBs. The diagram of the correlation with $\lambda$5849 versus $\lambda$5797 (Fig. \ref{fig:4-q5797_5849new}) shows that the reference C$_2$ DIBs follow a well-defined sequence very different  from the sequence spanned by the  $\zeta$, $\zeta\sigma$, and $\sigma$ main DIBs of Fan et al.\ (2022) and  by other not-C$_2$ DIBs that are strongly correlated with these main DIBs (Table \ref{tab:D1mainass}). It was therefore imposed that the candidates of Tables \ref{tab:B3robust} and \ref{tab:B4related} also pertain to the 5849-5797 C$_2$ sequence, in addition to meeting the conditions of Eqs.\ (\ref{eq:4knqC2}-\ref{eq:6R21C2}). Fig. \ref{fig:4-q5797_5849new} shows that the 12 solid C$_2$ candidates of Table \ref{tab:B3robust} strictly follow the C$_2$ sequence. An additional candidate that meets Eqs.\ (\ref{eq:4knqC2}-\ref{eq:6R21C2}) was discarded because it was located within the main-DIB sequence of Fig. \ref{fig:4-q5797_5849new}. Similarly, only DIBs that approximately followed the C$_2$ sequence were kept as C$_2$-related candidates listed in Table \ref{tab:B4related}. 
  
 Despite the strength of these correlation criteria, it is clear that all these new C$_2$ DIB candidates need some confirmation before they can definitely be considered as belonging to the C$_2$ DIB family or to a transition family bridging the gap between the C$_2$ and $\zeta$ families. 
  We might search for a confirmation like this through further correlations implying better data and additional sight lines, correlations with the bands of the interstellar molecule C$_2$, similar to Thorburn et al.\ (2003), or DIB-profile analyses extending the work of Elyajouri et al.\ (2018) and Ebenbichler et al.\ (2024). 
   
 As listed in Table \ref{tab:B5C2stat}, these 12 solid C$_2$ candidates display similar average properties as the group of highly correlated reference C$_2$ DIBs, as regards the degree of correlation with reference C$_2$ DIBs and anticorrelation with $\sigma$ DIBs. However, their average intensity is only about half that of the reference C$_2$ DIBs, which may explain why they were more difficult to identify. Similarly, Table \ref{tab:B5C2stat} shows that the possible C$_2$-related DIBs are also much weaker than the highly correlated reference C$_2$ DIBs and reference $\zeta$ DIBs.
  
Finally, practically all new C$_2$ DIB candidates and possible C$_2$-related DIBs (and reference C$_2$ DIBs) meet two key properties of C$_2$ DIBs (see Tables \ref{tab:B1thorburn}-\ref{tab:B4related}): 1) most are narrow (FWHM $<$ 0.9\,\AA , except for $\lambda$4699, 4726, 4987, 5262, and 6365, whose widths lie in the range 1.3-2.8\,\AA ), and 2) most are confined to short wavelengths ($\lambda$\,$\le$5945\,\AA , except for $\lambda\lambda$6093, 6365, 6729, and 6736).
  
It is remarkable that the whole extended C$_2$ family remains  well   distinct from the $\zeta$ family in the diagrams of  Figs. \ref{fig:1-kq_k6009new}-\ref{fig:3-R21_k6010new1}. However, the possible intermediate case of $\lambda$5849 and $\lambda$5828 is discussed in Appendix \ref{app:B.2groups}, and the tentative C$_2$-related DIBs seem to partially fill the gap between the  $\zeta$ and C$_2$  families (Fig. \ref{fig:3-R21_k6010new1}) as was noted by Fan et al.\ (2022), for example. On the other hand, the extended C$_2$ family strongly overlaps the possible  $\chi$ DIBs (Section \ref{sec:5peculiarsightlines}, Tables \ref{tab:C1chia}-\ref{tab:C2chib}), including reference C$_2$ DIBs, such as $\lambda\lambda$4726, 4984, 5512, 5541, 5546, and 5769.09. This is discussed in Section \ref{sec:5peculiarsightlines}.

Figs. \ref{fig:1-kq_k6009new}-\ref{fig:3-R21_k6010new1} and Figs. \ref{fig:10B1-kq_k6009longnew}-\ref{fig:11B2-R21_knqlongnew} show that the detailed behaviour of some of the 16 reference C$_2$ DIBs is
 diverse. The range of anticorrelation with $\sigma$ DIBs traced by kn$_{6009}$ extends from ${-0.8}$ to
0.0 and that of the correlation with the $\zeta$ DIB $\lambda$5797 q$_{5797}$ extends from 0.2 to 0.8, without a tight correlation between these two properties. The most extreme case is $\lambda$4363. It has the lowest correlation, knq$_{C2}$,  with the reference $C_{2}F22$ sample and the lowest correlation with  $\lambda$5797, while its UV sensitivity, traced by kn$_{6009}$ and R$_{21}$, is as high as normal. The DIBs $\lambda$4726 and $\lambda$5170 have values of kn$_{6010}$ and q$_{5797}$ close to $\zeta$ DIBs (Fig. \ref{fig:3-R21_k6010new1}), but   values of knq$_{C2}$ much higher than $\zeta$ DIBs. DIB $\lambda$5541 is also close to $\zeta$ DIBs for knq$_{C2}$ and q$_{5797}$, but not for kn$_{6009}$. Another group of most typical reference C$_2$ DIBs, such as $\lambda\lambda$4963, 4969, 4984, and 5512, has high correlation rates with $\lambda$5797 and $\lambda$5849.
  	
To summarise, adding these 12 solid C$_2$ candidates and 34 possible C$_2$-related DIBs (Tables \ref{tab:B3robust} and \ref{tab:B4related}) to the 16 reference C$_2$ DIBs (Table \ref{tab:B1thorburn}), plus $\lambda\lambda$5828, 5849, and 5769.9, yields a total number of 65 potential C$_2$ DIBs, 30 of which may be considered to be confirmed, including two  transition DIBs between C$_2$ and $\zeta$, $\lambda\lambda$5828 and 5849, in addition to 16 reference DIBs and 12 solid C$_2$ candidates. Although a large fraction of  C$_2$-related DIBs may be members of an intermediate family between C$_2$ and $\zeta$, rather than actual C$_2$ DIBs, this may be compared with $\sim$20 suggested C$_2$ DIBs before this work, 18 of which were confirmed. It is certain that a number of additional APO (weak) C$_2$-related DIBs, possibly $\sim$10-20\%, are still missed. 
Therefore, the number of C$_2$ candidates proposed here agrees with the statement of Fan et al.\ (2017) that 
the class of C$_2$ DIBs includes fewer than 10\% of the currently known DIBs\footnote{Despite the announcement in Fan et al.\ (2017), it seems that the justification of this statement was never published.}. However, the fraction of possible APO DIBs that are identified as C$_2$-like in Tables (\ref{tab:B1thorburn}-\ref{tab:B4related})  is higher at short wavelengths, namely, 53\%, 36\%, and 21\% in the ranges $\lambda <$ 5000\,\AA , 5000-5500\,\AA\, and 5500-6000\,\AA , respectively, and only 1\% at $\lambda >$ 6000\,\AA .

Although no C$_2$ DIB carrier has yet been identified, it is thought that this well-defined DIB family may present special clues for an 
identification of DIB carriers. Its limitation at short wavelengths,  $\lambda <$ 6000\,\AA\ (Fig. \ref{fig:12B3-lambda_allC2}), may be due to a high sensitivity of their carriers to UV photo-dissociation (e.g.\ Omont \& Bettinger 2020), and it might indicate neutral or anionic long carbon chains or rings as carriers (Section \ref{sec:6.2chainsrings}). It is remarkable that this limitation to $\lambda <$ 6000\,\AA\ is preserved by more than tripling the number of possible C$_2$ candidates.  

\begin{figure}
	\begin{center}
		\includegraphics[scale=0.55, angle=270]{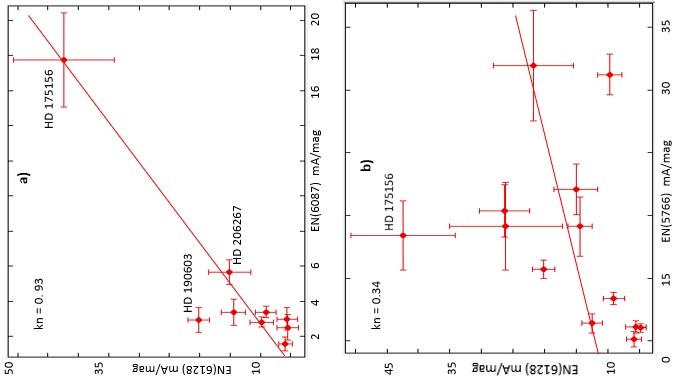}
		\caption{Examples of the  distribution of the normalised equivalent width (EN) of DIBs detected in various APO sight lines (with correlation fits and normalised Pearson coefficients kn) for three DIBs, including i) $(a)$ two DIBs, $\lambda$6128 and $\lambda$6087, that are enhanced in the  sight line of HD\,175156, compared with all other sight lines [R$_{\rm av}$(175156)\,=\,3.0 and 3.9, $\eta_{175}$\,=\,4.1 and 7, respectively, Eqs.\ (\ref{eq:7Rav},\ref{eq:8eta}), Table \ref{tab:C1chia}]; and ii) $(b)$ a more normal DIB, $\lambda$5766, classified as $\zeta$-type (Fan et al.\ 2022); the intensity of $\lambda$5766 in HD\,175156, EN(5766),  remains comparable with or lower than in  some other sight lines (R$_{\rm av}$(175156)\,=\,1.2, $\eta_{175}$\,=\,0.40). EN(6128) in HD\,175156 is out of the correlation fit by a factor $\sim$3.
		 The errors bars for EN$_i$ = EW$_i$/E(B-V) are built from the APO uncertainties on the EW and a systematic representative uncertainty on E(B-V) equal to 0.03\,mag.}
		 \label{fig:5-6128-6087-5766new}
	\end{center}
\end{figure}

\section{Set of  diffuse interstellar bands that are enhanced in peculiar sight lines}   
\label{sec:5peculiarsightlines}

\begin{figure}
	\begin{center}
		\includegraphics[scale=0.7, angle=270]{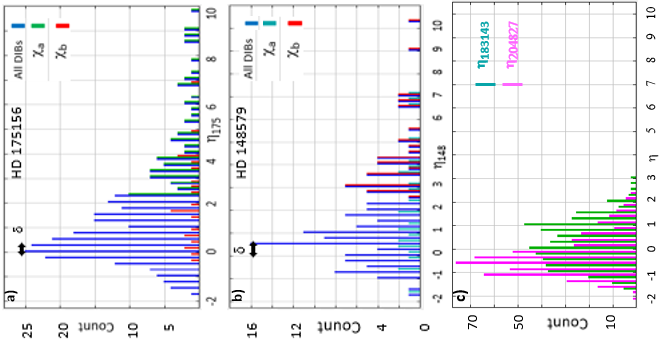}
		\caption{Distribution of the reduced enhancement $\eta_{175}$ (resp. $\eta_{148}$) (Eq.\ \ref{eq:8eta}) of the DIB normalised equivalent width, EN, of all DIBs, $\chi_a$ DIBs and $\chi_b$ DIBs, in the sight line of HD\,175156 (resp. HD\,148579). Both lines of sight have four DIBs with formal values of $\eta$\,$>$10.5 (Table \ref{tab:C2chib} and Fig. \ref{fig:9A1-histos_etanew1}). In panel {\bf c)}, these distributions are compared with more normal EN distributions in HD\,183143 and HD\,204827, (see also Fig. \ref{fig:9A1-histos_etanew1}). The shift $\delta$ of the central roughly Gaussian distribution from EN$_{\rm av}$ (Table \ref{tab:A1sightlines}) is  also displayed.
		}  
		 \label{fig:6-etanew1}
	\end{center}
\end{figure}

Generally, many lines of sight significantly contribute to the determination of the correlation coefficient between two strong DIBs. However, it was discovered that the correlations within a peculiar large set of DIBs may be dominated by a single APO sight line.
This section explores the properties of this set of DIBs and discusses the origin of their enhancement in peculiar sight lines and the nature of their carriers. 

\subsection{Strongly enhanced diffuse interstellar bands in HD\,175156 and HD\,148579}
\label{sec:5.1HD175-148}	
When we systematically searched for correlations  of C$_2$ DIBs with all other DIBs, we found that some poorly studied DIBs in various wavelength ranges displayed a high correlation with part of the C$_2$ DIBs.  
Moreover, these high correlations are dominated by substantial enhancements of these DIBs in some specific lines of sight, mostly HD\,175156 (see e.g. Fig. \ref{fig:5-6128-6087-5766new}) and HD\,148579 (see Fig. \ref{fig:14C2-6845_5862_5780new}). These two APO sight lines appear to have been little studied outside of the APO data (however, see Vos et al.\ 2011 and Krelowski et al.\ 2021 for  HD\,148579). This may explain that this enhancement was not reported previously.

The enhancement of DIB $i$ in sight line $s$ may be quantified by defining the ratio  
\begin{equation}	\label{eq:7Rav}	
	{\rm R_{av}}^i(s) = {\rm EN}_s^i /{\rm EN}_{\rm av}^i
\end{equation}
for each DIB, where EN$_{\rm av}^i$ is the average of EN$_s^i$ (Eq.\ \ref{eq:2EN}) over the sight lines in which the DIB $i$ is detected, or, better,  by the reduced variable 
\begin{equation}	\label{eq:8eta}
	\eta^i(s) = ({\rm EN}_s^i - {\rm EN}_{\rm av}^i)/\sigma_{\rm EN}(s),
\end{equation}
where $\sigma_{\rm EN}(s)$ is the standard deviation of the EN$_s^i$ distribution. Examples of striking enhancements in the sight line of HD\,175156 (HD\,148579) are displayed in Fig. \ref{fig:5-6128-6087-5766new}a (Fig. \ref{fig:14C2-6845_5862_5780new}a).
 
In order to analyse the case of these two exceptional sight lines, the distributions of $\eta^i(s)$ over $i$ for the 25 APO sight lines are investigated in Appendix \ref{app:A.2:eta} (see Fig. \ref{fig:9A1-histos_etanew1} and Table \ref{tab:A1sightlines}). It confirms that the distribution over $i$ of EN$^i_s$ (or $\eta^i(s)$) is approximately normal (roughly Gaussian) for most sight lines $s$ except for three, namely HD\,175156, HD\,148579, and HD\,23512. For these three sight lines, the number of substantially enhanced DIBs, for example, exceeding EN$_{\rm av}^i$ by more than 2$\sigma$, is much greater than a few percent. It reaches 27\%, 30\%, and 16\%, respectively, which corresponds to 79, 43, and 19 DIBs, respectively. The case of HD\,23512 is intermediate; the dozen DIBs in excess beyond 2$\sigma$, including $\lambda\lambda$5599, 5600, and 4259, does not allow a detailed study in this sight line. It would need a better detection rate than the APO Catalog (21\%, Table \ref{tab:A1sightlines}). 

The distribution of the values of  $\eta^i$(175156) and $\eta^i$(148579) for all APO DIBs is provided in Figs. \ref{fig:6-etanew1}a and \ref{fig:6-etanew1}b, respectively, with that of $\eta^i$(183143) and $\eta^i$(204827) for comparison in Fig. \ref{fig:6-etanew1}c. For HD\,175156 and HD\,148579, a substantial number of DIBs are in excess above $\eta$\,$\ga$\,2 (see Table \ref{tab:A1sightlines}). Practically all corresponding ratios R$_{\rm av}$(175156) and R$_{\rm av}$(148579) are greater than 2 (Tables \ref{tab:C1chia}-\ref{tab:C2chib},  Fig. \ref{fig:13C1-r175-148avnew}), meaning that the sight line has a dominant  weight in the correlations of these DIBs. Examples of these correlation diagrams for these two sight lines are displayed in  Figs. \ref{fig:5-6128-6087-5766new} and \ref{fig:14C2-6845_5862_5780new}, respectively.  It is not surprising that the correlation coefficient of two DIBs with high R$_{\rm av}$ values may reach high values up to kn\,$\ga$\,0.9 (Figs. \ref{fig:5-6128-6087-5766new}a and \ref{fig:14C2-6845_5862_5780new}a). On the other hand, kn remains generally low when one DIB has a low $\eta$ value (Figs. \ref{fig:5-6128-6087-5766new}b and \ref{fig:14C2-6845_5862_5780new}b). 

However, in extreme cases, where there is no significant correlation between the other sight lines (Fig. \ref{fig:15C3-examplesnew}), the formally high values of the Pearson correlation coefficients lose their usual meaning. With the limited number  of APO detections in these sight lines, especially HD\,148579 (Table \ref{tab:A1sightlines}), this is a major difficulty for exploiting these data and the information that they may hold about DIB carriers. This problem is discussed in Appendix \ref{app:C.2indeter}, where an empirical method is proposed for roughly estimating lower values for the correlation coefficients. 

The uncertainty of these estimates prevents us from answering the question whether this collection of enhanced DIBs might define a new DIB family. Nevertheless, we propose to call them "$\chi$" DIBs for the time being, for the purpose of investigating their properties and their connection with classical DIB families, as done in Section \ref{sec:5.2chiproperties}. More precisely, those that are enhanced by more than 2$\sigma$ in HD\,175156 (HD\,148579) are called $\chi_a$ ($\chi_b$). The 79 $\chi_a$ DIBs are listed in Table \ref{tab:C1chia}  and the 43 $\chi_b$ DIBs in Table \ref{tab:C2chib}. Eleven DIBs are both $\chi_a$ and $\chi_b$, namely, $\lambda\lambda$ 4975.97, 5512, 5706, 5922, 6093, 6102, 6103, 6468, 6755, 6862, and 7470. Such a high proportion of $\chi_b$ DIBs that are also $\chi_a$ and the similarity of many properties justify discussing $\chi_a$ and $\chi_b$ DIBs together as $\chi$ DIBs. $\chi_a$ DIBs are more numerous than $\chi_b$. A possible cause of this difference is the poor quality of APO observations of HD\,148579 at $\lambda$\,$\ga$\,6500\,\AA , where very few DIBs are detected.  

	Most of the 111 $\chi$ DIBs (Tables \ref{tab:C1chia}-\ref{tab:C2chib}) are relatively weak. The distribution of their average normalised equivalent width, EN$_{\rm av}$, over all sight lines (except for HD\,175156 and HD\,148579) is plotted in Fig. \ref{fig:17C5-ENav_FWHM}b and discussed in Appendix \ref{app:C.3prominent}.
	For more than 75\% of them, EN$_{\rm av}$ is lower than  10\,m\AA /mag. However, the $\chi_a$ ($\chi_b$) DIBs are significantly stronger in  HD\,175156 (HD\,148579), with typical EN\,$\ga$\,10\,m\AA /mag (Tables \ref{tab:C1chia}-\ref{tab:C2chib} and Fig. \ref{fig:17C5-ENav_FWHM}a). The few peculiar stronger $\chi$ DIBs are discussed in Appendix \ref{app:C.3prominent}. The most remarkable ones are $\lambda$4726 (a well-known very strong peculiar C$_2$ DIB), $\lambda$6591,  $\lambda$4439 (a $\zeta$ DIB), and $\lambda\lambda$6412, 5419, and 
	6128 (see Section \ref{sec:6.1coincidences}).  The distribution of the widths of the $\chi$ DIBs is displayed in Fig. \ref{fig:17C5-ENav_FWHM}c. It is comparable to that of all DIBs. Most $\chi_a$ and $\chi_b$ DIBs ($\sim$80\%) are narrow, with an FWHM\,$\la$\,1\,\AA . Only five are broader than 2\,\AA\ ($\lambda\lambda$4726, 5419, 5525, 6128, and 6412); their detection is more difficult, so that they need to be strong.

\subsection{Properties of enhanced diffuse interstellar bands in HD\,175156 and HD\,148579}
\label{sec:5.2chiproperties}

A first question is the degree of correlation of $\chi$ DIBs within each of the  $\chi_a$ and $\chi_b$ classes and between the two classes. This question is difficult to answer because most of the correlation diagrams between any two $\chi$ DIBs are somewhat similar to those of Figs. \ref{fig:15C3-examplesnew}  and \ref{fig:16C4-undeternew}, for which the formal values of the Pearson correlation coefficients are meaningless. The approximate estimated lower values of the correlation coefficients proposed in Appendix \ref{app:C.2indeter} are systematically used. However, they remain very uncertain. 

A few DIBs were first identified within each of the  $\chi_a$ and $\chi_b$ classes, which appear to be highly correlated with most other DIBs of their class. For all $\chi$ DIBs, Tables \ref{tab:C1chia}-\ref{tab:C2chib} report the average of these lower limits of the correlation coefficients (kn$_{175}$ and kn$_{148}$) over three such DIBs in each class, namely: $\lambda\lambda$6087, 6093, and 6128 for $\chi_a$ DIBs (Table \ref{tab:C1chia}, Eq.\ \ref{eq:C2kn175}) and $\lambda\lambda$5257, 6142, and 6485 for $\chi_b$ DIBs (Table \ref{tab:C2chib}, Eq.\ \ref{eq:C3kn148}). The histograms of Figs. \ref{fig:18C6-kn3}a, and \ref{fig:18C6-kn3}b show that these average limits, kn$_{175}$ and kn$_{148}$, are significantly high ($\ge$0.5) for a large fraction ($\ga$75\%) of the DIBs within each $\chi$  class (see Appendix \ref{app:C.2indeter}). However, the correlations with the other $\chi$ class are lower. 

Interestingly, the $\chi$ DIBs include three very broad DIBs that might match the strong bands of two long carbon chains and a ring that were recently measured by action spectroscopy (Marlton et al.\ 2022, 2023, 2024), namely $\lambda\lambda$6128, 6412, and 5419, which match the ring C$_{14}^+$ and the chains C$_{17}$H$^+$ and HC$_{11}$H$^+$, respectively (see Section \ref{sec:6.1coincidences}). Tables \ref{tab:C1chia}-\ref{tab:C2chib} report the average lower limits of the correlation coefficients (kn$_{\rm chr}$) of all $\chi$ DIBs with these three DIBs (Eq.\ \ref{eq:C1knchr}). The corresponding histograms are displayed in Fig. \ref{fig:18C6-kn3}c. The correlations are again substantial for these three DIBs with most of the $\chi_a$ and $\chi_b$ DIBs. 

Although correlations between the majority of $\chi$ DIBs appear to be substantial despite their simplified definition, the large uncertainty about the correlation coefficients seems to currently prevent the definition of a new DIB family. Two main problems seem difficult to overcome: i) The search for additional strongly correlated $\chi$ DIBs should be extended to lower values of $\eta_{175}$ and  $\eta_{148}$, 
$<$2; ii) the current definition of $\chi$ DIBs implies a large overlap with C$_2$ DIBs and with $\zeta$, $\zeta\sigma$, and $\sigma$ main DIBs, which seems difficult to reconcile with a consistent new family definition.

Among the 111 $\chi$ DIBs, Tables \ref{tab:C1chia}-\ref{tab:C2chib} identify a total of 27 DIBs that also belong to reference C$_2$ DIBs (6), candidate C$_2$ DIBs (3), or C$_2$-related DIBs (18). This is about 40\% of all 65 C$_2$-like DIBs, and more than half the subclass of 34 C$_2$-related DIBs (Table \ref{tab:B4related}). Reciprocally, Tables \ref{tab:C1chia} and \ref{tab:C2chib} show that about half $\chi_a$ and $\chi_b$ DIBs with $\lambda$\,$<$\,5950\,\AA\  are C$_2$ associated. This is also obvious in the various diagrams characterising the C$_2$-like DIBs and their UV sensitivity, such as R$_{21}$/kn$_{6010}$  (Fig. \ref{fig:19C7-6010_R21chinew}) and  q$_{5849}$/q$_{5797}$ (Fig. \ref{fig:20C8-5797_5849chinew}).

More generally, these two diagrams give an idea of the diversity of the properties of $\chi$ DIBs and their UV sensitivity. The values of the correlation (or anticorrelation) coefficient kn$_{6009}$ (and kn$_{6011}$, see Appendix \ref{app:B.3substitute}) extend over practically the whole possible range from -1 to +1 (Fig. \ref{fig:19C7-6010_R21chinew}). Similarly, in the diagram q$_{5849}$/q$_{5797}$ (Fig. \ref{fig:20C8-5797_5849chinew}), the distribution of the $\chi_a$ and $\chi_b$ DIBs not only strongly overlaps the sequence of C$_2$-associated DIBs, but also the sequence of classical $\zeta$, $\zeta\sigma$ and $\sigma$ main DIBs and their associates, and it covers the whole area between the two sequences. More quantitatively, the connection of the $\chi$ DIBs with the main classical DIBs is also reflected in the significant fraction of high values of their correlation coefficients q with representative 
classical main DIBs reported in Table \ref{tab:C3chimain}. This fraction is high for $\zeta$, $\zeta\sigma$, and $\sigma$ DIBs, such as $\lambda\lambda$5797, 6196, 6613, 6270, and 5780, where it reaches half of the $\chi$ DIBs for q\,$>$\,0.70, $\sim$20\% for q\,$>$\,0.85, and $\sim$10\% for q\,$>$\,0.90. It is lower for the extreme $\sigma$ DIB, $\lambda$6009. About 20\% of the $\chi$ DIBs have a level of correlation with the 47 main non-C$_2$ ($\zeta$, $\zeta\sigma$ and $\sigma$) DIBs of Fan et al.\ (2022), which is comparable with the minimum level of correlation of these main DIBs between themselves.

\begin{figure}
	\begin{center}
		\includegraphics[scale=0.52, angle=0]{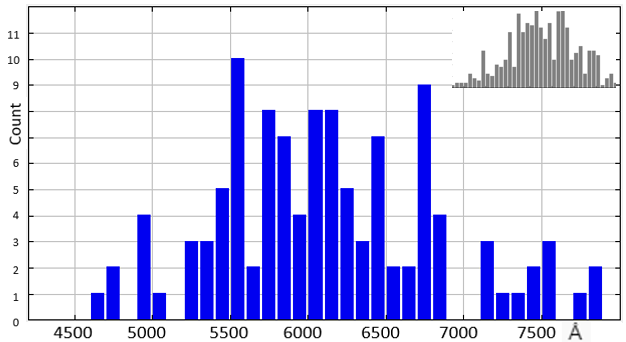} 
		\caption{Wavelength distribution in bins of 100\,\AA\  of the 111  $\chi$ DIBs  of Tables \ref{tab:C1chia}-\ref{tab:C2chib}, selected based on their enhanced intensity in the sight lines of HD\,175156 or HD\,148579. 
		For comparison, the inset shows the wavelength distribution for all 559 APO DIBs (see Figs. \ref{fig:21C9-chia_lambda} and \ref{fig:22C10-lambda_chib} for the wavelength distribution of $\chi_a$ and $\chi_b$ DIBs, respectively).}
		 \label{fig:7-chi_lambda}
	\end{center}
\end{figure}

Fig. \ref{fig:7-chi_lambda} displays the detailed wavelength distribution of the 111 $\chi$ DIBs  of Tables \ref{tab:C1chia}-\ref{tab:C2chib}.
(see also Figs\ \ref{fig:21C9-chia_lambda}-\ref{fig:22C10-lambda_chib}, showing the wavelength distributions of $\chi_a$ and  $\chi_b$ DIBs separately). 
 This distribution might be not random. There are some indications of evidence for a possible degree of periodic distribution with a period of about 300\,\AA . This does not seem conclusive, however. 

\subsection{Possible origin of the peculiarity of the HD\,175156 and HD\,148579 sight lines}
\label{sec:5.3origin}

Some special conditions in the ISM of the sight lines of HD\,175156 and HD\,148579 might explain the observed peculiarities of their DIBs. A search in CDS SIMBAD Astronomical Database shows that limited information is available about both sight lines except for that which is provided in Fan et al.\ (2019) and is summarised in Table \ref{tab:A1sightlines}. Nevertheless, both sight lines display a significant infrared excess that might be related to PAH emission. 

The {\it Gaia DR3} distance of HD\,175156 is 568$\pm$45\,pc. From the strong DIBs they observed, Benvenuti \& Porcedu (1989) identified its sight line as $\zeta$ (from Krelowski's private communication). The properties of this sight line were further described by Crawford (1992). He suggested that the observed strong single-component interstellar absorption occurs in outlying diffuse gas that is associated with the molecular cloud that causes the Aquila Ridge in the Milky Way. A {\it Spitzer} 5-14\,$\mu$m spectrum was obtained and analysed by Massa et al.\ (2022) in a strip $\sim$5"$\times$60'' around the star. The stacked spectrum shows a strong PAH emission, especially around 6.2\,$\mu$m. 

HD\,148579 is located in the well-studied region of star formation, $\rho$ Oph. Its {\it Gaia DR2} distance is only 139$\pm$3\,pc. The interstellar absorption was found to be dominated by a single cloud by Siebenmorgen et al. (2020) (see also the DIB profile study of Krelowski et al.\ 2021). Cotten \& Song (2016) quoted a strong infrared excess in the sight line of this star from the measured  22\,$\mu$m {\it WISE}\footnote{Wide-Field Infrared Survey Explorer} intensity.\footnote{ {\it Note added during the refereeing process.} A private communication from A. Witt confirms that both lines of sight show characteristics suggestive of strongly UV-irradiated dust clouds, where photo-processing may play a more significant role than would be typical for interstellar clouds subject to the more average interstellar radiation field. In addition, he suggests that, in both cases, a substantial part of the reddening comes from dust located in the vicinity of the star.} 

The possible enhanced PAH emission that is observed in the sight line of HD\,175156 and is perhaps associated with the infrared excess in HD\,148579 may lead us to wonder about two possibilities  for DIB carriers. They might be either  PAHs, perhaps photo-processed, or carbon rings (or chains) generated from PAHs in interstellar shocks, following the process proposed by Hrodmarsson et al.\ (2023, 2022). There appears to be no known evidence of shocks in these regions. However, the general context of star formation and the vicinity of massive young stars might favour their presence. Further studies of these regions and similar ones appear highly desirable in order to explore the origin of the enhanced abundance of $\chi$ DIB carriers.

\section{Possible signs indicating that carbon chains and rings might be carriers of some diffuse interstellar bands } 
\label{sec:6chains}

Prompted by the surprising detection of HC$_5$N and HC$_7$N in the ISM (Avery et al.\ 1976; Kroto et al.\ 1978), Douglas (1977) proposed that longer carbon chains might be the main carriers of diffuse interstellar bands. This was based on the properties of carbon chains, such as described by Pitzer \& Clementi (1959), especially their series of strong bands in the visible range that shift to the red with their size, and their expected stability in UV photon absorption due to internal conversion and infrared emission. 

The long-lasting DIB mystery and the soundness of this conjecture triggered an enormous effort for the synthesis and the visible spectroscopy of long chains, especially in the Basel laboratory led by John Maier (see the reviews by Maier 1998, Jochnowitz \& Maier 2008a,b who give a complete list of the compounds studied at the time, and  Zack \& Maier 2014a,b). However, it was disappointing not to achieve any DIB carrier confirmation except for the C$_{60}^+$ fullerene, so that Zack \& Maier (2014a) concluded that Douglas' proposition that 
the absorbing species are long chain carbon molecules, C$_n$ where n may lie in the range 5–15
was now excluded. They added that 
generally carbon chains up to a dozen atoms, their ions, and simple derivatives containing H or N are not responsible for the typical DIBs.

In almost all cases, measurements were only achieved  for n $\la$15, while it is notorious that these small hydrocarbons may be readily  photo-dissociated in 
the diffuse ISM (e.g. Tielens 2014). These laboratory results and various computations have brought striking confirmations of the  basic relation that  $\lambda$(n) is approximately proportional to n for many strong bands of carbon chains and rings, roughly in accordance with the particle-in-the-box model (see e.g. Anderson \& Gordon 2008 and references therein). In a few cases, calculations have  confirmed that their large oscillator strengths f$_{\rm n}$ are  also roughly proportional to n (e.g. Fischer \& Maier 1997, Reddy, Ghosh \& Mahapatra 2019). However,  until recently, most measurements were performed for chains that are too short for their strong bands to be in the DIB visible range (but see Section \ref{sec:6.1coincidences}). The behaviour of these bands in the visible DIB domain therefore has to rely on extrapolations, whose uncertainty precludes any DIB identification.  
It is also possible that the linear behaviour of $\lambda$ with n breaks down for some species, such as HC$_{\rm 2n}$H (Pino et al.\ 2001). 

As emphasised by Zack \& Maier (2014a,b), for example, a key and difficult question about the compatibility of long chains or rings with being DIB carriers is the width of their strong bands. If the lifetime of the excited state due to internal conversion is shorter than 0.1\,ps, the induced line width, $\ga$10-20\,\AA, is too broad to be  compatible with that of most detectable DIBs (around 1\,\AA\ or less). The internal conversion lifetime of excited states depends on the specific configuration of other excited states and their vibrational levels, so that it is hardly predictable (however, see Ghosh, Reddy \& Mahapatra 2019 for C$_{2p+1}$ chains, and also Pino et al.\ 2011 for internal conversion of PAHs). 
Double-resonance spectroscopy of HC$_{\rm 2n}$H$^+$ chains  has shown that the narrowness of their strong bands (1-10\,cm$^{-1}$) might be compatible with the observed DIB widths (Rice et al.\ 2010). Recent measurements of ring cations C$_{2n}^+$ have found, however that the widths are far too broad to be compatible (see references in Section \ref{sec:6.1coincidences}).
Therefore, the question remains generally  open whether the width of the strong optical bands of long chains or rings, corresponding to high excited levels, is compatible with the narrowness of most DIBs.

\subsection{Possible coincidences of diffuse interstellar bands with gas-phase bands of the C$_{14}^+$ ring, the C$_{17}$H$^+$ chain, and the HC$_{11}$H$^+$ chain}
\label{sec:6.1coincidences}

The spectroscopy of large molecules, including carbon clusters and other DIB carrier candidates, has made spectacular progress in the past few years, based on methods such as action spectroscopy, cavity ring-down spectroscopy, and double resonance. In addition to the confirmation of C$_{60}^+$ as the carrier of five near-infrared DIBs, gas-phase spectroscopy was recently achieved for whole series of strong bands of cation carbon rings and chains with a number of carbon atoms reaching up to $\sim$20-30. These outstanding results were obtained by the Melbourne group, led by E.J. Bieske, including  
C$_{2p}^+$ rings (Buntine et al.\ 2021, 2022; Marlton et al.\ 2023),   C$_{2p+1}$H$^+$ chains and rings (Marlton et al.\ 2022), and HC$_{2p+1}$H$^+$ chains (Marlton et al.\ 2024),
 and by Rademacher, Reedy \& Campbell (2022) for  C$_{2p}^+$ rings.
The quality of these data is such that it should allow us to identify the DIB carriers. 
However, these measurements have revealed that in most of the studied cases, the gas-phase widths of the bands, especially of C$_{2p}^+$ rings, are too broad to be compatible with DIB observations (Rademacher et al.\ 2022; Marlton et al.\ 2023). Nevertheless, the wavelengths and widths of three of the most remarkable $\chi$ DIBs might be compatible with these measurements of cation carbon rings or chains.
\smallskip
\smallskip

\subsubsection{Diffuse interstellar band $\lambda$6128 and the C$_{14}^+$ ring}
\smallskip

As noted by  Buntine et al.\ (2021) and Rademacher et al.\ (2022), and further analysed by Marlton et al.\ (2023), the relatively sharp main band of C$_{14}^+$ ring, measured to be at 6127.1$\pm$1\,\AA\ by two-colour photo-dissociation action spectroscopy by Marlton et al.\ (2023), is compatible within less than 1$\sigma$ with the relatively strong and broad $\chi_a$ DIB, $\lambda$6128.26$\pm$0.35. The  DIB width,  2.7\,\AA , is also compatible with the laboratory width, 6\,cm$^{-1}$ (2.3\,\AA ). 
Fig. \ref{fig:3-R21_k6010new1} of Rademacher et al.\ (2022) displays a number of vibronic transitions in the range of 200-500\,cm$^{-1}$ above the band origin (strongest peak of the band system).
 However, there is no match with observed DIBs in this range, especially since all DIB widths are narrow, $\la$1\,\AA , while all vibronic bands are much broader, $\ga$5\,\AA . These broad widths of the vibronic bands of the C$_{14}^+$ ring might explain why they are not detected as DIBs.
 In addition to the broad nature of the C$_{14}^+$ vibronic transitions, the fact that this spectral region of the DIB spectrum is congested further prevents their possible detection.

$\lambda$6128 is remarkable among the 79 $\chi_a$ DIBs because it is highly correlated with the whole $\chi_a$ set (Appendix \ref{app:C.3prominent}), and it is significantly stronger and broader than most other $\chi_a$ DIBs (Table \ref{tab:C1chia}, Fig. \ref{fig:17C5-ENav_FWHM}a). Its normalised equivalent width is EN\,=\,8.9 and 42\,m\AA /mag in the sight lines of HD\,183143 and HD\,175156, respectively. From these values, we may derive the possible fraction of interstellar carbon locked in the C$_{14}^+$ ring from the general equation (e.g. Cami 2014; Omont 2016)
\begin{equation}	\label{eq:9XCM}
	{\rm X_{CM}} = 3 \times 10^{-8} {\rm (EN/f) ~(N_C/20)}~ [5800/\lambda (\AA )]^2,
\end{equation}
where the total carbon abundance in the interstellar medium is assumed to be n$_{\rm C}$/n$_{\rm H}$ = 3.9 $\times$ 10$^{-4}$, and N$_{\rm C}$ is the number of carbon atoms of the molecule. For the fraction of interstellar carbon locked in its possible carrier, the C$_{14}^+$  ring, this yields
\begin{equation}	\label{eq:10XC14p}
{\rm X_{C}(C_{14}^+) = (1.7~ and~ 8) \times 10^{-6}\times (0.1/f_{6127})}
 \end{equation}
 in the sight lines of HD\,183143 and HD\,175156, respectively. However, it seems that there is no estimate for the oscillator strength of this band of C$_{14}^+$ (see Strelnikov et al.\ 2019; Buntine et al.\ 2021). 
  As for any molecule, the f-value of C$_{14}^+$ should be high enough to prevent an unlikely enormous abundance of this species, as stressed by Campbell \& Maier (2017), for instance. 
\smallskip
\smallskip

\subsubsection{Diffuse interstellar band $\lambda$6412 and the C$_{17}$H$^+$ chain}
\smallskip

\begin{figure}
	\begin{center}
		\includegraphics[scale=0.6, angle=0]{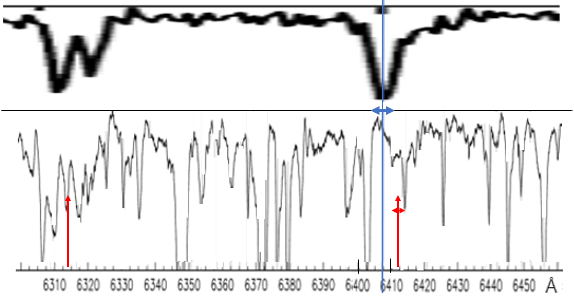}
	\caption{Comparison of the original band and the first vibronic band of C$_{17}$H$^+$ chain measured at 6407$\pm$3\,\AA\ and 6311-6321\,\AA , respectively, by Marlton et al.\ (2022, Fig.\ 5 ({\it top}); and, ({\it bottom}), i) the very broad DIB at 6412.37$\pm$0.96\,\AA , and ii) the possible very broad DIB observed at about  6314\,\AA\ (both are marked by red arrows).  The displayed DIB spectrum in the sight line of HD\,183143 is copied from the published DIB spectral atlas published by Hobbs et al.\ (2009). The vertical blue line marks the position of the wavelength of the original band  of C$_{17}$H$^+$ chain within the reported spectrum of HD 183143. The horizontal blue and red double arrows show the reported uncertainties on the measured values of the wavelengths of the origin band of C$_{17}$H$^+$  and of the APO DIB $\lambda$6412, respectively.} 
		 \label{fig:8-C17Hpnew1}
	\end{center}
\end{figure}

Marlton et al.\ (2022) reported 15607$\pm$8\,cm$^{-1}$ (6407$\pm$3\,\AA ) for the peak position of the strong original band of the C$_{17}$H$^+$ linear chain, measured by two-colour resonance-enhanced photo-dissociation, and  8.2\,\AA\ (20\,cm$^{-1}$) for its width. This is perhaps compatible with the very broad and strong $\chi_b$ DIB, $\lambda$6412, whose APO wavelength is 6412.37$\pm$0.96\AA\ (i.e.\ 15594.9$\pm$2.4\,cm$^{-1}$) and its very broad width of 7.6\,\AA\ (18\,cm$^{-1}$) (Fig. \ref{fig:8-C17Hpnew1}). In the APO Catalog, this  DIB  is quoted as "detected" in ten sight lines (including HD\,148579, but not HD\,175156) and "contaminated" in one sight line (HD\,183143). It is also quoted as possible by Sonnentrucker et al.\ (2018; see also Herbig \& Leka 1991; Jenniskens \& D\'esert 1994; Tuairisg et al.\ 2000). 
However, it is not reported in the HD\,183143 DIB catalogue of Hobbs et al.\ (2009), probably because of its contamination (Fig. \ref{fig:8-C17Hpnew1}).  
Therefore, it was missed by Marlton et al.\ (2022) in their search of possible DIB coincidences with their laboratory measurements. 
This 
coincidence might be credible when considering the matched broad widths, $\sim$8\,\AA , the difficulty in accurately characterising these very broad DIBs, and the rarity of the latter (only a dozen in the whole visible range).
If confirmed, this match might be key in showing the presence of a first long carbon chain as the carrier of a strong DIB, with an EW\,=\,119\,m\AA\ and 91\,m\AA\ 
in HD\,183143 and HD\,148579, respectively.

In addition to the original band at 6407\,\AA , the high-quality laboratory spectrum of the C$_{17}$H$^+$ chain displays a spectacular double vibronic progression, corresponding to at least two stretching modes at $\sim$230\,cm$^{-1}$ and $\sim$1780\,cm$^{-1}$ (see Fig.\ 4 of Marlton et al.\ 2022). All vibronic companions are at least twice broader than the original band at 6407\,\AA , which could have made their detection as DIBs almost hopeless. Nevertheless, there is a possible match between the first vibronic band of C$_{17}$H$^+$ at 6311-6321\,\AA\ and a possible very broad DIB (Fig. \ref{fig:8-C17Hpnew1}). This feature was interpreted as a very broad DIB by Herbig (1975) and Tuairisg et al.\ (2000), but it was not included in the preliminary and final APO DIB catalogues (Hobbs et al.\ 2009, Fan et al.\ 2019) and it was only quoted as possible by Sonnentrucker et al.\ (2018). All its properties are very uncertain in any case because it is impossible to determine exactly where it begins and ends, and where the underlying continuum is located. Herbig quoted 6314\,\AA\ for its centre, and Tuairisg et al.\ proposed a FWHM\,=\,23\,\AA\ and EW\,=\,390\,m\AA\ in HD\,183143. 
The approximate agreement in position, width, and even profile with the vibronic band of C$_{17}$H$^+$ is striking (Fig. \ref{fig:8-C17Hpnew1}). However, the above value for the EW, 3.3 times greater than the APO value for $\lambda$6412, would not be compatible with Fig.\ 4 of Marlton et al.\ (2022, partly reproduced in Fig. \ref{fig:8-C17Hpnew1}), where the vibronic band is weaker than twice the original band, but any estimate of the DIB EW is very uncertain because of contamination by strong stellar lines (see the spectrum of HD\,183143 in Fig. \ref{fig:8-C17Hpnew1}, from Hobbs et al.\ 2009). 

The other vibronic bands of C$_{17}$H$^+$ (Marlton et al.\ 2022) are even broader and often weaker than the first band at 6311-6321\,\AA . It is therefore not surprising that no matching DIB can be identified among APO DIBs close to the measured positions. 
Other smaller chains, C$_{2p+1}$H$^+$, that 
were measured by Marlton et al.\ (2022) seem to have a width of their origin band that is comparable to C$_{17}$H$^+$. The absence of strong broad DIBs that match the position of their origin band within the $\pm$8\,cm$^{-1}$ uncertainty of the measurements might be explained again by the difficulty of detecting very broad DIBs that are comparable to $\lambda$6412 and possibly weaker than it, and/or by the destruction of these smaller chains in the ISM.

The possible identification of an associated vibronic DIB significantly enhances the credibility of C$_{17}$H$^+$ as the carrier of $\lambda$6412.  The APO normalised  equivalent width of $\lambda$6412 is estimated as EN\,=\,94 and 267\,m\AA /mag in the sight lines of HD\,183143 and HD\,148579, respectively. From these values, we may derive the possible fraction of interstellar carbon locked in C$_{17}$H$^+$ chain from Eq.\ (\ref{eq:9XCM}), which yields  X$_{\rm C17H+ }$\,$\sim$\,3$\times$10$^{-6}$/f$_{6407}$ and 8$\times$10$^{-6}$/f$_{6407}$ in the sight lines of HD\,183143 and HD\,148579, respectively, where f$_{6407}$ is the oscillator strength of the origin band of  C$_{17}$H$^+$. Marlton et al.\ (2022) calculated a total oscillator strength f$_{\rm T}$ = 5.2 for the 1$^1\Sigma^+$--X$^1\Sigma^+$ 
 transition of  C$_{17}$H$^+$ based on $\omega$B97X-D/cc-pVDZ TD-DFT\footnote{time-dependent density functional theory}-level calculations. Their Figure 4 shows that the origin band contributes only a small fraction of this value, maybe $\sim$5-10\%. This would yield 
 \begin{equation}	\label{eq:11XC17Hp}
  {\rm X_{C}(C_{17}H^+) \sim  (0.5-1)\times 10^{-5}}
  \end{equation}
for the fraction of interstellar carbon locked in the C$_{17}$H$^+$ chain in the sight line of HD\,183143. This may be compared with the possible abundance in the same sight line of C$_{14}^+$,  
X$_{\rm C}$(C$_{14}^+$)\,=\,1.7 $\times$10$^{-6}$$\times$(0.1/f$_{6127}$), 
if it were the carrier of $\lambda$6128 (Eq.\ \ref{eq:10XC14p}
) and  of C$_{60}^+$, ~
X$_{\rm C}$(C$_{60}^+$) $\sim$ 10$^{-4}$ (Walker et al.\ 2015, 2016).

$\lambda$6412 is well correlated with most $\chi_b$ DIBs and with a number of $\chi_a$ DIBs (Appendix \ref{app:Cchi}). 
 It is only mildly correlated with the 
 reddening, E(B-V) (q=0.56). It is slightly more correlated with the $\zeta$, $\zeta\sigma$, and $\sigma$ major DIBs (q\,$\sim$\,0.65-0.75).
\smallskip
\smallskip

\subsubsection{Diffuse interstellar bands $\lambda$5450 and $\lambda$5419 and the HC$_{11}$H$^+$ chain}
\smallskip

Marlton et al.\ (2024) have recently reported the two-colour resonance enhanced photo-dissociation spectroscopy of HC$_{2p+1}$H$^+$ (p\,=\,2–6) chains. They quoted a possible match between two strong bands of HC$_{11}$H$^+$ chain with two strong very broad DIBs, namely, i) the origin band of the \~{C}$^2\Pi_u$--\~{X}$^2\Pi_g$ transition at $\lambda$\,=\,5451$\pm$3\,\AA\ (FWHM\,=\,15\,\AA ), with the DIB $\lambda$5450 (APO $\lambda$\,=\,5450.1\,\AA , FWHM\,=\,9.9\,\AA ); and ii) its mixed first vibronic band, shifted by 95\,cm$^{-1}$, at $\lambda$\,=\,5423$\pm$4\,\AA\ (FWHM\,=\,10\,\AA ), with the DIB  $\lambda$5419 (APO $\lambda$\,=\,5419.6\,\AA , FWHM\,=\,7\,\AA ). However, various arguments may question these identifications (Marlton et al.\ 2024).
In addition, $\lambda$5450 and $\lambda$5419 are poorly correlated (kn = 0.46, q = 0.61), $\lambda$5450 is not a $\chi$ DIB, it is poorly correlated with $\chi$ DIBs and strongly correlated with the main DIBs (q$_{6196}$ = 0.90), while it is the reverse for $\lambda$5419, which is a major $\chi_a$ DIB. Therefore, it seems probable that the match between the two DIBs and the bands of HC$_{11}$H$^+$ is not perfect, and that one of the DIBs at least is contaminated by some blend. This is confirmed by the APO Catalog, where $\lambda$5450 is quoted as "contaminated" for most sight lines, and $\lambda$5419 for half of the sight lines. 

\subsection{Further discussion of chains and rings as possible carriers of $\chi$ and C$_2$ diffuse interstellar bands}
\label{sec:6.2chainsrings}

A few  arguments in favour of long carbon chains and rings as possible carriers of $\chi$ and C$_2$ DIBs may be summarised as follows.

The confirmation that 94\% C$_2$-like DIBs have a short wavelength, $\lambda$ $<$ 5950\,\AA , reinforces the conclusions of Omont \& Bettinger (2020) about the correlation between wavelength, size, and UV resilience of the carriers of weak and narrow DIBs, and the likelihood of elongated carriers for these DIBs. 
Neutral and anionic carbon chains or rings are the most obvious and simplest candidates for these carriers. 

 Although no identification is confirmed, the possible match of five strong very broad DIBs with strong bands of long carbon chains or ring is impressive. It seems unlikely that all are mere coincidence. In particular, it is striking that they include four of the dozen DIBs that are broader than 7\,\AA . 
The possible identifications of very broad DIBs carried by the C$_{17}$H$^+$ chain and the C$_{14}^+$ ring seem relatively alluring for $\lambda$6412 and $\lambda$6128, but they also need confirmation by further good-quality observations in various sight lines.  
If confirmed, these would be the first identifications of long carbon chains or rings in the diffuse ISM.  It might open the way to considering the possibility of other types of long molecules as DIB carriers. It would be logical to first consider the possible presence  in the diffuse ISM of different charge states of other long members of chain series whose short members are observed in dense clouds, such as C$_n$ , HC$_n$, HC$_n$H, HC$_{2p+1}$N, and C$_{2p+1}$N. Various conditions should be addressed, however, including stability, chain or ring transformation, chemistry, formation, band width and profile, and vibronic bands.

 Three of these very broad DIBs, $\lambda\lambda$6128, 6412, and 5419, are members of the possible $\chi$ class, whose DIBs are enhanced in the HD\,175156 or HD\,148579 sight lines.  If the possible semi-periodic $\lambda$ distribution,  
with a period of $\sim$300\,\AA , of $\chi$  DIBs were confirmed, it might be  compatible with chain or ring carriers that display this periodicity. 

 Although they are mostly weak, the sum of the normalised equivalent widths of all C$_2$-like and $\chi$ DIBs represents a significant fraction of the total equivalent width of all DIBs. For instance, in the sight line of HD\,204827, EN$_{tot}$(C$_2$) $\sim$  900\,m\AA /mag, and EN$_{tot}(\chi)$ $\sim$  1300\,m\AA /mag ($\sim$800\,m\AA /mag for narrow $\chi$ DIBs). These values may be compared with those for all DIBs, except for $\lambda$4429, in HD\,204827,  
EN$_{tot}$(Allwo4429) $\sim$  5100\,m\AA /mag (2900\,m\AA /mag, for all narrow DIBs) (Omont \& Bettinger 2020, Table 2). This yields $\sim$18\% and $\sim$25\% for the contributions of C$_2$ and $\chi$ DIBs to the total equivalent width of all DIBs (31\% and 28\% for all narrow DIBs), respectively. The contribution is much smaller in the $\sigma$ sight lines, such as HD\,183143.

 The possibility that carbon chains and rings contribute to the carriers of major DIBs might be considered as well, for instance the  $\sim$50 DIBs considered by Fan et al.(\ 2022), which are known to be strongly correlated, and their associates (Appendix \ref{app:D.1associated}). For instance, the extrapolation up to C$_{27}$ of the matrix spectra of C$_{2p+1}$ chains, measured by Wyss et al.\ (1999) and  Forney et al.\ (1996) up to n\,=\,2p+1\,=\,21, might match $\lambda\lambda$5797, 6196, and 6613 DIBs for  C$_{23}$,  C$_{25}$, and  C$_{27}$, respectively (Appendix \ref{app:D.2C2p+1}). In addition to the uncertainty of the extrapolations, however, it seems that these chains could hardly survive to cyclisation in the diffuse ISM, and the expected $^1\Sigma_{\rm u}^+$ - X$^1\Sigma_{\rm g}^+$ strong band of linear C$_{27}$ seems hardly compatible with the likely presence of a Q band in the DIB profile of $\lambda$6613 (see e.g. MacIsaac et al.\ 2022), so that such an identification appears questionable. 

 Various indications, including the detailed view of vibronic bands of C$_{2p+1}$H$^+$ and HC$_{2p+1}$H$^+$ chains obtained by Marlton et al.\ (2022, 2024), tend to show that the chains of interest as DIB carriers probably have  low-frequency optically active vibration modes in the range of 100-400\,cm$^{-1}$.
As is well known, correlations should be key for identifying DIBs with a common carrier, including vibronic DIBs. 
In principle, different DIBs with the same carrier and the same lower level should be 100\% correlated. In practice, the data are always noisy, and lower correlation coefficients may be considered, perhaps  q $\ga$ 0.95 for strong DIBs with good-quality data, and $\ga$ 0.90 for data of poorer quality. However, similarly high correlations may be reached with similar but different carriers.  

As shown in particular by several studies using APO data (Fan et al.\ 2017, 2022; Smith et al.\ 2021), many strong DIBs meet the above requirements for possible vibronic DIBs. Tables \ref{tab:D3vib5797-6196}-\ref{tab:D4vib6613} list  a number of possible vibronic DIBs that are highly correlated with $\lambda\lambda$5797, 6196, or 6613 and are blueshifted by a few hundred cm$^{-1}$, which might be compatible with carbon chains, such as C$_{2p+1}$. Their number is limited to some DIBs for $\lambda\lambda$5797 and 6196 (Table \ref{tab:D3vib5797-6196}). Nevertheless,  the high number ($\sim$15-20) of possible vibronic DIBs associated with $\lambda$6613 (Table \ref{tab:D4vib6613}) is impressive. Although it may me partly fortuitous, it probably implies a different carrier structure from $\lambda$5797 and $\lambda$6196, whichever it is.  

 If the identification of C$_{17}$H$^+$ as the carrier of $\lambda$6412 were confirmed, the estimate of the fraction of interstellar carbon that it contains, X$_{\rm C}$(C$_{17}$H$^+$) $\sim$ (0.5-1)$\times$10$^{-5}$ (Eq.\ \ref{eq:11XC17Hp}), would give a lower limit of the interstellar carbon that is locked in chains or rings in the diffuse ISM. Including chains and rings of different nature, length, and charge (whether DIB carriers or not) might yield an abundance that is one or two orders of magnitude higher for all  chains and rings. This would remain lower by an order of magnitude at least  than the total carbon in PAHs ($\sim$4$\times$10$^{-2}$ in normal galaxies; see e.g. Shivaei et al.\ 2024), but it would still be significant. 

In conclusion, each of the above arguments is fragile, and none of them is decisive in proving that long chains and rings are important DIB carriers. Nevertheless, their accumulation is impressive. This conjecture  would also require  that they have efficient formation processes in the diffuse ISM and that they are stable against photo-dissociation. However, the convergence of  arguments in favour of chains and rings  justifies further efforts  to confirm this half-century conjecture.

\section{Summary and conclusion}
\label{sec:7conclusion}

 The systematic analysis of DIB correlations has been extended through the remarkable  capabilities of the APO Catalog. It is shown that important results can be derived, even with weak DIBs, despite the limited statistics.
 
 The number of C$_2$-like DIB candidates has been substantially increased. 
 In addition to the $\sim$20 known confirmed C$_2$ DIBs, 12 new C$_2$ candidates and 34 possible C$_2$-related DIBs were identified, mostly at $\lambda$\,$<$\,6000\,\AA . 
 Some of these candidates need to be confirmed, for example by profile analysis or correlations with bands of the interstellar C$_2$ molecule. A more thorough analysis of C$_2$-like DIBs may lead to a continuous series of DIBs with sensitivity to UV-quenching ranging from reference C$_2$ to $\zeta$ DIBs. With these candidates, the census of C$_2$ DIBs appears to be not far from being complete, in agreement with the prediction of Fan et al.\ (2017) that the total number of C$_2$ DIBs probably does not exceed $\sim$10\% of all DIBs. The group of C$_2$-like DIBs remains key in view of DIB-carrier identification because of their limitation to short wavelengths, which might indicate long carbon chains or rings.
  
The strong enhancement of two subsets of DIBs in two sight lines, HD\,175156 and HD\,148579, was a surprise. Their significant  correlations justified discussing them as a single class that was tentatively denoted $\chi$. It might include part of the C$_2$ DIBs. However, it is clear that this new feature of the DIB panorama needs further analysis. A deeper statistical analysis should aim at confirming the consistency of this class, and whether it might lead to the definition of a new DIB family.  Finding other similar sight lines in EDIBLES, for example, would be mandatory for this purpose. Possible processes at the origin of the enhancement of these DIB carriers should be investigated  by screening possible peculiar interstellar conditions in these sight lines.  It would be important to confirm that the interstellar material that causes this is located close to these stars and how it might be affected by this vicinity.  

A few other signs were identified that may point to chains or rings as major DIB carriers, including the confining of C$_2$ DIBs at short wavelengths and the possible tentative identification of C$_{17}$H$^+$ and HC$_{11}$H$^+$ chains (each of which include two very broad bands) and C$_{14}^+$ ring (one broad band) as carriers of strong DIBs, including three $\chi$ bands. 
If confirmed, this would make various carbon chains and rings, with about 15 to 30 carbon atoms, significant members of the panel of cosmic molecules. However, even in the most extreme estimates, the possible total abundance of chains and rings in the diffuse ISM will probably remain modest, 
at least an order of magnitude lower than the total carbon in PAHs.  Trying to confirm the identification of C$_{17}$H$^+$ chain and C$_{14}^+$ ring with better data and other sight lines is a first priority for validating this conjecture. If the presence of the C$_{14}^+$ ring were confirmed, an accurate estimate of its f-value would also allow 
one to better estimate the total abundance of all C$_{2n}^+$ rings, most of which are not DIB carriers because their bands are too broad.  
 
The ability of chains and rings as short as C$_{17}$H$^+$, C$_{14}^+$, or HC$_{11}$H$^+$ to be DIB carriers is conditioned by a survival against photo-dissociation much better than that of PAHs of similar size. It seems possible that this might result from an efficient optical fluorescence that might induce a cooling that is much faster than infrared cooling (Iida et al.\ 2022; Lacinbala et al.\ 2022). This remains to be modelled in detail for these molecules, however. 
Various chemical processes that include chains and rings should be explored to assess  their relative abundances. Finally, at least one viable formation process of interstellar chains and rings should be identified, such as PAH shattering in shocks
(Duley 2000; Jones 2016; Hrodmarsson et al.\ 2022, 2023), or chains attached to PAHs or PAH clusters (Zanolli et al.\ 2023).  

Various extensions of this work may be considered. Two  main goals should be to address the nature and origin of the enhancement of the so-called $\chi$ DIBs in peculiar sight lines, as quoted above, and to further explore the possible C$_2$-related subfamily (Table \ref{tab:B4related}) that is intermediate between C$_2$ and $\zeta$ DIBs. 

A systematic theoretical exploration of optical spectra of possible long chains and rings might be considered, for instance by extending the treatment of Marlton et al.\ (2022), to pursue the identification of series of strong bands in the DIB domain. Further studies of internal conversion similar to that of Ghosh, Reddy \& Mahapatra (2019) might be helpful.
The highest priority should be given to sophisticated laboratory spectroscopy, including neutral and anionic species, 
using various methods of gas-phase spectroscopy, which is mandatory for an identification of  DIBs.

\smallskip

\begin{acknowledgements}
	I thank the referee, N. Cox, for his very helpful comments and suggestions.  
	I especially thank Ugo Jacovella  and Holger Bettinger for continuous discussions and their important suggestions during the preparation of the paper. 
	I am indebted to Pierre Cox and Patrick Boiss\'e for their careful reading of the manuscript and their suggestions for improving it. I want to especially thank A. Witt for providing an unpublished analysis about the properties of the ISM in the sight lines of HD\,175156 and HD\,148579. 
	I thank the whole APO team and especially Haoyu Fan for working out the unique APO Catalog. 
	I thank S.\ Marlton for kindly providing detailed spectral data, and O.\ Bern\'e, E.\ Dartois, L.\ Hobbs, C.\ Joblin, R.\ Lallement, J. Maier, T.\ Pino, D.\ Welty and He Zhao for various help and the discussions we have had on the contents of the paper.
	Finally, I made great use of the TOPCAT tool,  https://www.star.bris.ac.uk/~mbt/topcat/ ; "TOPCAT \& STIL: Starlink Table/VOTable Processing Software" (Taylor 2005) - and would like here to thank its developers.
\end{acknowledgements}

{\bf References}  
\smallskip

Allamandola, L.~J., Tielens, A.~G.~G.~M., \& Barker, J.~R.\ 1985, \apjl, 290, L25. doi:10.1086/184435

Allamandola, L.~J., Hudgins, D.~M., Bauschlicher, C.~W., et al.\ 1999, \aap, 352, 659

Anderson, B.~D., \& Gordon, C.~M.\ 2008, J. Chem.\ Education, 85, 1279

Andrews, H., Boersma, C., Werner, M.~W.\ et al.\ 2015, \apj, 807, 99

Avery, L.~W., Broten, N.~W., MacLeod, J.~M., et al.\ 1976, \apjl, 205, L173. doi:10.1086/182117

Baron, D., Poznanski, D., Watson, D., et al.\ 2015, \mnras, 451, 332. doi:10.1093/mnras/stv977

Benvenuti, P. \& Porceddu, I.\ 1989, \aap, 223, 329

Buntine, J.~T., Carrascosa, E., Bull, J.~N., et al.\ 2022, Rev.\ Sci.\ Instrum., 93, 043201

Buntine, J.~T., Cotter, M.~I., Jacovella, U., et al.\ 2021, J.\ Chem.\ Phys.\ 155, 214302

Burkhardt, A.~M., Long Kelvin Lee, K., Bryan Changala, P., et al.\ 2021, \apjl, 913, L18. doi:10.3847/2041-8213/abfd3a


Cami, J.\ 2014, The Diffuse Interstellar Bands, IAU Symposium, 297, 370. doi:10.1017/S1743921313016141

Cami, J., Sonnentrucker, P., Ehrenfreund, P., et al.\ 1997, \aap, 326, 822

Cami, J. \& Cox, N.~L.~J.\ 2014, The Diffuse Interstellar Bands, IAU Symposium, 297

Cami, J., Cox, N.~L., Farhang, A., et al.\ 2018, The Messenger, 171, 31

Campbell, E.~K., Holz, M., Gerlich, D., \& Maier, J.~P.\ 2015, \nat, 523, 322 

Campbell, E.~K. \& Maier, J.~P., 2017, J.\ Chem.\ Phys., 146, 160901


Candian, A., Gomes Rachid, M., MacIsaac, H., et al.\ 2019, \mnras, 485, 1137. doi:10.1093/mnras/stz450

Castellanos, R., Najarro, F., Garcia, M., et al.\ 2024, \mnras. doi:10.1093/mnras/stae1472

Cernicharo, J., Ag{\'u}ndez, M., Cabezas, C., et al.\ 2021, \aap, 649, L15. doi:10.1051/0004-6361/202141156


Chown, R., Sidhu, A., Peeters, E., et al.\ 2024, \aap, 685, A75. doi:10.1051/0004-6361/202346662

Cordiner, M.~A., Linnartz, H., Cox, N.~L.~J., et al.\ 2019, \apjl, 875, L28. doi:10.3847/2041-8213/ab14e5

Cotten, T.~H. \& Song, I.\ 2016, \apjs, 225, 15. doi:10.3847/0067-0049/225/1/15

Cox, N.~L.~J., Cami, J., Kaper, L., et al.\ 2014, \aap, 569, A117. doi:10.1051/0004-6361/201323061

Cox, N.~L.~J., Cami, J., Farhang, A., et al.\ 2017, \aap, 606, A76. doi:10.1051/0004-6361/201730912


Crawford, M.~K., Tielens, A.~G.~G.~M., \& Allamandola, L.~J.\ 1985, \apjl, 293, L45 

Crawford, I.~A.\ 1992, The Observatory, 112, 161

Douglas, A.~E.\ 1977, \nat, 269, 130

Draine, B.~T. \& Li, A.\ 2007, \apj, 657, 810. doi:10.1086/511055

Duley, W.~W. \& Kuzmin, S.\ 2010, \apjl, 712, L165. doi:10.1088/2041-8205/712/2/L165


Ebenbichler, A., Postel, A., Przybilla, N., et al.\ 2022, \aap, 662, A81. doi:10.1051/0004-6361/202142990

Ebenbichler, A., Smoker, J.~V., Lallement, R., et al.\ 2024, \aap, 686, A50. doi:10.1051/0004-6361/202348871

Elyajouri, M., Lallement, R., Cox, N.~L.~J., et al.\ 2018, \aap, 616, A143. doi:10.1051/0004-6361/201833105

Ensor, T., Cami, J., Bhatt, N.~H., et al.\ 2017, \apj, 836, 162. doi:10.3847/1538-4357/aa5b84

Fan, H., Welty, D.~E., York, D.~G., et al.\ 2017, \apj, 850, 194. doi:10.3847/1538-4357/aa9480 

Fan, H., Hobbs, L.~M., Dahlstrom, J.~A., et al.\ 2019, \apj, 878, 151. doi:10.3847/1538-4357/ab1b74 

Fan, H., Hobbs, L.~M., Dahlstrom, J.~A., et al.\ 2020, VizieR Online Data Catalog, 187

Fan, H., Schwartz, M., Farhang, A., et al.\ 2022, \mnras, 510, 3546. doi:10.1093/mnras/stab3651 

Fan, H., Rocha, C.~M.~R., Cordiner, M., et al.\ 2024, \aap, 681, A6. doi:10.1051/0004-6361/202243910


Fischer, G., \& Maier, J.~P.\ 1997, Chemical Physics 223, 149

Foing, B.~H., \& Ehrenfreund, P.\ 1994, \nat, 369, 296

Foing, B.~H., \& Ehrenfreund, P.\ 1997, \aap, 317, L59 

Forney, D., Freivogel, P., Grutter, M., et al.\ 1996, J.\ Chem.\ Phys.\ 104, 4954; doi.org/10.1063/1.471127


Friedman, S.~D., York, D.~G., McCall, B.~J., et al.\ 2011, \apj, 727, 33. doi:10.1088/0004-637X/727/1/33

Fulara, J., Lessen, D., Freivogel, P., et al.\ 1993, \nat, 366, 439. doi:10.1038/366439a0

Fulara, J. \& Kre{\l}owski, J.\ 2000, \nar, 44, 581. doi:10.1016/S1387-6473(00)00108-1




Geballe, T.~R., Najarro, F., Figer, D.~F., et al.\ 2011, \nat, 479, 200. doi:10.1038/nature10527

Ghosh, A., Reddy, S.~R., \& Mahapatra, S.\ 2019, J. Chem. Phys. 151, 054304. doi:10.1063/1.5108726

Hamano, S., Kobayashi, N., Kawakita, H., et al.\ 2022, \apjs, 262, 2. doi:10.3847/1538-4365/ac7567

Heger, M.~L.\ 1922a, Lick Observatory Bulletin, 10, 141. doi:10.5479/ADS/bib/1922LicOB.10.141H
 
Heger, M.~L.\ 1922b, Lick Observatory Bulletin, 10, 146

Herbig, G.~H.\ 1975, \apj, 196, 129. doi:10.1086/153400


Herbig, G.~H.\ 1995, \araa, 33, 19 

Herbig, G.~H. \& Leka, K.~D.\ 1991, \apj, 382, 193. doi:10.1086/170708

Hobbs, L.~M., York, D.~G., Snow, T.~P., et al.\ 2008, \apj, 680, 1256 

Hobbs, L.~M., York, D.~G., Thorburn, J.~A., et al.\ 2009, \apj, 705, 3

Hrodmarsson, H.~R., Bouwman, J., Tielens, A.~G.~G.~M., \& Linnartz, H.\ 2022, International Journal of Mass Spectrometry, 476, 116834

Hrodmarsson, H.~R., Bouwman, J., Tielens, A.~G.~G.~M., \& Linnartz, H.\ 2023, International Journal of Mass Spectrometry, 485, 116996



Iida, S., Hu, W., Zhang, R., et al.\ 2022, \mnras, 514, 844. doi:10.1093/mnras/stac1349


Jenniskens, P. \& Desert, F.-X.\ 1994, \aaps, 106, 39

Joblin, C., Maillard, J.~P., D'Hendecourt, L., et al.\ 1990, \nat, 346, 729



Jochnowitz, E.~B. \& Maier, J.~P.\ 2008a, Annual Review of Physical Chemistry, 59, 519. doi:10.1146/annurev.physchem.59.032607.093558

Jochnowitz, E.~B. \& Maier, J.~P.\ 2008b, Molecular Physics, 106, 2093. doi:10.1080/00268970802208588

Jones, A.~P.\ 2016, Royal Society Open Science, 3, 160223. doi:10.1098/rsos.160223

Ka{\'z}mierczak, M., Schmidt, M.~R., Bondar, A., et al.\ 2010, \mnras, 402, 2548. doi:10.1111/j.1365-2966.2009.16065.x

Ka{\'z}mierczak, M., Schmidt, M., Weselak, T., et al.\ 2014, The Diffuse Interstellar Bands, IAU Symposium, 297, 121. doi:10.1017/S1743921313015718

Kos, J. \& Zwitter, T.\ 2013, \apj, 774, 72. doi:10.1085/0004-637X/774/1/72

Krelowski, J., \& Walker, G.~A.~H.\ 1987, \apj, 312, 860

Krelowski, J.\ 1989, Astronomische Nachrichten, 310, 255. doi:10.1002/asna.2113100403

Krelowski, J., Snow, T.~P., Seab, C.~G., et al.\ 1992, \mnras, 258, 693. doi:10.1093/mnras/258.4.693

Kre{\l}owski, J., Galazutdinov, G.~A., Gnaci{\'n}ski, P., et al.\ 2021, \mnras, 508, 4241. doi:10.1093/mnras/stab2774

Kroto, H.~W., Kirby, C., Walton, D.~R.~M., et al.\ 1978, \apjl, 219, L133. doi:10.1086/182623




Lacinbala, O., Calvo, F., Dartois, E., et al.\ 2023, \aap, 671, A89. doi:10.1051/0004-6361/202245421

 Leger, A. \& Puget, J.~L.\ 1984, \aap, 137, L5

L\'eger, A., \& D'Hendecourt, L.\ 1985, \aap, 146, 81 

Linnartz, H., Cami, J., Cordiner, M., et al.\ 2020, Journal of Molecular Spectroscopy, 367, 111243. doi:10.1016/j.jms.2019.111243

MacIsaac, H., Cami, J., Cox, N.~L.~J., et al.\ 2022, \aap, 662, A24. doi:10.1051/0004-6361/202142225

Maier, J.~P.\ 1998, Journal of Physical Chemistry A, 102, 3462. doi:10.1021/jp9807219

Maier, J.~P., Walker, G.~A.~H., \& Bohlender, D.~A.\ 2004, \apj, 602, 286. doi:10.1086/381027


Marlton, S.~J.~P., Jack T. Buntine, J.~T., Liu, C., et al.\ 2022, J.\ Phys.\ Chem.\ A, 126, 6678. doi.org/10.1021/acs.jpca.2c05051

Marlton, S.~J.~P., Jack T. Buntine, J.~T., Watkins, P., et al.\ 2023, J.\ Phys.\ Chem.\ A, 127, 1168. doi.org/10.1021/acs.jpca.2c07068

Marlton, S.~J.~P., Liu, C., Watkins, P., \& Bieske, E.~J.\ 2024, Phys.\ Chem.\ Chem.\ Phys., 26, 12306.  doi:10.1039/d4cp00625a

Massa, D., Gordon, K.~D., \& Fitzpatrick, E.~L.\ 2022, \apj, 925, 19. doi:10.3847/1538-4357/ac3825

McCabe, M.\ 2019, Astronomy and Geophysics, 60, 4.29. doi:10.1093/astrogeo/atz164

McCall, B.J., \& Griffin, R.E.\ 2013, ds. Proc.\ R.\ Soc.\ A 469:2012.0604.dx.doi.org/10.1098/rspa.2012.060

McGuire, B.~A., Loomis, R.~A., Burkhardt, A.~M., et al.\ 2021, Science, 371, 1265. doi:10.1126/science.abb7535

McGuire, B.~A.\ 2022, \apjs, 259, 30. doi:10.3847/1538-4365/ac2a48

Merrill, P.~W.\ 1934, \pasp, 46, 206 

Merrill, P.~W.\ 1936, \pasp, 48, 179 

Merrill, P.~W. \& Wilson, O.~C.\ 1938, \apj, 87, 9. doi:10.1086/143897

Omont, A.\ 2016, \aap, 590, A52. doi:10.1051/0004-6361/201527685


Omont, A. \& Bettinger, H.~F.\ 2020, \aap, 637, A74. doi:10.1051/0004-6361/201937071

Omont, A. \& Bettinger, H.~F.\ 2021, \aap, 650, A193. doi:10.1051/0004-6361/202140675

Pino, T., Ding, H., G{\"u}the, F., \&  Maier, J.~P.\ 2001, J.\ Chem.\ Phys., 114, 2208. doi: 10.1063/1.1338530

Pino, T., Carpentier, Y., F{\'e}raud, G., et al.\ 2011, EAS Publications Series, 46, 355. doi:10.1051/eas/1146037

Pitzer, K.~S., \& Clementi, E.\ 1959, J\. Am.\ Chem.\ Soc., 81, 4477

Puget, J.~L. \& Leger, A.\ 1989, \araa, 27, 161. doi:10.1146/annurev.aa.27.090189.001113


Rademacher, J., Reedy, E.~S., \& Campbell, E.~K.\ 2022, J.\ Phys.\ Chem., A 126, 2127

Reddy, S.~R., Ghosh, A., \& Mahapatra, S.\ 2019, J. Chem. Phys. 151, 054303. doi: 10.1063/1.5108725

Rice, C.~A., Rudnev, V., Dietsche, R., et al.\ 2010, \aj, 140, 203. doi:10.1088/0004-6256/140/1/203

Rice, C.~A. \& Maier, J.~P.\ 2013, Journal of Physical Chemistry A, 117, 5. doi:10.1021/jp401833m

Salama, F., Bakes, E.~L.~O., Allamandola, L.~J., \& Tielens, A.~G.~G.~M.\ 1996, \apj, 458, 621 

Salama, F., \& Ehrenfreund, P.\ 2014,  The Diffuse Interstellar Bands, IAU Symposium, 297, 364

Salama, F., Galazutdinov, G.~A., Kre{\l}owski, J., et al.\ 2011, \apj, 728, 154. doi:10.1088/0004-637X/728/2/154

Sarre, P.~J.\ 2006, Journal of Molecular Spectroscopy, 238, 1

Shivaei, I., Alberts, S., Florian, M., et al.\ 2024, arXiv:2402.07989, submitted to \aap. 

Siebenmorgen, R., Kre{\l}owski, J., Smoker, J., et al.\ 2020, \aap, 641, A35. doi:10.1051/0004-6361/202037511

Sita, M.~L., Changala, P.~B., Xue, C., et al.\ 2022, \apjl, 938, L12. doi:10.3847/2041-8213/ac92f4

Smith, W.~H., Snow, T.~P., \& York, D.~G.\ 1977, \apj, 218, 124. doi:10.1086/155664

Smith, F.~M., Harriott, T.~A., Majaess, D., et al.\ 2021, \mnras, 507, 5236. doi:10.1093/mnras/stab2444

Smith, E.~R., Smith, F.~M., Harriott, T.~A., et al.\ 2022, Research Notes of the American Astronomical Society, 6, 82. doi:10.3847/2515-5172/ac680f

Smoker, J.~V., M{\"u}ller, A., Monreal Ibero, A., et al.\ 2023, \aap, 672, A181. doi:10.1051/0004-6361/202142267

Snow, T.\ 1995a, The Diffuse Interstellar Bands, 202, 325. doi:10.1007/978-94-011-0373-2\_32

Snow, T.\ 1995b, The Diffuse Interstellar Bands, 202, 379. doi:10.1007/978-94-011-0373-2\_38

Snow, T.~P. \& McCall, B.~J.\ 2006, \araa, 44, 367. doi:10.1146/annurev.astro.43.072103.150624

Sonnentrucker, P., York, B., Hobbs, L.~M., et al.\ 2018, \apjs, 237, 40. doi:10.3847/1538-4365/aad4a5


Strelnikov, D.~V., Link, M., \& Kappes, M.~M.\ 2019, J.\ Phys.\ Chem.\ A, 123, 5325

Taniguchi, K., Gorai, P., \& Tan, J.~C.\ 2024, \apss, 369, 34. doi:10.1007/s10509-024-04292-9

Taylor, M.~B.,  2005, Astronomical Data Analysis Software and Systems XIV, eds. P.\ Shopbell et al., ASP Conf. Ser. 347, 29


Thaddeus, P.\ 1995, The Diffuse Interstellar Bands, 202, 369. doi:10.1007/978-94-011-0373-2\_37

Thorburn, J.~A., Hobbs, L.~M., McCall, B.~J., et al.\ 2003, \apj, 584, 339 

Tielens, A.~G.~G.~M.\ 2008, \araa, 46, 289. doi:10.1146/annurev.astro.46.060407.145211

Tielens, A.~G.~G.~M.\ 2013, Reviews of Modern Physics, 85, 1021 

Tielens, A.~G.~G.~M.\ 2014, The Diffuse Interstellar Bands, IAU Symposium, 297, 399. doi:10.1017/S1743921313016207

Tielens, A.~G.~G.~M., \& Snow, T.~P.\ 1995, The Diffuse Interstellar Bands, {\it Astrophysics and Space Science Library}, 202, Kluwer 

Tuairisg, S. {\'O}., Cami, J., Foing, B.~H., et al.\ 2000, \aaps, 142, 225. doi:10.1051/aas:2000148


van der Zwet, G.~P. \& Allamandola, L.~J.\ 1985, \aap, 146, 76

Vos, D.~A.~I., Cox, N.~L.~J., Kaper, L., Spaans, M., \& Ehrenfreund, P.\ 2011, \aap, 533, A129 

Walker, G.~A.~H., Bohlender, D.~A., Maier, J.~P., et al.\ 2015, \apjl, 812, L8. doi:10.1088/2041-8205/812/1/L8

Walker, G.~A.~H., Campbell, E.~K., Maier, J.~P., et al.\ 2016, \apj, 831, 130. doi:10.3847/0004-637X/831/2/130


Wyss, M., Grutter, M., \& Maier, J.~P.\ 1999, Chem..\ Phys.\ Lett., 304, 35

Xiang, F.~Y., Li, A., \& Zhong, J.~X.\ 2017, \apj, 835, 107. doi:10.3847/1538-4357/835/1/107

York, D.~G.\ 2024, Research in Astronomy and Astrophysics, 24, 016001. doi:10.1088/1674-4527/acf35d

Zack, L~N.\ \& Maier, J.~P.\ 2014a, The Diffuse Interstellar Bands, IAU Symposium, 297, 237. doi:10.1017/S1743921313015949

Zack, L.~N. \& Maier, J.~P.\ 2014b, Chem. Soc. Rev., 43, 4602

Zanolli, Z., Malc{\i}o{\u{g}}lu, O.~B., \& Charlier, J.-C.\ 2023, \aap, 675, L9. doi:10.1051/0004-6361/202245721


Zhang, Y., Sadjadi, S., \& Hsia, C.-H.\ 2020, \apss, 365, 67. doi:10.1007/s10509-020-03779-5


\appendix

\section{APO sight-lines and DIB intensities}
\label{app:Asightlines}	

\subsection{APO sight-lines properties} 
\label{app:A.1properties}	

The properties of the 25 stars of the APO Catalog and their sight lines are briefly described in Fan et al.\ (2019). The information about their main properties is reproduced in Table \ref{tab:A1sightlines}. The sight lines, most frequently quoted in the present paper, include:

- HD\,204827 (Cepheus) and HD\,183143 (HT Sge), whose DIB properties were listed in the preliminary version of the APO Catalog of Hobbs et al.\ (2008, 2009). HD\,183143 is the reference for typical $\sigma$ sight lines. HD\,204827 is substantially UV-shielded; it is often used as reference both for $\zeta$ and C$_2$ DIBs (e.g.\ Thorburn et al.\ 2003).
The ratio R$_{21}$\,=\,EW(HD204827)/EW(HD183143) (Eq.\ \ref{eq:1R21}) provides a straightforward information about the behaviour of a DIB with respect to UV intensity, and therefore, about its belonging to DIB families, $\sigma$ (R$_{21}$\,$\la$\,0.5), $\zeta\sigma$ (0.5\,$\la$\,R$_{21}$\,$\la$\,0.7), $\zeta$ 
(0.8\,$\la$\,R$_{21}$\,$\la$\,1.5) and C$_2$ (R$_{21}$\,$\ga$\,1.5).

- HD\,175156 (Scutum) and HD\,148579 (Ophiucus) display specific enhancements of $\chi_a$ and $\chi_b$ DIBs, respectively, that are at the origin of the definition of the tentative $\chi$ class (Section \ref{sec:5peculiarsightlines}). See Section \ref{sec:5.3origin} for the limited knowledge about the properties of these sight lines.

- HD\,23512 (Taurus) might also have some enhancements of specific DIBs.

- HD\,37061 (NU Ori) has a very strong UV intensity.

- HD\,166734  (Serpens) and HD\,168625 (V4030 Sgr) display high normalised equivalent widths for $\sigma$ DIBs. HD\,166734 has the highest detection rate of all sight lines. 

- VI Cyg 12 and VI Cyg 5 have a high extinction.

\begin{table*}
	\caption{Properties of APO sight-lines.}
	\label{tab:A1sightlines}	
	\begin{tabular}{lccccccccc}
		\hline \hline
		Star      & E(B-V)   &f$_{\rm H2}$& Ndet & \%det$^a$  & R$_{\zeta\sigma}^b$& Type$^c$ & $\delta^d$  &  N2$\sigma^e$ & \%$2\sigma^f$ \\
		\hline 
		&&&&&&&&& \\   
		HD20041   &   0.72    &  0.42  &351 &  63\%      &   0.33  &  $\sigma$ & +0.5 & 14&  4\%  \\
		Cernis52  &   0.90    &$>$0.78 & 121 & 21\%      &   0.91  &  $\zeta$ & +0.5 &6 &  5\%  \\
		HD23180   &   0.31    &  0.55  & 279 & 50\%      &   0.63  &  $\zeta$ & +0.5 & 11 &  4\%  \\
		HD281159  &   0.85    & 0.50  & 369 & 26\%      &   0.35  &  $\sigma$ &--0.5 &0 &  0\%  \\
		HD23512   &   0.36     & 0.62  & 118 & 21\%      &   0.29  &  $\sigma$ & -0.7 & 19&  16\%  \\
		HD24534   &   0.59    &  0.76  &226 &  40\%      &   0.75  &  $\zeta$ &  0.0 &14 & 6\%  \\
		HD24912   &   0.33     & 0.38  & 300 & 53\%      &   0.23  &  $\sigma$ & +0.5 &18 &   6\% \\
		HD28482   &   0.48    &  0.66  & 146 & 26\%      &   0.26  &  $\sigma$ & +0.5 &12 &  8\%  \\
		HD37061   &  0.52     & 0.02  & 144 & 25\%      &   0.11  &   $\sigma$ & --0.5 &3 &  2\%  \\
		HD37903   &   0.35    &  0.53  &164 &  29\%      &   0.15  &  $\sigma$ & 0.0 &15 &  9\%  \\
		HD43384   &   0.58    & 0.44  &365 &  65\%      &   0.32  &  $\sigma$ & +0.5 &6 & 2\%   \\
		HD147084  &   0.73     & 0.59  &109 &  19\%      &   0.42  & $\sigma$ & +0.5 & 2& 2\%  \\
		HD147889  &   1.07    & 0.45  & 376 & 67\%      &   0.47  &  $\zeta\sigma$ & 0.0 &27 &  7\%  \\
		HD148579  &   0.34    &  0.45  & 143&  25\%      &   0.42  & $\sigma$ & +0.5 &43 &  30\% \\
		HD166734  &   1.39    &  0.39  & 498 & 89\%      &   0.42  & $\sigma$ & 0.0 &30 & 6\%  \\
		HD168625  &   1.48     & 0.33  &351  & 63\%      &   0.29  &  $\sigma$ & +0.3 & 12&  3\%  \\
		HD175156  &   0.31    &  0.31  &290 &  52\%      &   0.57  &  $\zeta$ & +0.3 &79 &  27\%  \\
		HD183143  &   1.27    &  0.31  & 472 & 84\%      &   0.30  &  $\sigma$ & +0.3& 9 &  2\%  \\
		HD190603  &   0.72     & 0.16  &321 &  57\%      &   0.29  &  $\sigma$ & +0.5 &10 &  3\%  \\
		HD194279  &   1.20    &  0.30  &343 &  61\%      &   0.35  &  $\sigma$ &  0.0 & 2 &  1\%  \\
		VICyg 5   &   1.99    &  0.47  &469  & 84\%      &   0.32  &  $\sigma$ &  --0.5 & 2&  1\%   \\
		VICyg 12 &   3.31    &$>$0.48 &439 &  78\%      &   0.37  & $\sigma$ & --0.5  &2 & 1\%  \\
		HD204827  &   1.11    &  0.67  & 491 & 88\%      &   0.76 &  $\zeta$ & +0.3  &30 &  6\%  \\
		HD206267  &   0.53    &  0.42  & 374 & 67\%      &   0.36 &  $\sigma$ &  +0.3 &16 &  4\%  \\
		HD223385  &   0.67    &  0.12  &275 &  49\%      &   0.25 &  $\sigma$ & +0.3 &14 & 5\%   \\
		\hline
		\hline
	\end{tabular}
	{\tiny ~~~~~~~  \\
		\\
		$a$. Percentage of detection of the 559 APO DIBs in the considered line of sight.  \\
		$b$. Ratio of the equivalent widths of $\lambda$5797 to $\lambda$5780 DIBs in the line of sight, that is believed to correlate well with the amount of UV radiation (e.g. Ensor et al.\ 2017). \\
		$c$. UV-type of the sight line, inferred from the ratio R$_{\zeta\sigma}$, following the definition of Fan et al.\ (2019), putting the limit between the $\sigma$ and $\zeta$ types at R$_{\zeta\sigma}$\,$\sim$\,0.5. Type "$\zeta\sigma$" denotes a sight line intermediate between $\zeta$ and $\sigma$, as defined by Fan et al.\ (2022).\\
		$d$. Reduced shift of the central approximate Gaussian distribution from EN$_{\rm av}$ (Section \ref{sec:5.1HD175-148} and Fig. \ref{fig:9A1-histos_etanew1}). \\
		$e$. Number of DIBs whose EN exceeds the centre of the approximate Gaussian distribution by more than 2$\sigma_{\rm EN}$.\\ 
		$f$. Percentage of DIBs whose EN exceeds the centre of the approximate Gaussian distribution by more than 2$\sigma_{\rm EN}$. }
\end{table*}

\subsection{Distribution of DIB intensities in APO sight lines}
\label{app:A.2:eta}

\begin{figure*}
	\begin{center}
		\includegraphics[scale=0.73, angle=0]{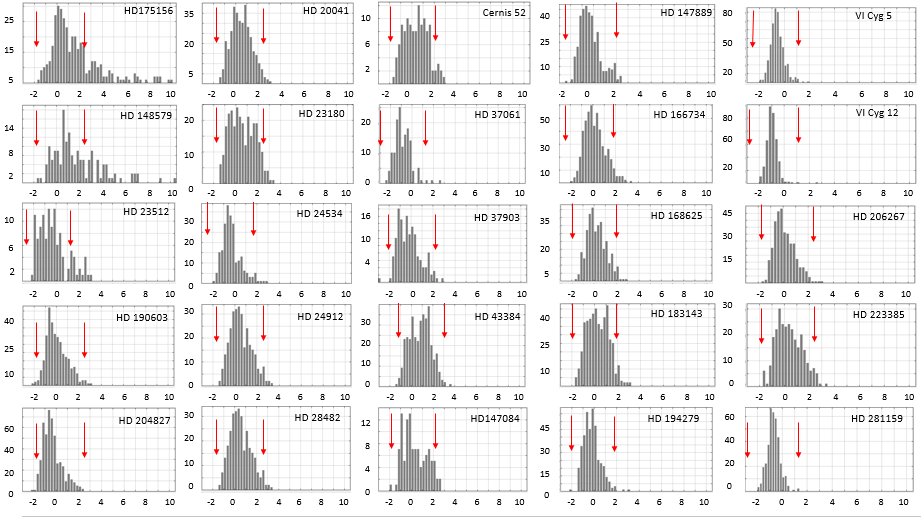}
		\caption{Distribution of the reduced enhancement $\eta$ (Eq.\ \ref{eq:8eta}) of the DIB normalised equivalent width, EN, in all 25 APO sight lines. In each histogram, red arrows mark the position of $\eta$-$\delta$\,=\,$\pm$2, where 
		 $\delta$ is the shift of the central roughly Gaussian distribution from EN$_{\rm av}$ (Table \ref{tab:A1sightlines}).  The line of sight of HD 175156 has four DIBs with formal values of $\eta_{175}$ -- 14, 15, 17 and 35 (Table \ref{tab:C1chia}) --  outside of the range displayed here (-3\,$<$\,$\eta$\,$<$10.5). The line of sight of HD 148579 has four DIBs with formal values of $\eta_{148}$ -- 11, 13, 16 and 32 (Table \ref{tab:C2chib}) --  outside of this range.}
		\label{fig:9A1-histos_etanew1}	
	\end{center}
\end{figure*}

Fig. \ref{fig:9A1-histos_etanew1} displays the histograms of $\eta$ values (reduced normalised equivalent widths, Eq.\ [\ref{eq:8eta}]) for the 25 APO lines of sight. For most sight lines (except HD\,175156, HD\,148579 	and HD\,23512), the $\eta$ distribution is compact and it may roughly be approximated by a Gaussian, whose centre is shifted by $\delta$. Approximate values of $\delta$ were inferred from the histograms and they are reported in Table \ref{tab:A1sightlines}. After correction for this shift, the histograms were used to approximately estimate the number of detected DIBs whose EN/EN$_{\rm av}$ exceeds 2$\sigma$ (Table \ref{tab:A1sightlines}), for supporting the discussion of Section \ref{sec:5.1HD175-148}.

\FloatBarrier
\section{Individual data of C$_2$ DIBs}
\label{app:BC2}

\subsection{Extended tables and figures}

\begin{table*}
	\caption{Data of 16 reference C$_2$-DIBs listed in Thorburn et al.\ (2003)}
	 \label{tab:B1thorburn}
	\begin{tabular}{lcccccccccccc}
		\hline \hline
		$\lambda^a$  & {\small EN204$^b$} & {\small EN147$^b$} & {\small R$_{21}^c$} & {\small N*$^d$} & kn$_{C2}^e$ & q$_{C2}^{e1}$ & knq$_{C2}^{e2}$  & {\small QC$_2^f$} &  {kn$_{6009}^g$} & {\small FWHM$^h$} & {\small R175av$^i$} & {$\chi^j$}  \\
		\AA  & {\tiny m\AA\ /mag} & {\tiny m\AA\ /mag} & & & & & & & &   \AA  & &\\
		\hline
		4363.83 &   13.4  &    8.1  &    $>$5  &    12    &     0.56 & 0.56 & 0.56  &     3  &     -0.45  &     0.6  &     0.70   &     \\
		{\bf 4726.98}$^k$ &    245 &    278 &{\bf 1.9}  &    23    &{\bf 0.81}  & 0
		91 & 0.86 &   4  &{\bf -0.19}  &    2.8  &     2.17  &  $\chi_a$   \\
		4734.77 &    13.5  &    13.1    &    $>$5  &    17    &     0.72 & 0.86 & 0.79  &     4  &     -0.59  &       0.4  &     1.06  &     \\
		{\bf 4963.92}$^k$ &    52.5 &    56.5  &{\bf 2.4}  &  24  &{\bf 0.89} & 0.95 &   & 4  &{\bf -0.52}  &    0.7  &     2.09  &     \\
		4969.12 &    14.3  &    14.5   &     2.2  &    14    &     0.66 & 0.94 & 0.80  &     4  &     -0.54  &      0.8  &     ""    &     \\
		4979.62 &    12.4  &    10.9   &     6  &    16    &     0.90 & 0.85 & 0.87   &     4  &     -0.70  &      0.6  &     1.05  &     \\
		{\bf 4984.78}$^k$ &   28.8  &    22.6 &{\bf $>$5} &   18   &{\bf 0.82} & 0.93 & 0.88 &   4  &{\bf -0.74}  &      0.5  &     2.62  &   $\chi_a$  \\
		5003.58 &    14.1  &    7.3  &     $>$5  &    12    &     0.58 & 0.75 & 0.67  &     4  &     -0.54  &    0.6  &     1.47  &     \\
		5170.49 &    8.9   &    8.4   &     $>$2.5  &    14    &     0.64 & 0.87 & 0.75  &     3 &     -0.08  &    0.4  &     1.8   &     \\
		5176.0  &    25.0  &    23.1  &     6  &    17    &     0.84 & 0.92 & 0.88  &     4  &     -0.79  &     0.5  &     1.47  &     \\
		 {\bf 5418.87}$^k$ &    43.8  &  37.9  &{\bf 4.5}  &    19   &{\bf 0.86} & 0.94 & 0.90 &  4  &{\bf -0.75}  &     0.7  &     1.71  &     \\
		{\bf 5512.68}$^k$ &    19.2  &   18.3 &{\bf 3.0}  & 19    &{\bf 0.83} & 0.90 & 0.86  &   4  &{\bf -0.74}  &    0.5  &     2.54  &  $\chi_{ab}$   \\
		5541.92 &    12.7  &    10.0  &    1.5  &    12    &     0.60 & 0.76 & 0.68  &     4  &     -0.74  &    0.6  &     2.98  &  $\chi_a$   \\
		{\bf 5546.46}$^k$ &    18.1  &   19.5 &{\bf 2.9}  &  20   &{\bf 0.87} & 0.92 & 0.89 &   4  &{\bf -0.60}  &      0.7   &     1.91  &  $\chi_{b}$   \\
		5762.68 &    11.4  &    12.9   &    2.9  &    17    &     0.68 & 0.61 & 0.65  &     4  &     -0.54  &    0.5  &     2.10   &     \\
		5769.09 &    18.7  &   18.2   &    $>5$ &    19    &     0.77 & 0.89 & 0.83  &     4   &     -0.73 &    0.6  &     2.37  & $\chi_a$   \\ 
		\hline
		\hline  
	\end{tabular}
	~~~~~~~~~~~~~~~~~~~~~~~~~~~~~~~~~~~~~~\\
	{\small $^a$ DIB wavelength. \\
		$^b$ Normalised APO DIB equivalent width, EW/E(B-V), in the lines of sight  of  HD\,204827 and HD\,147889, respectively. \\
		$^c$ Ratio of DIB equivalent widths in the sight lines  of HD\,204827 and HD\,183143. For DIBs undetected in HD\,183143, see Appendix \ref{app:B.3substitute}. \\
		$^d$ Number of sight lines in which the DIB is detected in the APO Catalog. \\
		$^e$ Average normalised correlation Pearson coefficient of the DIB (dropping the sight lines of HD\,175156 and 148579) with the sample of six major C$_2$ DIBs from Fan et al.\ (2022) (these DIBs are identified in bold in the table$^k$). \\ 
		$^{e1}$ Same as $^e$ for the regular Pearson coefficient.\\
		$^{e2}$ knq$_{C2}$ = [kn$_{C2}$ + q$_{C2}$]/2 (Eq.\ \ref{eq:3knq}).\\
		$^f$ Robustness index of the identification as C$_2$ DIB, as described in Section \ref{sec:4.2newC2} (4 is the highest robustness).\\
		$^g$ Normalised correlation Pearson coefficient of the DIB (dropping the sight lines of HD\,175156 and 148579) with the major $\sigma$ DIB, $\lambda$6009.\\
		$^h$ Measured width of the DIB from the APO Catalog (Fan et al.\ 2019).  \\
		$^i$ Ratio of the DIB equivalent width in the sight line HD\,175156 to the average of its equivalent width in all sight lines (Eq.\ \ref{eq:7Rav}).\\
		$^j$ Membership of the $\chi$ family (Section \ref{sec:5peculiarsightlines}, Tables \ref{tab:C1chia}-\ref{tab:C2chib}).\\
		$^k$ Member of the the sample of six major C$_2$ DIBs from Fan et al.\ (2022).\\
	}
\end{table*}

\begin{table*}
	\caption{Data of C$_2$-DIB candidates from the literature}
	\label{tab:B2literature}	
	\begin{tabular}{lccccccccccccc}
		\hline \hline
		$\lambda^a$  & {\small EN204$^b$} & {\small EN147$^b$} & {\small R$_{21}^c$} & {\small N*$^d$} & kn$_{C2}^e$ & q$_{C2}^{e1}$ & knq$_{C2}^{e2}$  & {\small QC$_2^f$} &  {kn$_{6009}^g$} & {\small FWHM$^h$} & {\small R175av$^i$} & {$\chi^j$} & ref \\
\hline
5061.50 &14.1  & 16.3 & 2.2  & 17  & 0.69 & 0.90 & 0.79  & 3 & -0.58  &0.53 & 2.9  &  & 5 \\
5245.44   & 8.5 & 6.4 & 1.5 & 13 & 0.38 & 0.85 & 0.62  & 2 & -0.46 & 0.5 & 3.07 & & 5 \\
 5547.43 & 3.7  &  4.3 &  1.3 & 18 & 0.46 & 0.70 & 0.58  & 1 & -0.16   & 0.45 & 0.95  & &   2 \\
 5769.91 & 4.1  & 4.5 &  1.9 & 17 & 0.48 & 0.66 & 0.57 & 0.5 & -0.05  & 0.51 & 1.75  &  & 2 \\ 
 5793.24 & 18.1 & 18.7 & 1.3  & 21 & 0.04  & 0.54 & 0.29 &  0   & 0.30&    0.82 & 1.52  &  & 4\\ 
 5828.50  & 10.5 & 16.8 &  1.1 & 20 & 0.69 & 0.84 & 0.77 & 3   &  0.06 & 0.71  & 3.56  & $\chi_a$ & 3  \\
 5849.82 &  83.1 & 64.0 & 1.5 & 25 & 0.70 & 0.84 & 0.77  & 2 & -0.01 & 0.8 & 2.1 & $\chi_a$ & 6\\
 5910.57 & 18.6  & 17.9 & 2.3  & 19  & 0.67 & 0.78 & 0.72 & 2 & -0.25  & 0.8 & 2.04 &  & 5 \\
 6729.22 & 15.2 & 13.6 &  2.0 & 22 & 0.53 & 0.78 & 0.72 & 2   & -0.38 & 0.51 & 0.66  &   & 1 \\ 	
		\hline
		\hline  
	\end{tabular}
	~~~~~~~~~~~~~~~~~~~~~~~~~~~~~~~~~~~~~~\\
	{\small Same notes as in Table \ref{tab:B1thorburn}. \\
	References (ref): 1 Thorburn et al.\ (2003); 2 Elyajouri et al.\ (2018); 3 Fan et al.\ (2022); 4 Fan et al.\ (2022, Table 1); 5 Ebenbichler et al.\ (2024);  6 Ka{\'z}mierczak et al.\ (2014).}
\end{table*}

\begin{figure*}
	\begin{center}
		\includegraphics[scale=0.77, angle=0]{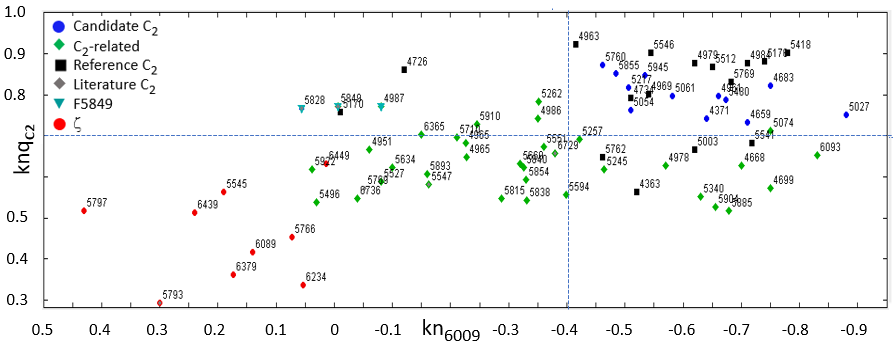}
		\caption{Correlation  diagram for candidate C$_2$-DIBs (blue dots, Table \ref{tab:B3robust}) and  possible C$_2$-related DIBs (green diamonds, Table \ref{tab:B4related}), comparing them to reference C$_2$-DIBs (black squares, Table \ref{tab:B1thorburn}), candidate C$_2$-DIBs from the literature (grey diamonds, Table \ref{tab:B2literature}), 
		possible DIBs similar to $\lambda$5849 (F5849, cyan triangles) and a comparison sample of $\zeta$ DIBs (red dots).   
			The horizontal axis displays the  normalised anticorrelation Pearson coefficient, kn$_{6009}$, with the $\sigma$ DIB $\lambda$6009 (excluding the sight lines of HD\,175156 and HD\,148579). The vertical axis displays the average correlation coefficient, knq$_{C2}$ (Eq.\ \ref{eq:3knq}), with a subsample of six reference C$_2$ DIBs ($\lambda\lambda$4726, 4963, 4984, 5418, 5512 and 5546). It is seen that the selection criteria (Eqs.\ (\ref{eq:4knqC2},\ref{eq:6R21C2})) discriminate well C$_2$  DIBs from $\zeta$ DIBs. Most  C$_2$  DIBs and related DIBs are significantly anticorrelated with $\sigma$ DIBs (kn$_{6009}$ $<$
			0). See Fig. \ref{fig:1-kq_k6009new} for a reduced version of this figure without DIB labels.}
	\label{fig:10B1-kq_k6009longnew}		
	\end{center}
\end{figure*}

\begin{figure*}
	\begin{center}
		\includegraphics[scale=0.77, angle=0]{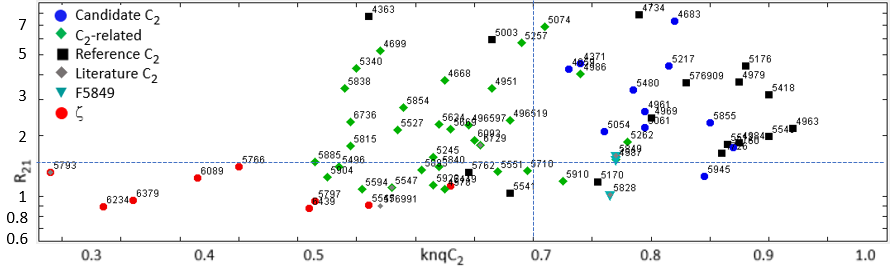}
		\caption{ 
			Same definitions as symbols as Fig. \ref{fig:10B1-kq_k6009longnew} for the correlation  diagram of solid candidate C$_2$-DIBs and possible C$_2$-related DIBs.  
			The horizontal axis displays the average normalised Pearson correlation factor, knq$_{C2}$ (Eq.\ \ref{eq:3knq} and Fig. \ref{fig:10B1-kq_k6009longnew}). 
			The vertical axis displays the ratio of the DIB equivalent widths in the sight line of HD\,204827 to HD\,183143 (Eq.\ \ref{eq:1R21}; actual values or equivalent values, see Appendix \ref{app:B.3substitute}). It is seen that the selection criteria (Eqs.\ [\ref{eq:4knqC2},\ref{eq:5kn6009C2}], dotted blue lines) discriminate well C$_2$  DIBs from $\zeta$ DIBs. Most C$_2$ DIBs and related DIBs have R$_{21}$ $\ge$ 1.5. See Fig. \ref{fig:2-R21_knqnew} for a reduced version of this figure without DIB labels.}
		 \label{fig:11B2-R21_knqlongnew}
	\end{center}
\end{figure*}

\begin{table*}
	\caption{Data of 12 confirmed new C$_2$-DIB candidates}
	\label{tab:B3robust}
	\begin{tabular}{lcccccccccccc}
		\hline \hline
		$\lambda^a$  & {EN204$^b$} &  {EN147$^b$} &  {R$_{21}^c$} & N*$^d$ & kn$_{C2F22}^e$  & q$_{C2}^{e1}$ & knq$_{C2}^{e2}$  &  QC$_2^f$ & {kn$_{6009}^g$} & {\small FWHM$^h$} & R175av$^i$ & {$\chi^j$}  \\
 \AA & {\small m\AA\ /mag} &  {\small m\AA\ /mag}  & & & & & & & & \AA & & \\
 \hline
4371.60 &(27) &  30.5   & (4) & 7   & 0.67 & 0.81 & 0.74  & 3 & -0.64  &0.70 &      &   \\
4659.86 & 5.5   & 9.9  & 4.2  & 11  & 0.68 & 0.78 & 0.73  & 4 & -0.71  &0.47 & 2.33 &   \\
4683.03 & 18.9   & 18.5  & 7.0 & 19  & 0.80 & 0.84 & 0.82  & 4 & -0.75  &0.46 & 2.14 &   \\
4961.93 & 4.5   & 3.6  & 2.6  & 9   & 0.72 & 0.87 & 0.79 & 4 & -0.66  &0.40 &      &   \\
5027.57 & 9.7   &       & $>$3.2  & 6   & 0.66 & 0.84 & 0.75 & 3 & -0.88  &0.56 &    &   \\
5054.85 & 10.8 &   6.1  & 2.1  & 11  & 0.65 & 0.87 & 0.76  & 3 & -0.51  &0.57 & 1.59 &   \\
5061.50 & 14.1  & 16.3  & 2.2  & 17  & 0.69 & 0.90 & 0.79  & 3 & -0.58  &0.53 & 2.9  & $\chi_a$  \\
5217.85 & 4.8 &  2.7 &  4.4  & 8   & 0.78 & 0.85 & 0.81  & 3 & -0.50  &0.41 & 1.76 &   \\
5480.79 & 7.0 &  6.1   & 3.3  & 13  & 0.76 & 0.81 & 0.78  & 3 & -0.67  &0.50 & 0.88 &   \\
5760.48 & 6.3  & 5.7  & 1.7 & 14  & 0.87 & 0.87 & 0.87 & 3 & -0.46  & 0.62 & 1.74 & $\chi_b$ \\
5855.62 & 5.5  & 4.1  & 2.3  & 15  & 0.82 & 0.88 & 0.85  & 4 & -0.48  &0.47 & 1.62 &   \\
5945.54 & 8.7   & 9.3 &  1.2  & 11  & 0.90 & 0.79 & 0.84  & 3 & -0.53  & 0.46 &      &  \\	
				\hline
		\hline  
	\end{tabular}
~~~~~~~~~~~~~~~~~~~~~~~~~~~~~~~~~~~~~~\\
	{\small Same notes as in Table \ref{tab:B1thorburn}, except the following. \\
	$^b$ Normalised APO DIB equivalent width (EW/E(B-V)) in the sight lines of HD\,204827 and HD\,147889, respectively. For $\lambda$4371, which is not detected in HD\,204827, an approximate value of EN204 is inferred from that of EN147 (see Appendix \ref{app:B.3substitute})\\
	}
\end{table*}

\begin{table*}
	\caption{Data of 34  C$_2$-related candidates}
	\label{tab:B4related}
	\begin{tabular}{lcccccccccccc}
		\hline \hline
				$\lambda^a$  & {EN204$^b$} & {EN147$^b$} & {R${21}^c$} & N*$^d$ & kn$_{C2}^e$  & q$_{C2}^{e1}$ & knq$_{C2}^{e2}$  &  QC$_2^f$ & {kn$_{6009}^g$}&  {\small FWHM$^h$} & R175av$^i$ & {$\chi^j$}  \\
		\AA  & {\small m\AA\ /mag} & {\small m\AA\ /mag} & & & & &  & & & \AA & & \\
		\hline
4668.66 & 9.3  & 18.7  & 4.2 & 14  & 0.52 & 0.73 & 0.63 & 2 & -0.70  & 0.6 & 3.34 & $\chi_a$ \\ 
4699.29 &(33)& 37.2  & (5) & 6   & 0.58 & 0.56 & 0.57 & 2 & -0.75  & 1.3 &      &   \\
4951.13 & 9.5  & 6.6  & 3.4 & 11  & 0.62 & 0.71 & 0.66 & 2 & -0.06  & 0.6 &        & \\
4965.19 & 6.1  & 9.3  & 1.8 & 16  & 0.56 & 0.80 & 0.68 & 2 & -0.23   & 0.5 & 2.75 & $\chi_a$ \\ 
4965.97  & 5.0 & 4.7 & 3.7 & 14 & 0.47 & 0.82 & 0.65 & 1 & -0.23 & 0.5 & 2.31  &  $\chi_{ab}$\\
4978.06 & 1.4  & 3.2  & $>$1.8 & 7   & 0.69 & 0.56 & 0.63 & 1 & -0.57  & 0.5 &      &  \\ 
4986.77 &8.8   & 6.1  & 4.7  & 16  & 0.63 & 0.85 & 0.74  & 2 & -0.35  & 0.8 & 1.91 &   \\
4987.53 &21.7  & 20.6 & 1.1  & 15  & 0.74 & 0.80 & 0.77  & 2 & -0.08 & 1.6 & 2.07 &   \\
5074.48 & 16.8 & 7.7 & 5.2 & 6 &  0.78 & 0.64 & 0.71 & 1 & -0.75 & 0.4 & & \\
5245.44   & 8.5 & 6.4 & 1.5 & 13 & 0.38 & 0.85 & 0.62  & 2 & -0.46 & 0.5 & 3.07 & $\chi_a$\\
5257.44 & 11.4 &  & $>$5 & 12  & 0.72 & 0.66 & 0.69 & 1 & -0.42   & 0.8 &      & $\chi_b$ \\
5262.44  & 4.8 & 3.7 & 1.8 & 14 & 0.72 & 0.84 & 0.78 & 2 &-0.35 & 2 & 0.6 &$\chi_b$ \\
5340.38 & 7.2  & 7.1  &  3.6 & 14  & 0.53 & 0.57 & 0.55 & 1 & -0.63  & 0.8 & 2.29  & $\chi_b$  \\
5496.23 & 4.6  & 2.8 & 1.4 & 12  &  0.41 & 0.66  & 0.53  & 1  & 0.03  &  0.5 & 3.06  &  $\chi_a$ \\
5527.49 & 5.7 &  & 2.6 & 10 & 0.43 & 0.74 & 0.58 & 1 & -0.08 & 0.6  & & \\
5547.43 & 3.7 & 4.3  & 1.1  & 18  & 0.46  & 0.70  &  0.58 & 1  & -0.16  & 0.5  & 0.95  &   \\
5551.01  & 2.1 & 3.1 & 1.3  & 10 & 0.49 & 0.85 & 0.67  & 1 & -0.36 & 0.4 & 1.59 & \\
5594.58 & 6.7  & 9.3  & 1.1 & 18  & 0.39  & 0.72  & 0.57 & 1 & -0.40 & 0.6 & 2.23  &  $\chi_a$ \\
5634.98 & 5.4 & 6.3 & 2.2 & 8 & 0.47 & 0.77 & 0.62 & 1 & -0.10 & 0.8 &  & \\
5669.12 & 6.8  & 3.1  & 3.8 & 14  & 0.59 & 0.67 & 0.63 & 2 & -0.32   & 0.7 &      &  \\ 
5710.58 & 11.2 & 6.8  & $>$5. & 15  & 0.63 & 0.76 & 0.69 & 2 & -0.21   &  0.6  & 5.3  & $\chi_a$ \\  
5815.72 & 5.1  & 6.1  & 1.8 & 13  & 0.52 & 0.57 & 0.54 & 1 & -0.29   & 0.7 & 1.79 &  \\ 
5838.04  & 6.8 & 6.0 & 3.4 & 13 & 0.49 & 0.59 & 0.54 & 1 & -0.33 & 0.5 & 1.84 &$\chi_b$ \\
5840.65 & 8.6   & 6.7  & 3.9  & 19   & 0.61 & 0.63 & 0.62  & 2 & -0.33  & 0.5 & 1.53 &   \\
5854.58 & 6.8 & 5.2 & 2.7 & 13 & 0.39 & 0.79 & 0.59 & 1 & -0.33 & 0.6 & 1.41 & \\
5885.40 & 5.0 & 6.1  & 1.5 & 7 & 0.49 & 0.54  & 0.51  & 1 & -0.68 & 0.6 & 2.53  &   \\
5893.52 & 8.6  & 10.0 & 1.3  & 14  &  0.40 & 0.80  &  0.60 & 1  & -0.16  & 0.8  & 2.17  &  $\chi_a$ \\
5904.58  & 3.7  &  & 1.2 & 9 & 0.43 & 0.62 & 0.53  & 1 & -0.66 & 0.9 & & \\
5910.57 & 18.6  & 17.9 & 2.3  & 19  & 0.67 & 0.78 & 0.72 & 2 & -0.25  & 0.8 & 2.04 &  $\chi_a$ \\
5922.32 &  4.3 & 5.3  & 1.13  & 15  &  0.48 & 0.75  & 0.61  & 1  & 0.04  & 0.5  & 2.57  & $\chi_{ab}$  \\
6093.33 & 4.1  &   & 2.1 & 9   & 0.65 & 0.65 & 0.65 & 1 & -0.83  & 0.8 & 2.82 & $\chi_{ab}$  \\ 
6365.02 & (7) & 8.0 & & 8 & 0.54 & 0.70 & 0.62 & 1 & -0.15 & 1.3 &  & \\	
6729.22  & 15.2 & 13.6 & 1.8 & 22 & 0.53 & 0.78 & 0.66 & 2 & -0.38 & 0.6 & 1.17 & \\	
6736.27 & 5.3  &   & 2.3  & 14  & 0.56  & 0.53  & 0.54  &  1 & -0.04  & 0.6  &   &   \\
		\hline
		\hline  
	\end{tabular}
	~~~~~~~~~~~~~~~~~~~~~~~~~~~~~~~~~~~~~~\\
	{\small Same notes as in Table B.1, except the following. \\
		$^b$ Normalised APO DIB equivalent width (EW/E(B-V)) in the sight lines of HD\,204827 and HD\,147889, respectively. For $\lambda\lambda$4699 \& 6365, that are not detected in HD\,204827, an approximate value of EN204 is inferred from that of EN147 (see Appendix \ref{app:B.3substitute})\\
	}
\end{table*}

\begin{table}[!htbp]
	\tiny
	\caption{Average properties of various C$_2$-DIB groups$^a$}
	\label{tab:B5C2stat}
	\begin{tabular}{lcccccc}
		\hline 
		 C$_2$DIB group & {\tiny N$_{\rm DIB}^b$}  &  {\tiny EN$_{204}^c$}  &  {\tiny kn$_{C2}^d$} & {\tiny knq$_{C2}^{dd}$} & {\tiny kn$_{6009}^e$} &  R$_{21}^f$  \\
		\hline
		&&&&&&\\
		RefAll$^g$ 	             & 16  &  35  & 0.75 & 80 &  -0.54 &    3.2  \\
		{\small Refwo4726$^h$}   & 15  &  20  &	0.74 & 0.78 &  -0.57 &    3.3  \\
		{\small RefHighcorr$^i$} & 11  &  23  & 0.77 & 0.81 &  -0.56 &    3.1  \\
		{\small RefMedcorr$^j$}	 & 4   &  13  & 0.60 & 0.66 &  -0.58 &    4.0  \\
		C$_2$ candidates$^k$    &  12    &  10  & 0.75 & 0.79 &  -0.63   &  3.3 \\
		C$_2$-related$^l$	    &   34   &   11  & 0.55 & 0.63 &  -0.33   & 2.3  \\ 
		$\zeta$F22$^n$         &  10   &  41  & 0.31 & 0.41  &  0.17   & 1.1  \\
		\hline
		\hline
	\end{tabular}
	{\small ~~~~~~~  \\
		$^a$ See Section \ref{sec:4.2newC2}. \\
		$^b$ Number of DIBs in the group. \\
		$^c$ Average normalised equivalent width  in the  line  of sight of HD\,204827. \\
		$^d$ Average normalised correlation Pearson coefficient with the sample of six major C$_2$ DIBs from Fan et al.\ (2022). \\
		$^{dd}$ Corresponding average value of knq$_{C2}$ = [kn$_{C2}$ + q$_{C2}$]/2 (Eq.\ \ref{eq:3knq}). \\
		$^e$ Average normalised correlation Pearson coefficient with the  $\sigma$ DIB $\lambda$6009.\\
		$^f$ Average ratio of DIB equivalent widths in the sight  lines of HD\,204827 to HD\,183143. For DIBs undetected in HD\,183143, see Appendix \ref{app:B.3substitute}.\\  
		$^g$ All 16 reference C$_2$ DIBs (Table \ref{tab:B1thorburn}). \\
		$^h$ 16 reference C$_2$ DIBs except $\lambda$4726.\\
		$^i$ Highly correlated  reference C$_2$ DIBs (knq$_{C2}$ $\ge$ 0.8). \\
		$^j$ Moderately correlated  reference C$_2$ DIBs (0.55 $\le$ knq$_{C2}$ $\le$ 0.7). \\
		$^k$ Solid C$_2$-DIB candidates (Section \ref{sec:4.2newC2}, Table \ref{tab:B3robust}).\\
		$^l$ C$_2$-related DIB candidates (Section \ref{sec:4.2newC2}, Table \ref{tab:B4related}).\\
		$^n$ Reference $\zeta$ DIBs from Fan et al.\ (2022).\\
	}
\end{table}

Table \ref{tab:B1thorburn} recalls the most standard list of 16 reference C$_2$-DIBs, established by Thorburn et al.\ (2003), and their properties. This list  was basically confirmed by Ka{\'z}mierczak et al.\ (2010) and Elyajouri et al.\ (2018). 

Additional C$_2$ DIBs previously proposed by various authors are listed in Table \ref{tab:B2literature}. They include: $i)$ $\lambda$6729 (Thorburn et al.\ 2003; Fan et al.\ 2022) and $\lambda\lambda$5061, 5245 \& 5910 (Ebenbichler et al.\ 2024), that are also included in Tables \ref{tab:B3robust} or  \ref{tab:B4related}
; $ii)$ $\lambda$5828 (Fan et al.\ 2022) and $\lambda$5849 (Ka{\'z}mierczak et al.\ 2014), that are significantly anticorrelated with $\lambda$6009 and might be in the transition between the C$_2$ and $\zeta$ families; $iii)$ $\lambda$5547 and $\lambda$5769.91, that were tentatively proposed as C$_2$ DIBs by Elyajouri et al.\ (2018); however, they hardly fit the criteria discussed here for C$_2$-DIB candidates, although $\lambda$5547 is included in the list of C$_2$-related DIBs (Table \ref{tab:B4related}); $iv)$  $\lambda$5793 is definitely not a C$_2$ DIB, although it was listed as so by error in Table 1 of Fan et al.\ 2022, while it is rightly quoted as a $\zeta$ DIB in the rest of this paper. 

As discussed in Section \ref{sec:4C2}, the parameters of classical C$_2$ DIBs of Table \ref{tab:B1thorburn} are the reference for establishing the criteria of Eqs.\ (\ref{eq:4knqC2}-\ref{eq:6R21C2}), that are used for identifying possible additional members of the extended C$_2$ family. These criteria include: i) knq$_{C2}$ = (kn$_{C2}$ + q$_{C2}$)/2 (Eq.\ \ref{eq:3knq}), the average correlation coefficient of the DIB with the sample of six major C$_2$ DIBs from Fan et al.\ (2022); ii) the anticorrelation coefficient with $\lambda$6009, one of the most extreme $\sigma$ DIBs; and iii) the ratio R$_{21}$ = EW(204827)/EW(183143) (Eq.\ \ref{eq:1R21}). They are combined for defining a quality factor, Q$_{C2}$, for belonging to the C$_2$ family, from their proximity to the criteria of Eqs.\ (\ref{eq:4knqC2}-\ref{eq:6R21C2}) (see Figs\ B.1-\ref{fig:11B2-R21_knqlongnew}).

As described in Section \ref{sec:4.2newC2}, the values of Q$_{C2}$ allow the selection of 12 solid  C$_2$-DIB candidates, with Q$_{C2}$ = 4 or 3. They are listed in Table \ref{tab:B3robust}, with the same parameters as for reference C$_2$-DIBs in Table \ref{tab:B1thorburn}. 34  additional tentative candidates, with Q$_{C2}$ = 2 or 1, are similarly listed in Table \ref{tab:B4related}, as C$_2$-related DIBs.

Figures B.1 and B.2 are the extensions of Fig. \ref{fig:1-kq_k6009new} \& Fig. \ref{fig:2-R21_knqnew} with source labels. They display the positions of all C$_2$-like DIBs (reference, candidates and related), 
with respect to the C$_2$ selection criteria of Eqs.\ (\ref{eq:4knqC2}-\ref{eq:6R21C2}). Reference $\zeta$ DIBs (Fan et al.\ 2022) have been added for comparison. Fig. \ref{fig:10B1-kq_k6009longnew} plots knq$_{C2}$ $versus$ kn$_{6009}$. Fig. \ref{fig:11B2-R21_knqlongnew} plots R$_{21}$ (Eq.\ \ref{eq:1R21}) $vs$ knq$_{C2}$. 

\subsection{Groups and peculiarities of C$_2$ and related DIBs }
\label{app:B.2groups}
As quoted, the degree of correlation between the 16 reference C$_2$ DIBs (Table \ref{tab:B1thorburn}, Thorburn et al. 2003) is diverse. From Fig. \ref{fig:10B1-kq_k6009longnew}, 
one may distinguish a main group (denoted "HighCorr") of 12 DIBs, highly correlated (knq$_{C2}$\,$\ga$\,0.8) with the six C$_2$ DIBs of Fan et al. (2022), and a group ("MidCorr") of four DIBs ($\lambda\lambda$5003, 5541, 5762 \& 4363), mildly correlated (knq$_{C2}$\,$\la$\,0.65).
 It is seen that most solid C$_2$ candidates may be associated with the HighCorr group, and none with the MidCorr group. Most possible C$_2$-related DIBs are intermediate between the MidCorr group and $\zeta$ DIBs. 

Table \ref{tab:B5C2stat} 
summarises some average properties of the different groups of C$_2$ DIBs and candidates. It is seen that the solid C$_2$ candidates are comparable to the HighCorr reference group for  knq$_{C2}$ and kn$_{6009}$, but they have a much weaker intensity, e.g.\ EN(204827).  
The possible C$_2$-related DIBs have a comparable intensity with the C$_2$ candidates, but significantly lower correlation coefficients.

Note that the 27 C$_2$ DIBs or candidates, that are tentatively classified as $\chi$ DIBs, are discussed in Appendix \ref{app:C.4chimain}
 (see Tables \ref{tab:C1chia}-\ref{tab:C2chib}, \ref{tab:B3robust}-\ref{tab:B4related} and Fig. \ref{fig:19C7-6010_R21chinew}). They include most $\chi$ DIBs with kn$_{6009}$\,$<$\,-0.2, and a significant number of Thorburn reference C$_2$-DIBs, namely: $\lambda\lambda$ 4726, 4984, 5512, 5541, 5546, 5769.09.

In the diagram, knq$_{C2}$/kn$_{6009}$ (Fig. \ref{fig:10B1-kq_k6009longnew}), two reference C$_2$ DIBs, $\lambda\lambda$ 4726, 5170, occupy a peculiar position in the upper left part of the figure (see also Fig. \ref{fig:3-R21_k6010new1}). They seem to be associated with the major DIB $\lambda$5849, that is known to be at the transition between the $\zeta$ and C$_2$ families. Such an association might be confirmed by the high correlation coefficients of these two C$_2$ DIBs with $\lambda$5849 (Fig. \ref{fig:4-q5797_5849new}). Another C$_2$ DIB, $\lambda$5828 (Fan et al.\ 2022, Table \ref{tab:B2literature}), and a C$_2$-related DIB of Table \ref{tab:B4related}, $\lambda$4987, lie close to these three DIBs in Fig. \ref{fig:10B1-kq_k6009longnew}. In order to see whether they might be associated with $\lambda\lambda$ 5849, 4726, 5170, they are identified with the same symbol as $\lambda$5849 in Figs. \ref{fig:10B1-kq_k6009longnew}- \ref{fig:11B2-R21_knqlongnew} and \ref{fig:1-kq_k6009new}-\ref{fig:2-R21_knqnew}. The association seems possible for $\lambda$5828, but more doubtful for $\lambda$4987.

\begin{figure}
	\begin{center}
		\includegraphics[scale=0.52, angle=0]{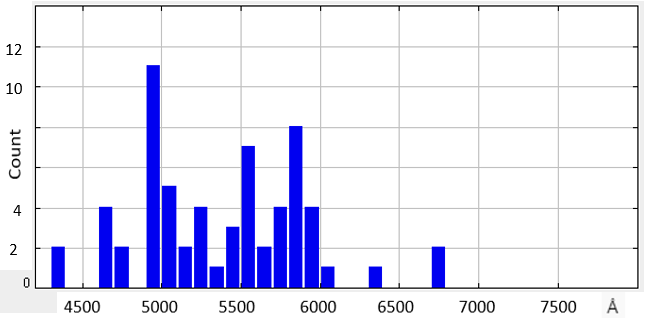}
		\caption{Wavelength distribution of all 65 C$_2$ DIBs and candidates (Tables \ref{tab:B3robust}-\ref{tab:B4related}), showing their limitation to $\lambda$ $<$ 6000\,\AA\ for most of them. 
		}
		 \label{fig:12B3-lambda_allC2}
	\end{center}
\end{figure}

\subsection{Substitute parameters of C$_2$ DIBs and others} \label{app:B.3substitute}	
Because of the incomplete detections of DIBs in any sight line, some of the key parameters, used for characterising or selecting C$_2$ DIBs, would lack in Tables \ref{tab:B3robust}-\ref{tab:B4related}. In most cases such lacking parameters may simply be replaced by substituted values inferred from similar sight lines or DIBs that are present in the APO Catalog.

\smallskip
$\bullet$ {\it EN(204827)} 

The normalised equivalent width, EN(204827), is used to 
characterise the strength of C$_2$ (and $\zeta$) DIBs and in the definition of the ratio R$_{21}$ = EN(204827)/EN(183141) (Eq.\ \ref{eq:1R21}).
When a DIB is not detected in the sight line of HD\,204827, the value used for EN(204827) may be inferred from the detection of the DIB in HD\,147889 and the high correlation  of the strength of C$_2$ DIBs  in these two sight lines. In the case of C$2$-like DIBs, the relation EN(204827)\,$\sim$\,0.89\,$\times$\,EN(147889) is used [while for all DIBs, EN(204827)\,$\sim$\,0.75\,$\times$\,EN(147889)].

\smallskip
$\bullet$ {\it EN(183143)}

EN(183143) is used to characterise 
the strength of DIBs in $\sigma$ sight lines and in the definition of the ratio R$_{21}$. In case of non detection, it may be inferred from a detection in the highly correlated sight line of HD\,166734, using  EN(183143)\,$\sim$\,1.38\,$\times$\,EN(166734).

\smallskip
$\bullet$ {\it R$_{21}$}
	
In most cases, R$_{21}$ is calculated from the actual or inferred values of EN(204827) and EN(183141). However, in case the C$_2$-like DIB is undetected in both HD\,183143 and HD\,166734, an approximate lower value of R$_{21}$ may be inferred from the upper limits of EN in these two sight lines. 

\smallskip
$\bullet$ {\it Correlation coefficients with $\lambda$6009 DIB}

$\lambda$6009 is one of the most extreme $\sigma$ DIBs, with a high detection rate, so that its anticorrelation with C$_2$ DIBs is used to characterise 
their sensitivity to UV  quenching (Figs.\ \ref{fig:1-kq_k6009new} and \ref{fig:10B1-kq_k6009longnew}). It has a twin DIB, $\lambda6011$, of similar intensity, that is correlated at 96\%, so that it is likely that they have the same carrier. When calculating or discussing correlations or anticorrelations with other DIBs, it is equivalent or slightly better to use EN(6010) = [EN(6009)+EN(6011)]/2 or EW(6010) = [EW(6009)+EW(6011)]/2) (as it is done, e.g. in Figs. \ref{fig:3-R21_k6010new1} \& \ref{fig:19C7-6010_R21chinew}). 
\FloatBarrier
\FloatBarrier
\section{Individual data of $\chi$  DIBs}
\label{app:Cchi}	

\subsection{Extended tables and figures of $\chi$ DIBs}
\label{app:C.1tabsfigs}	

\begin{figure}
 	\begin{center}
 		\includegraphics[scale=0.55, angle=0]{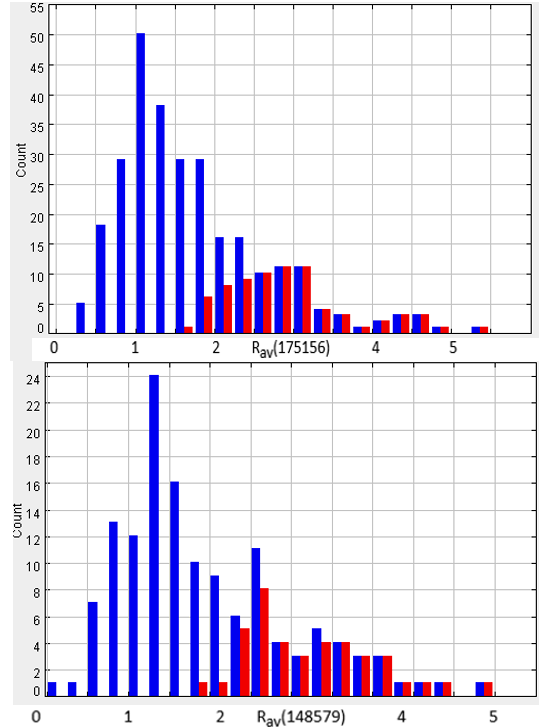}
 		\caption{Distribution of the ratio R$_{\rm av}$(175156) (Eq.\ \ref{eq:7Rav}) [resp. R$_{\rm av}$(148579)] of the DIB normalised equivalent widths in the sight  line of HD\,175156 (resp. HD\,148579) to their average in all sight lines, for all detected DIBs (blue), and for $\chi_a$ and $\chi_b$ DIBs (red), respectively (Tables \ref{tab:C1chia}-\ref{tab:C2chib}).
 		}  
 	\label{fig:13C1-r175-148avnew}			
 	\end{center}
 \end{figure}

Table \ref{tab:C1chia}  lists the 79 $\chi_a$ DIBs, that are enhanced in HD\,175156, with their main
 properties (Section \ref{sec:5.1HD175-148}), including: the normalised equivalent width, EN(175156), and its reduced value $\eta_{175}$ (Eq.\ \ref{eq:8eta}); the
  enhancement ratio, R$_{\rm av}$(175156) (Eq.\ \ref{eq:7Rav}, Fig. \ref{fig:13C1-r175-148avnew});  the correlation coefficients with $\lambda\lambda$6009, 5797 \& 5512, plus R$_{21}$, that characterised their UV behaviour; and average correlation coefficients with three main $\chi_a$ DIBs, three main  $\chi_b$ DIBs (Eqs.\ \ref{eq:C2kn175}-\ref{eq:C3kn148}) and with the three strong very broad $\chi$ DIBs whose wavelengths might match strong bands of two carbon chains and a ring  (Sections 5.2 \& 6.1), 
  \begin{equation}	\label{eq:C1knchr}
  	kn_{\rm chr} = (kn_{6412} + kn_{5419} + kn_{6128})/3, 
  \end{equation}  
  
  Table \ref{tab:C2chib} lists similar quantities for 43  $\chi_b$ DIBs, that are enhanced in HD\,148579, and their properties in this sight line.
  
  Fig. \ref{fig:13C1-r175-148avnew} displays the distribution of the ratio R$_{\rm av}$(175156) [resp. R$_{\rm av}$(148579)] (Eq.\ \ref{eq:7Rav}), for  $\chi_a$ (resp. $\chi_b$) DIBs, compared with all detected DIBs. 
  
\begin{figure}
	\begin{center}
		\includegraphics[scale=0.65, angle=270]{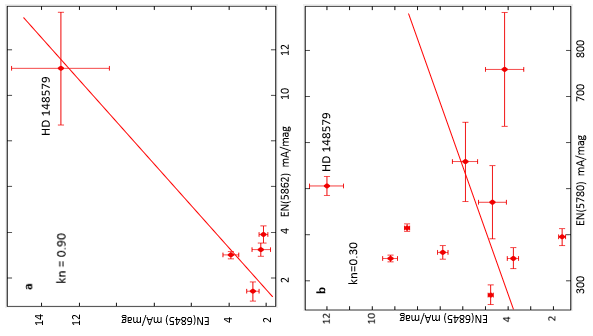}
		\caption{Examples of the intensity distribution of DIBs detected in the various sight  lines (similar to Fig. \ref{fig:5-6128-6087-5766new}, with correlation fits and normalised Pearson coefficients kn) for three DIBs including: a) two DIBs, $\lambda$6845 and $\lambda$5862, that are enhanced in the sight  line of HD\,148579, compared with all other sight  lines (R$_{\rm av}$(148579)\,=\,3.3 and 2.8, respectively); and b) a more normal DIB, $\lambda$5780, classified as $\sigma$-type (Fan et al.\ 2022); the intensity of $\lambda$5780 in HD\,148579 remains comparable with or lower than  some other sight  lines (R$_{\rm av}$(148579)\,=\,0.87).}
	\label{fig:14C2-6845_5862_5780new}		
	\end{center}
\end{figure}

  An example of peculiar enhancement in the sight line of HD\,148579 is provided in Fig. \ref{fig:14C2-6845_5862_5780new}a for two $\chi_b$ DIBs. They are much stronger than in all other sight lines, by a factor, R$_{\rm av}$(148579) $\sim$ 3, compared with the sight-line average. On the other hand, Fig. \ref{fig:14C2-6845_5862_5780new}b  displays the case of a normal ($\sigma$) DIB, whose intensity in HD\,148579 remains comparable to or lower than some other sight lines [R$_{\rm av}$(148579) = 0.87]. 
  
\begin{figure}
	\begin{center}
		\includegraphics[scale=0.72, angle=270]{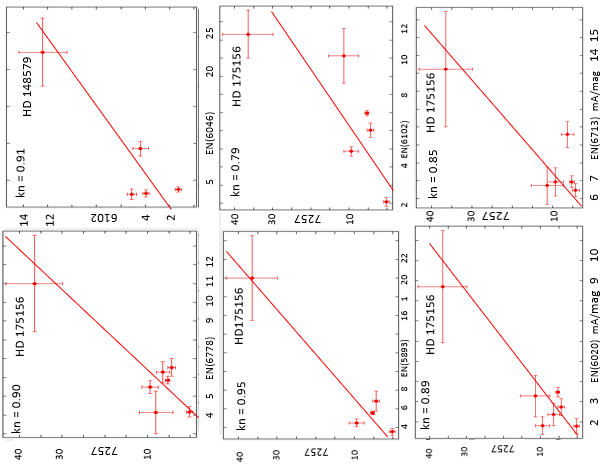}
		\caption{Examples of significant, but indeterminate, correlations in the sight  lines of HD\,175156 and HD\,148579 (with correlation fits and formal Pearson coefficients kn $>\sim$ 0.8). In all cases, it is estimated that a  conservative lower limit of the actual value of kn is 0.5.}
	\label{fig:15C3-examplesnew}			
	\end{center}
\end{figure}

\begin{figure}
	\begin{center}
		\includegraphics[scale=0.56, angle=0]{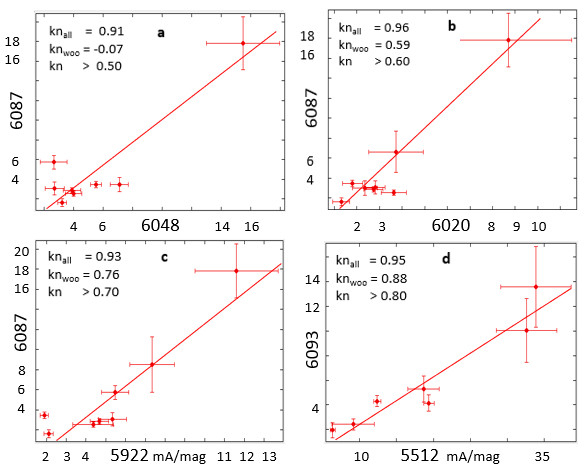}
		\caption{Examples of estimated lower limits for indeterminate normalised Pearson correlation coefficients, in the insert: kn$_{\rm all}$, formal coefficient with all sight lines (displayed fit); kn$_{\rm woo}$, formal coefficient excluding the sight lines of HD\,175156 and HD\,148579; kn, estimated lower limit (Tables \ref{tab:C1chia}-\ref{tab:C2chib}). Note the large differences between the values of kn$_{\rm all}$ and kn$_{\rm woo}$.} 
	\label{fig:16C4-undeternew}			
	\end{center}
\end{figure}

  Fig. \ref{fig:15C3-examplesnew} \& Fig. \ref{fig:16C4-undeternew} show examples of significant, but indeterminate, correlations in the sight  lines of HD\,175156 and HD\,148579 (with  formal Pearson coefficients kn $\ga$ 0.8). In Fig. \ref{fig:16C4-undeternew}, examples are given about the way approximate lower limits of kn were systematically estimated in such cases. Such limits are reported in Tables \ref{tab:C1chia}-\ref{tab:C2chib} for kn$_{5512}$, and they were used for kn$_{6087}$, kn$_{6093}$, kn$_{6128}$, kn$_{5257}$, kn$_{6412}$, kn$_{6845}$ and kn$_{5419}$, which enter in the calculation of the average coefficients kn$_{175}$,  kn$_{148}$ and kn$_{\rm chr}$  that are reported in Tables \ref{tab:C1chia}-\ref{tab:C2chib} (Section \ref{sec:5.1HD175-148}, Eqs.\ (\ref{eq:C1knchr}-\ref{eq:C3kn148}).
  
\begin{figure}
	\begin{center}
		\includegraphics[scale=0.65, angle=0]{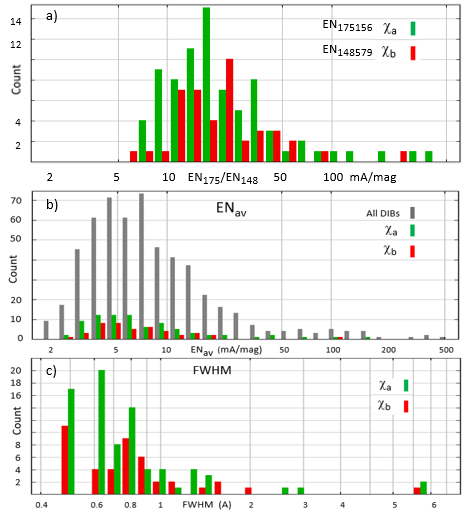}
		\caption{Distribution of normalised equivalent width
			and width, FWHM, of $\chi_a$ and $\chi_b$ DIBs: {\bf a)} Distribution of EN$_{175156}$ of $\chi_a$ DIBs in the sight line of HD\,175156 and of EN$_{1148579}$ of $\chi_b$ DIBs in the sight line of HD\,148579; {\bf b)} Distribution of average value of EN, EN$_{\rm av}$, over all sight lines, for all DIBs and for $\chi_a$ and $\chi_b$ DIBs; {\bf c)} Distribution of the widths of $\chi_a$ and $\chi_b$ DIBs. 
	}
	\label{fig:17C5-ENav_FWHM}
	\end{center}
\end{figure}

  Fig. \ref{fig:17C5-ENav_FWHM} displays the distribution of EN(175156), EN$_{\rm av}$ and the DIB width, for $\chi_a$ DIBs and HD\,175156, and similar quantities for $\chi_b$ DIBs and HD\,148579. 
  
\begin{figure}
	\begin{center}
		\includegraphics[scale=0.55, angle=0]{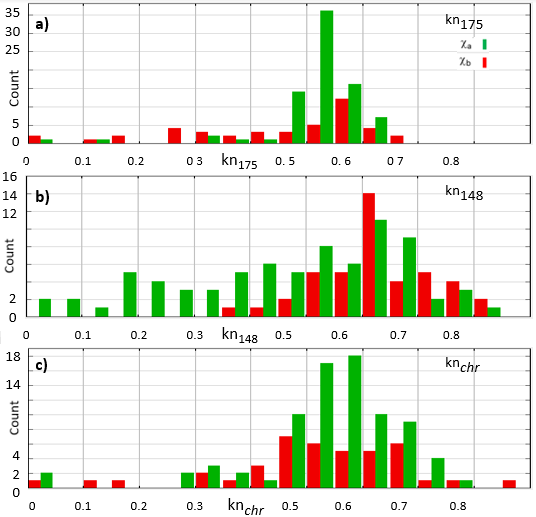}
		\caption{Average lower-limit values of normalised correlation coefficients kn, for the $\chi_a$ DIBs (green bars) and the $\chi_b$ DIBs (red bars), with: {\bf a)} three $\chi_a$ DIBs $\lambda\lambda$6087, 6093, 6128 (kn$_{175}$, Table \ref{tab:C1chia}, Eq.\ \ref{eq:C2kn175});  {\bf b)} three $\chi_b$ DIBs $\lambda\lambda$5257, 6412, 6845 (kn$_{148}$, Table \ref{tab:C2chib}, Eq.\ \ref{eq:C3kn148}); {\bf c)} three $\chi$ DIBs that might be carried by a carbon chain or ring, $\lambda\lambda$5419, 6128 \& 6412 (kn$_{\rm chr}$, Tables \ref{tab:C1chia}-\ref{tab:C2chib}, Eq.\ \ref{eq:C1knchr}).}
	\label{fig:18C6-kn3}
	\end{center}
\end{figure}

  Fig. \ref{fig:18C6-kn3} displays the distribution of the lower limits of the average values of the correlation coefficients of  $\chi_a$  and $\chi_b$ DIBs with three groups of three $\chi$ DIBs (Eqs.\ \ref{eq:C1knchr}-\ref{eq:C3kn148}). 
  
\begin{figure}
	\begin{center}
		\includegraphics[scale=1, angle=0]{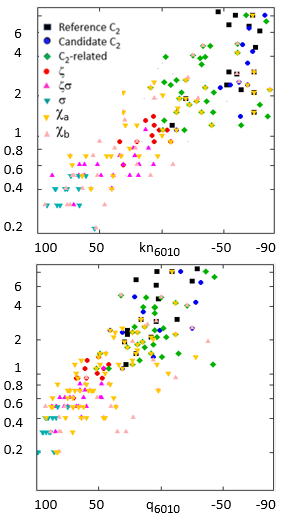}
		\caption{Correlation diagram for 79 $\chi_a$ DIBs (Table \ref{tab:C1chia}) and 43 $\chi_b$ DIBs (Table \ref{tab:C2chib}), compared with C$_2$ reference and candidate DIBs and C$_2$-related DIBs (Tables \ref{tab:B3robust}-\ref{tab:B4related}), and reference main $\zeta$, $\zeta\sigma$ and $\sigma$ DIBs from Fan et al.\ (2022).
		 The vertical axis displays the ratio R$_{21}$ of the DIB equivalent widths in the sight line of HD\,204827 to HD\,183143 (actual values or equivalent values, see Appendix \ref{app:B.3substitute}).
	 	 In the top (resp. bottom) panel, the horizontal axis displays the anticorrelation factor, kn$_{6010}$, equivalent to kn$_{6009}$ (resp. q$_{6010}$, equivalent to q$_{6009}$) (excluding the sight lines of HD\,175156 and HD\,148579). The comparison of the top and bottom panels visualises
	 	  the difference between kn$_{6010}$ and q$_{6010}$.} 
	\label{fig:19C7-6010_R21chinew}		
	\end{center}
\end{figure}

\begin{figure}
	\begin{center}
		\includegraphics[scale=0.47, angle=0]{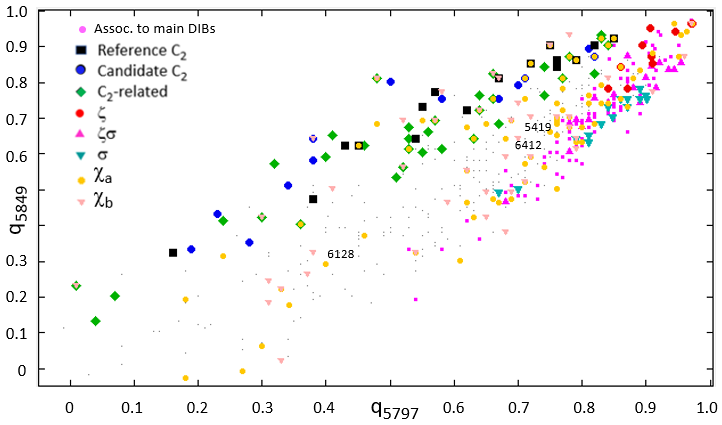}
		\caption{Regular correlation coefficients q  with $\lambda$5797 $versus$ $\lambda$5849 (excluding the sight lines of HD\,175156 and HD\,148579) for: i) $\chi_a$ and $\chi_b$ DIBs (Tables \ref{tab:C1chia}-\ref{tab:C2chib}); ii) the same comparison samples as Fig. \ref{fig:19C7-6010_R21chinew}; and iii) the additional sample of 106 DIBs of Table \ref{tab:D1mainass} that are strongly correlated with the main DIBs (Appendix \ref{app:D.1associated}, pink small dots). } 
	\label{fig:20C8-5797_5849chinew}
	\end{center}
\end{figure}

  Fig. \ref{fig:19C7-6010_R21chinew}  shows the diagrams of kn$_{6010}$ $vs$ R$_{21}$, for $\chi_a$  and $\chi_b$ DIBs and comparison samples.  Fig. \ref{fig:20C8-5797_5849chinew}  shows a similar diagram  for q$_{5797}$ $vs$ q$_{5849}$.
  
  Table \ref{tab:C3chimain} reports some statistics about the highest correlation coefficients of $\chi$ DIBs with classical strongest DIBs (Appendix \ref{app:C.4chimain}
  ). 
 
\begin{figure}
	\begin{center}
		\includegraphics[scale=0.55, angle=0]{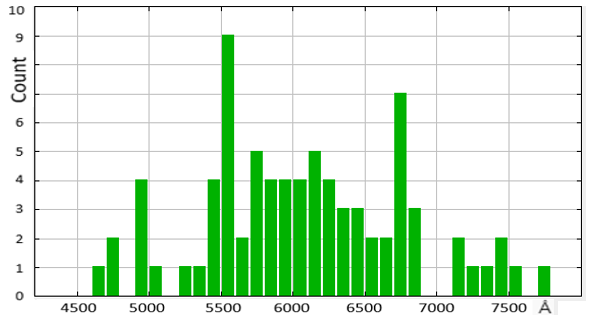}
		\caption{Same as Fig. \ref{fig:7-chi_lambda}, for the wavelength distribution of the 79 $\chi_a$ DIBs of Table \ref{tab:C1chia}. }
	\label{fig:21C9-chia_lambda}
	\end{center}
\end{figure}

\begin{figure}
	\begin{center}
		\includegraphics[scale=0.55, angle=0]{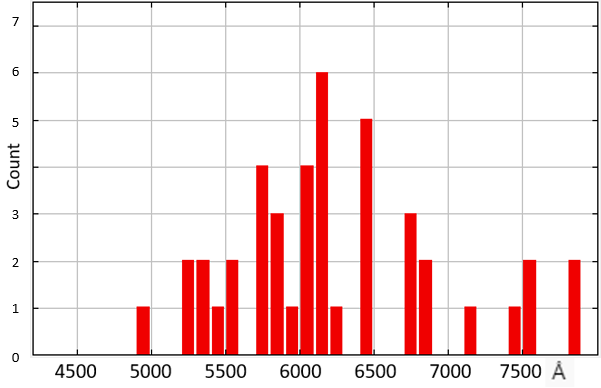}
		\caption{Same as Figs.\ 7 \& \ref{fig:21C9-chia_lambda}, for the wavelength distribution of the 43 $\chi_b$ DIBs  of Table \ref{tab:C2chib}}
		\label{fig:22C10-lambda_chib}
	\end{center}
\end{figure}

  Fig. \ref{fig:21C9-chia_lambda} \& Fig. \ref{fig:22C10-lambda_chib} display the wavelength distribution of $\chi_a$ and  $\chi_b$ DIBs, respectively. 
  
\begin{table*}
	\caption{Data of 79  $\chi_a$ DIBs}
	\label{tab:C1chia}
	\begin{tabular}{lccccccccccccll}
		\hline  
		 &&&&&&&&&&&&&& \\
		$\lambda^a$ & $\eta_{175}^{aa}$  & {EN175$^b$}  &R$_{175av}^c$ & N*$^d$ &  kn$_{175}^{e1}$ &  kn$_{148}^{e2}$ & kn$_{\rm chr}^{e3}$  &  {kn$_{6009}^f$} & R$_{21}^g$  &{\small FWHM$^h$} &   kn$_{5512}^e$ & q$_{5797}^e$ & C$_2$m$^i$ & $\chi_b^j$\\
		\AA &  & {\tiny m\AA\ /mag}   & & & &&&& & \AA & &  &  &   \\  
		\hline
		 &&&&&&&&&&&&&& \\
		4668.66   & 4.7  & 30  & 3.3  & 14  & 0.57& 0.44 & 0.52   & -0.70  & 4.2  & 0.6  &  0.81 & 0.36  & C$_2^r$  &  \\
		4726.98	   & 2.8  & 340  & 2.2  & 24  & 0.50& 0.54 & 0.60  & -0.12  & 1.9  & 2.8  & 0.73  & 0.79  & {\bf C$_2^{th}$}  &   \\	
		4780.10   &  2.6 & 89  & 2.0  & 18  & 0.45& 0.07 & 0.31  & +0.33  &  0.6 & 1.5 & 0.04  & 0.79  &   &  \\
		4959.52   & 4.6  & 12  & 3.0  & 7  &  0.48&  0.18 &  0.55 & -0.19  &   & 0.8  & $>$60  &0.66  &   &  \\
		4965.19 & 3.3  &14. & 2.8 &16  & 0.50& 0.54 & 0.58 &   -0.23 & 1.8   & 0.5 & 0.70 & 0.53 & C$_2^r$  &      \\
		4965.97&  7.4 	& 7.4 &2.3 &16  & 0.50& 0.65 & 0.51 &   -0.23 &2.6 &     0.5  & 0.77 &  0.57  & C$_2^r$ & $\chi_b$ \\
		4984.78   & 3.5  & 38  & 2.6  &  19 & 0.31& 0.22 & 0.47  & -0.71  &  2.1 &   0.5   & 0.62  & 0.72 & {\bf C$_2^{th}$}  &  \\
		5061.5&   4.1  &26.  & 2.7& 17 &  0.50& 0.49 & 0.58  &        -0.58      & 2.4 &     0.5 & 0.71 & 0.71 &C$_2^c$ & \\
		5245.44& 5.4 &  18. &3.1 &13 & 0.27 & 0.40  & 0.63&  -0.46 &1.5 &       0.5 & $>$65  &  0.63 & C$_2^r$ &  \\
		5342.52& 8.0   &12.  & 3.7    & 9  &  0.63& 0.64  & 0.57 &     +0.39   & 0.8 &     0.6 & $>$65  & 0.67 &  & \\
		5419.57   & 6.0  & 388  & 3.3  & 9  & 0.62&  0.77 & 0.75  & +0.58  &  1.3 & 7  & $>$55  & 0.75  &   &  \\
		5490.3& 3.1   &6.8  & 2.6   & 15 &  0.63&  0.61 & 0.57 &      +0.40      & 1.4 &   0.5 & 0.77  & 0.77  &  &   \\
		5494.10&  2.5   &36.  & 1.7    & 20 & 0.53& 0.62  & 0.69 &       +0.19      & 0.7 &   0.6 & 0.57 & 0.95 & m$\zeta$ &  \\
		5496.23 & 4.3 &  8.7 & 3.1 & 12 &  0.50& 0.69  & 0.53 &    +0.03 & 1.1 &  0.5 & 0.74 & 0.77 & C$_2^r$  &     \\
		5502.84 &9.8 &  19.  & 3.1 & 10 &  0.50& 0.17  & 0.41 &    +0.02  & 1.0 &   1.1  & $>$60  & 0.89 & m$\zeta$ &  \\
		5504.32 & 4.7 &  8.7  & 2.8    & 14 &  0.63& 0.62  & 0.67 &    +0.04   & 1.3 &     0.5 & $>$65  & 0.71 & & \\
		5512.68 &3.8 &32.  & 2.3    & 20 &  0.58& 0.77  &   0.57   &   -0.65  & 1.8 &  0.5 &  x & 0.75 & {\bf C$_2^{th}$} &$\chi_b$ \\
		5525.07 &6.2 & 226 & 3.5 & 13 &  0.53& 0.66  & 0.68 &    -0.20 & 1.0 &   7.5 &  $>$55 & 0.81 &   &    \\
		5530.10 & 15. &  18 &  6.4 &9  &  0.52&  0.20 &   &   +0.02 & 1.3 & 0.6 & $>$55 & 0.24  &  &  \\
		5541.92 & 5.7 &  26.  & 3.0    & 13 & 0.57& 0.07  & 0.31 &       -0.72  & 1.0 &    0.6 & $>$75  & 0.85  & {\bf C$_2^{th}$} & \\
		5545.08 &3.5 &  55.  &   2.4     & 21 &  0.53& 0.57  & 0.59 &       +0.19 & 0.9 &   0.8 &  0.60 & 0.97  & m$\zeta$ &  \\
		5559.91   &  17. & 23  & 4.5  & 7  & 0.45&  0.43 &  0.45 &  -0.15 & 1.0  & 1.4  & $>$50  & 0.96  &  m$\zeta$ &  \\
		5594.58   &  3.2 & 16  & 2.2  & 18  &  0.50& 0.65 & 0.61  & -0.40  &  1.2 & 0.6  & 0.57  & 0.95  & C$_2^r$  &  \\
		5643.70   &  3.3 & 10  & 2.6  &  13 & 0.48&  0.56 & 0.59  & -0.10  &  1.2 & 0.8  & 0.64  & 0.77  &   &  \\
		5645.70   & 14.  & 31  & 8.1  & 12  & 0.50&  0.74 & 0.64  & -0.20  &  0.8 &  0.6 &  $>$55 & 0.83  &   &  \\
		5706.50  & 3.9 & 8.7  & 2.4  & 8  & 0.50& 0.75  &  0.63 & +0.30  &   & 0.5  &  $>$80 & 0.62  &   &$\chi_b$ \\
		5710.58 & 6.1  & 37  & 5.3  & 15  & 0.48& 0.38  &  0.50 & -0.21  & 1.5  & 0.6  & 0.65  & 0.78  & C$_2^r$  &     \\
		5749.18 & 8.9&26.  & 4.4    & 15 & 0.50&  0.32  & 0.60 &   -0.21     & 0.6 &     0.8  & $>$55 & 0.68 &  &   \\
		5756.10  & 4.0  &8.1  & 2.6   & 8  &  0.53&  0.52 & 0.57 &     +0.28   & 3.0 &     0.8 & $>$60 &$>$0.55 &  & \\
		5769.09 & 2.5 &22.  & 2.2    & 20 & 0.54& 0.39  &  0.45  &    -0.68      & 12. &  0.6  & 0.87 & $>$0.45 & {\bf C$_2^{th}$}  & \\
		5828.50  & 4.9 &33.  & 3.3    & 20 &  0.58& 0.68  & 0.63 &     +0.06      & 1.1 &     0.8 & 0.85 & 0.82 & C$_2^c$ & \\
		5844.89 & 6.3 &15.  & 3.1    & 19 &  0.50& 0.68  & 0.53 &      +0.32      & 0.7 &     0.5 & $>$55 & 0.91 & m$\zeta$ & \\
		5849.8  & 2.4 &102  & 1.8    & 25 &   0.50& 0.51  & 0.55 &     -0.01  & 1.5 &   0.8  & 0.78  & 0.96  & C$_2^r$ & m$\zeta$ \\
		5893.52 & 3.0 &20.  & 2.2    & 14 &  0.53& 0.43  & 0.57 &       -0.16      & 1.9 &     0.8   & 0.60 & 0.66 & C$_2^r$ &  \\
		5910.57   & 2.5  & 24  & 2.0  & 19  &  0.38& 0.40 & 0.45  & -0.25  &  2.3 & 0.8  & 0.67  & 0.83  & C$_2^r$  &  \\
		5922.32 & 3.9 &11.  & 2.2    & 15 &  0.63& 0.58  & 0.52 &     +0.04      & 2.3 &    0.5 & $>$85 & 0.64 & C$_2^r$ &$\chi_b$ \\
		5927.61 & 6.9 & 18.  & 3.0    & 12 & 0.50&  0.38 & 0.50 &  -0.11  & 0.7 &    0.6  & $>$55  &  0.83 & C$_2^r$  &  \\
		\hline
	\end{tabular}
	~~~~~~~~~~~~~~~~~~~~~~~~~~~~~~~~~~~~~~\\
	\\
	{\small 	$^a$ DIB wavelength. \\
		$^{aa}$  Reduced normalised equivalent width in the sight line HD\,175156 (Eq.\ \ref{eq:8eta}). \\
		$^b$ Normalised equivalent width in HD\,175156. \\
		$^c$ R$_{\rm av}$(175156), ratio of the DIB equivalent width in the sight line HD\,175156 to the average of its equivalent width in all sight lines (Eq.\ \ref{eq:7Rav}). \\
		$^d$ Number of sight lines in which the DIB is detected in the APO Catalog. \\
		$^e$ Correlation coefficients with $\lambda_{5512}$ or $\lambda_{5797}$. \\
		$^{e1}$ Lower limit of the average of the normalised correlation Pearson coefficients of the DIB with $\lambda$6087, $\lambda$6093 and $\lambda$6128 (using all APO sight lines) (Eq.\ \ref{eq:C2kn175}). \\
		$^{e2}$ Idem with $\lambda$5257, $\lambda$6412 and $\lambda$6845  (Eq.\ \ref{eq:C3kn148}). \\
		$^{e3}$ Idem with $\lambda$6412, $\lambda$6128 and $\lambda$5419  (Eq.\ \ref{eq:C1knchr}). \\
		$^f$ Normalised Pearson coefficient  with $\lambda$6009.\\
		$^g$ Ratio of DIB equivalent widths in the sight lines  of HD\,204827 and HD\,183143 (Eq.\ \ref{eq:1R21}). For DIBs undetected in HD\,183143, see Appendix \ref{app:B.3substitute}. \\
		$^h$ Measured width of the DIB from the APO Catalog (Fan et al.\ 2019).  \\
		$^i$ Membership of the C$_2$ family [C$_2^{th}$, reference C$_2$ DIB (Table \ref{tab:B1thorburn}); C$_2^{c}$, candidate C$_2$-DIB (Table \ref{tab:B3robust}); C$_2^{r}$, C$_2$-related DIB (Table \ref{tab:B4related}); or DIB associated with the non-C$_2$ main DIBs of Fan et al.\ (2022)
		 (m$\zeta$, m$\zeta\sigma$, m$\sigma$) (Table \ref{tab:D1mainass})].  \\
		$^j$ Membership of the $\chi_b$ family (Table \ref{tab:C2chib}).\\ 
	}
\end{table*}

\addtocounter{table}{-1}

\begin{table*}[!htbp]
	\caption{continued - Data of 79  $\chi_a$ DIBs}
	\begin{tabular}{lccccccccccccll}
		\hline 
		&  &  &  &  &  &  &  &  &  &  &   &   &   &\\
		$\lambda^a$ & $\eta_{175}$  & {EN175$^b$} & R$_{175av}^c$ & N*$^d$ &  kn$_{175}^e$ &   kn$_{148}^e$ &  kn$_{\rm chr}^e$  &  {kn$_{6009}^f$} & R$_{21}^g$  &{\small FWHM$^i$} &   kn$_{5512}^e$ & q$_{5797}^e$ & C$_2$m$^i$ & $\chi_b^j$\\
		\AA   &  & {\small m\AA\ /mag}  & & & & & & &     &\AA &   & & &\\
		\hline 
		&  &  &  &  &  &  &  &  &  &  &   &   &  &\\
5975.71  & 7.3  &  21.9   &  4.3   &  17    &  0.58  & 0.56  &  0.63      &  -0.15     &  1.6   &  0.5   &  0.71   &   $>$0.60  &  & \\ 
6020.44  & 4.0  &  8.7    &  2.6   &  13    &  0.55  &  0.45 &  0.58      &  -0.01    &  2.0     &  0.6  &  0.48  &   $>$0.60  &  & \\ 
6048.42  & 6.9  &  15.5   &  4.1   &  11  &  0.52  & 0.53  &  0.53   &  +0.67   &  0.7    &  0.9   &  0.69   &   $>$0.55  & m$\zeta\sigma$  &\\ 
6087.5 & 7.2  &  17.7   &  4.4    &  14    &  0.50  & 0.28  &  0.51      &  +0.58      &  0.7     &  0.9    &  0.18      &   $>$0.65  &  & \\ 
6093.3 & 4.7  &  10.0   &  2.8   &  9     &  0.50  & 0.52  &  0.50      &  -0.83   &  2.2  &  0.8  &  0.48  &   $>$0.80  & C$_2^r$ & $\chi_b$  \\ 
6102.39  & 2.8 &  10.3   &  2.9    &  15    &  0.60  & 0.63  &  0.70 &  +0.64      &  0.5     &  0.6    &  0.54      &   0.62 &  & $\chi_b$  \\ 
6103.41  & 10. &  17.1   &  6.9   &  11    &  0.53  & 0.62  &  0.53      &  +0.35      &  0.7     &  0.6    &  0.33   &   $>$0.60 & & $\chi_b$  \\ 
6128.26  & 3.8 &  42.3   &  3.1    &  15    &  0.50  & 0.80  &  0.65      &  +0.43      &  0.7     &  2.7    &  0.40 &   0.57   &  & \\ 
6161.95  & 3.0  &  18.1   &  2.3   &  19    &  0.45  &  0.15 &  0.45      &  +0.12      &  1.0     &  0.5    &  0.81 &   0.34   & &  \\ 
6163.5   & 35. &  69.0   &  19.  &  15    &  0.50  & 0.68  &  0.53      &  +0.29      &  2.3     &  0.5    &  0.65      &   $>$0.50  & &  \\ 
6211.69  & 2.3  &  16.8   &  2.0   &  21    &  0.00  &  -0.13  &  0.27    &  +0.32  &  0.6   &  0.6    &  0.95      &   0.15   &  m$\zeta\sigma$ &\\ 
6244.41  & 3.4  &  10.6   &  2.7   &  14    &  0.48   & 0.34 &  0.45      & +0.08   &  1.3     &  0.6    &  0.86      &   0.49   &  & \\ 
6259.65  & 3.1  &  11.3   &  2.8   &  14    &  0.60   & 0.66 &  0.70    &  +0.17    &  1.2     &  0.7    &  0.62     &   $>$0.50  &  & \\ 
6277.83  & 3.1 &  46.5   &  2.6   &  13    &  0.53   & 0.10 &  0.39      &  -0.21        &  1.3     &  1.2    &  0.46      &   $>$0.60  & &  \\ 
6358.36  & 4.5  &  15.8   &  3.0   &  12    &  0.58  &  0.60 &  0.60    &  +0.57    &  0.7     &  0.7    &  0.82      &   $>$0.50  & &  \\ 
6369.28  & 8.5   &  34.8   &  3.0   &  9     &  0.48  & 0.17  &  0.47   &  -0.17   &  0.4     &  1.3    &  0.80   &   $>$0.55  & m$\sigma$ &  \\ 
6384.91  & 9.1    &  39.7   &  9.8   &  13    &  0.50  & 0.47  &  0.51   &  -0.07  &  0.6     &  0.6    &  0.30      &   $>$0.45  &  & \\ 
6425.68  & 2.7 &  35.5   &  2.2   &  21    &  0.55   & 0.36 &  0.56      &  +0.43   &  0.7   &  0.7    &  0.86   &   0.32   & m$\zeta\sigma$ & \\ 
6468.71  & 2.4  &  16.1   &  1.9   &  19    &  0.57  &  0.72 &  0.65   &  +0.32  &  1.5     &  0.9    &  0.69      &   0.74 &  & $\chi_b$  \\ 
6498.0   & 3.2  &  12.3   &  2.0   &  16    &  0.47  & 0.04  &  0.36   &  +0.76  &  0.5     &  0.7    & 0.87   &   0.16   &  m$\zeta\sigma$ & \\ 
6548.99  & 2.9 &  15.5   &  2.5   &  14    &  0.45  &  0.63 &  0.51      &  +0.26     &  0.5     &  0.6    &  0.74   &   0.37   & m$\sigma$ & \\ 
6599.99  & 8.7  &  14.8   &  4.5   &  17    &  0.48  & 0.63  &  0.53     &  -0.08     &  0.4     &  0.5    &  0.76      &   $>$0.45  & &  \\ 
6660.67  & 3.8 &  131.  &  3.7   &  24    &  0.10   & -0.10 &  0.30      &  +0.16   &  0.4     &  0.6    &  0.80      &   0.20   & m$\sigma$ & \\ 
6699.28  & 3.2  &  63.5   &  2.6   &  22    &  0.30  &  -0.19 &  0.29    & +0.16  &  0.5   &  0.7   &  0.84      &   0.13   &  m$\zeta\sigma$ & \\ 
6713.79  & 3.5  &  13.2   &  2.5   &  14    &  0.52  & 0.27  &  0.61   &  +0.67   &  0.2     &  1.0    &  0.76      &   0.26   &  m$\sigma$  &\\ 
6718.33  & 2.3  &  11.9   &  3.0   &  12    &  0.53   & 0.58 &  0.65    &  +0.24    &  0.4     &  0.6    &  0.18      &   0.56   &  & \\ 
6727.64  & 9.1  &  18.7   &  4.1   &  18    &  0.57  & 0.51  &  0.66    &  +0.50   &  0.6     &  0.7    &  0.76      &   $>$0.50  & &  \\ 
6733.25  & 3.0  &  17.1   &  2.2   &  16    &  0.47  & 0.21  &  0.45    &  +0.15     &  1.4     &  1.3    &  0.76      &   $>$0.55  & &  \\ 
6747.82  & 3.5  &  19.0   &  2.9    &  13    &  0.47  & 0.55  &  0.58    &  +0.30   &  0.4     &  1.0    &  0.61    &   0.48   &  m$\sigma$ & \\ 
6755.95  & 2.6 &  8.7    &  1.7   &  9     &  0.52   & 0.48 &  0.48      &  +0.47    &     &  0.9    &  0.76      &   $>$0.45 & & $\chi_b$  \\ 
6786.39  & 2.3  &  7.7    &  1.7   &  12    &  0.42  &  0.41 &  0.61    &  +0.18      &  0.5     &  0.8    &  0.76      &   0.29   &  & \\ 
6841.61  & 2.3 &  9.0    &  2.0   &  15    &  0.50  & 0.26  &  0.51      &  +0.60   &  0.5     &  0.6    &  0.78      &   0.16   &  m$\sigma$ &\\ 
6847.73  & 5.4  &  11.9   &  3.3    &  11    &  0.57  & 0.62  &  0.73    &  +0.72    &  0.8     &  0.8    &  0.27   &   $>$0.50  & &  \\ 
6862.49  & 6.9  &  25.2  &  4.8 &  17  &  0.57  & 0.53  &  0.55    &  +0.63  &  0.4  & 0.5 &  0.71  &  $>$0.50  & m$\zeta\sigma$ & $\chi_b$ \\ 
7124.74  & 2.5   &  14.8   &  2.0   &  6     &  0.45  &  0.24 &  0.47  &  -0.47   &    &  1.3    &       &   $>$0.45  & m$\zeta\sigma$ & \\ 
7139.89  & 4.5  &  16.5   &  3.6   &  15    &  0.55  & 0.45  &  0.72   &  +0.52      &  0.2     &  0.7    &  0.63      &   0.43   & &  \\ 
7257.46  & 8.9  &  41.3   &  3.5   &  9     &  0.50  & 0.62  &  0.65     &  -0.31    &      &  1.0    &  0.88      &   $>$0.50  &  m$\sigma$ &\\ 
7366.05  & 2.4  &  24.8   &  2.1   &  17    &  0.50  & 0.32  &  0.50      &  +0.09       &  0.3     &  0.7    &  0.78      &   0.20   &  & \\ 
7451.36  & 3.2  &  20.6   &  2.3   &  16    &  0.53  & 0.35  &  0.55      &  +0.45      &  0.5     &  1.0    &  0.71      &   0.51   &  & \\ 
7470.38  & 3.5 &  15.5   &  2.4    &  20    &  0.57  & 0.59  &  0.59      &  +0.67      &  0.7     &  0.8    &  0.72      &   0.51 &  &  $\chi_b$ \\ 
7553.75  & 7.1   &  17.7   &  3.1   &  6     &  0.50  &  0.00 &  0.00       &  -0.09      &        &  1.4    &       &   $>$0.45  & &  \\ 
7706.76  & 4.8      &  18.4   &  3.1    &  13    &  0.55  &  0.19 &  0.56      &  +0.47      &  0.7 & 0.8 &  &0.62 & & \\
	\hline
\end{tabular}
~~~~~~~~~~~~~~~~~~~~~~~~~~~~~~~~~~~~~~\\
{\small See notes of Table \ref{tab:C1chia} in previous page.}
\end{table*}

\begin{table*}
	\caption{Data of 43 $\chi_b$ DIBs}
	\label{tab:C2chib}
	\begin{tabular}{lccccccccccccll}
		\hline 
		&&&&&&&&&&&&&&\\
		$\lambda^a$ & $\eta_{148}^{ab}$  & {EN148$^k$} & R$_{148av}^m$ & N*$^d$ &  kn$_{175}^{e1}$ &  kn$_{148}^{e2}$ &  kn$_{\rm chr}^{e3}$  &  {kn$_{6009}^f$} & R$_{21}^g$  &{\small FWHM$^i$} &   kn$_{5512}^e$ & q$_{5797}^e$ & C$_2$m$^i$ & $\chi_a^q$  \\
		\AA   &  & {\small m\AA\ /mag} &  & & & & & &  & \AA &  &  &  &\\
		\hline 
		&&&&&&&&&&&&&& \\ 
		4965.97   &    6.8       &  11.8    &    3.6    &    16     &    0.57  &  0.65  &    0.68       &    -0.22      &    2.6      &    0.5     &    0.77    &    0.57       &  C$_2^r$ & $\chi_a$    \\ 
		5257.44   &    3.0       &  20.6    &    2.9    &    12     &    0.27  &  0.81  &    0.63       &    -0.42      &             &    0.8     &    0.95    &    0.52       &    C$_2^r$  &  \\ 
		5262.44   &    7.2       &  15.0    &    4.3     &    14     &    0.55   & 0.69  &    0.69       &    -0.35      &    4.7      &    0.6     &    0.87    &    0.66       &   C$_2^r$  &  \\ 
		5340.38   &    3.0       &  24.1    &    3.5    &    15     &    0.55  &  0.72 &    0.60       &    -0.63      &    4.9      &    0.8     &    0.80    &    0.30       &   C$_2^r$  &  \\ 
		5371.12   &    16.      &  89.4    &    9.0    &    11     &    0.56   & 0.64  &    0.66       &    +0.19       &    0.9      &    2.0     &     $>$0.50   &    0.31       &    &   \\ 
		5497.46   &    6.7       &  30.0    &    3.8    &    8      &    0.32   &  0.50 &    0.41       &    +0.11       &    2.7      &    1.0     &    $>$0.60   &    0.78       &    &   \\ 
		5512.68   &    4.0       &  33.8    &    2.6    &    20     &    0.58   & 0.77  &    0.57       &    -0.65      &    3.0      &    0.5     &      x      &    0.75       &  {\bf C$_2^{th}$} & $\chi_a$    \\ 
		5546.46   &    2.8       &  27.4    &    2.4    &    21     &    0.57  & 0.79   &    0.60       &    -0.54      &    2.9      &    0.7     &    0.90    &    0.67       &     {\bf C$_2^{th}$} &   \\ 
		5706.5    &    3.0       &  7.6     &    2.1     &    8      &    0.50  &  0.75  &    0.63       &    +0.30       &              &    0.5     &     $>$0.80  &    0.62       &  &  $\chi_a$   \\ 
		5760.48   &    6.9       &  16.8    &    4.1    &    14     &    0.43  &  0.60  &    0.49       &    -0.46     &    2.5      &    0.7     &     $>$0.70   &    0.38       &    C$_2^{c}$ &   \\ 
		5769.91   &    4.3       &  12.6    &    2.7    &    17     &    0.58   & 0.74  &    0.55       &    -0.05       &    1.9      &    0.5     &    0.25    &    0.72       &    &   \\ 
		5785.0    &    3.2       &  22.1    &    2.4    &    20     &    0.27  & 0.61   &    0.31       &    -0.10      &    0.6      &    1.0     &    0.43    &    0.78       &    &   \\ 
		5838.04   &    3.2       &  14.4    &    4.1     &    15     &    0.47  &  0.63  &    0.50       &    -0.33      &    4.0      &    0.5     &    0.68    &    1.0        &   C$_2^r$  &  \\ 
		5862.21   &    4.2       &  11.2    &    2.7    &    11     &    0.62   & 0.48  &    0.02        &    +0.28       &    1.3      &    0.5     &     $>$0.65   &    0.41       &   &    \\ 
		5866.32   &    2.9       &  17.4    &    2.4    &    9      &    0.00    &  0.42 &    0.13       &    -0.29      &    1.9      &    0.6     &    $>$0.60   &    0.58       &    &   \\ 
		5922.32   &    5.2       &  14.1    &    3.1    &    15     &    0.63   &  0.58 &    0.52       &    +0.04        &    2.3      &    0.5     &     $>$0.85   &    0.64       & &  $\chi_a$    \\ 
		6046.78   &    5.6       &  22.4    &    3.2    &    11     &    0.58   & 0.67  &    0.67       &    +0.16       &              &    1.4     &    $>$0.60  &    0.38       &   &    \\ 
		6066.71   &    11.      &  22.1    &    5.2    &    14     &    0.37   &  0.63 &    0.47       &    +0.63       &    0.4      &    0.5     &     $>$0.50   &    0.59       &   &    \\ 
		6071.33   &    3.1       &  14.4    &    2.6    &    14     &    0.25   & 0.62  &    0.34       &    +0.72        &    0.3      &    1.1     &    0.20    &    0.68       &     &  \\ 
		6084.92   &    4.9       &  20.0    &    3.0    &    21     &    0.33   & 0.57  &    0.42       &    +0.47       &    0.7      &    0.9     &    0.34    &    0.78       &     &  \\ 
		6093.3  &    7.2       &  13.5    &    3.8    &    9      &    0.50   & 0.52  &    0.50       &    -0.83      &    2.2      &    0.8     &     $>$0.80   &    0.48       & C$_2^r$ & $\chi_a$  \\ 
		6102.39   &    3.7       &  12.4    &    3.4    &    15     &    0.60   & 0.63  &    0.70       &    +0.64       &    0.5      &    0.6     &    0.62    &    0.54       &  & $\chi_a$   \\ 
		6103.41   &    2.8       &  6.5     &    2.6    &    11     &    0.53  &  0.62  &    0.53       &    +0.35       &    0.7      &    0.6     &    $>$0.60   &    0.33    &   &    $\chi_a$   \\ 
		6142.21   &    32.      &  48.2    &    10.   &    11     &    0.13  &  0.35  &    -0.16      &    +0.28       &    0.6      &    0.9     &     $>$0.50   &    0.96       &   m$\zeta\sigma$  &  \\ 
		6170.51   &    4.3       &  46.5    &    3.5    &    10     &    0.25   &  0.61 &    -0.18      &    +0.38       &    0.3      &    1.6     &     $>$0.50   &    0.33       &   &    \\ 
		6189.52   &    3.7       &  12.6    &    3.0    &    20     &    0.34  &  0.60  &    0.39       &    +0.29       &    0.5      &    0.8     &    0.20    &    0.65       &    &   \\ 
		6210.07   &    4.4       &  12.9    &    4.1    &    19     &    0.65  &  0.63  &    0.41       &    +0.20        &    0.8      &    0.5     &     $>$0.50   &    0.37       &    &   \\ 
		6412.37   &    2.9       &  267.   &    2.6    &    11     &    0.58   &  0.81 &    0.75       &    +0.15       &             &    7.6     &    0.80    &    0.70       &    &   \\ 
		6439.51   &    2.7       &  33.2    &    1.9    &    24     &    0.42  &  0.56  &    0.55       &    +0.24       &    0.9      &    0.8     &    0.54    &    0.97       &   m$\zeta$ &  \\ 
		6462.24   &    4.6       &  20.0    &    3.8    &    14     &    0.51  &  0.70  &    0.67       &    +0.30           &    0.3      &    0.7     &     $>$0.50   &    0.31       &    &   \\ 
		6463.65   &    4.1       &  38.8    &    2.6    &    21     &    0.36  &  0.60  &    0.48       &    +0.12       &    0.8      &    0.9     &    0.38    &    0.84       &    &   \\ 
		6468.71   &    4.1       &  20.9    &    2.5    &    19     &    0.57   & 0.72  &    0.65       &    +0.32       &    1.5      &    0.9     &    0.74    &    0.69       &  &  $\chi_a$   \\ 
		6719.19   &    3.7       &  9.4     &    2.7     &    15     &    0.61   & 0.62  &    0.60       &    +0.70       &    0.5      &    0.5     &     $>$0.55   &    0.79       &  &     \\ 
		6755.95   &    13.      &  23.8    &    4.8    &    9      &    0.52   & 0.48  &    0.48       &    +0.47       &             &    0.9     &    $>$0.45   &    0.76   &    &   $\chi_a$    \\ 
		6788.83   &    3.7       &  15.6    &    2.8    &    19     &    0.18   &  0.50 &    0.45       &    +0.40       &    0.2      &    0.8     &    0.14    &    0.79       &   m$\zeta\sigma$  &  \\ 
		6845.37   &    4.4       &  12.9    &    3.3    &    12     &    0.68   & 0.77  &    0.85       &    +0.37        &    0.5      &    0.7     &    $>$0.60   &    0.65       &    m$\sigma$  & \\ 
		6862.49   &    6.6       &  24.1    &    4.6    &    17     &    0.57   &  0.53 &    0.55       &    +0.63      &    0.4      &    0.5     &    $>$0.50   &    0.71       & m$\zeta\sigma$ & $\chi_a$    \\ 
		7136.79   &    3.1       &  21.8    &    2.3    &    15     &    0.40   & 0.65  &    0.45       &    +0.71       &    0.3      &    1.1     &    0.29    &    0.69       &    &   \\ 
		7470.38   &    3.6       &  15.6    &    2.4    &    20     &    0.57   & 0.59  &    0.59       &    +0.67       &    0.7      &    0.8     &    0.51    &    0.72      &  &  $\chi_a$   \\ 
		7568.05   &    9.2       &  59.1    &    7.5     &    10     &    0.45   &  0.57 &    0.46       &    -0.28      &             &    0.8     &     $>$0.55   &    0.52       &   &    \\ 
		7570.32   &    4.2       &  23.2    &    3.7     &    10     &    0.16   & 0.53  &    0.16       &    -0.13      &    0.5      &    0.9     &    $>$0.60   &    0.70       &   &    \\ 
		7822.88   &    3.1       &  45.6    &    3.4    &    12     &    0.46   &  0.70 &    0.50       &    +0.36       &    0.3      &    1.6     &            &    0.68       &    m$\sigma$  & \\ 
		7832.88   &    5.1       &  54.7    &    2.9     &    20     &    0.00    &   0.62 &    0.51       &    +0.39       &    0.4      &    0.8     &            &    0.91       &    m$\zeta\sigma$  & \\
		\hline  
	\end{tabular}
	~~~~~~~~~~~~~~~~~~~~~~~~~~~~~~~~~~~~~~\\
	{\small Same notes as Table \ref{tab:C1chia}, except as follows.\\
		$^{ab}$  Reduced normalised equivalent width in the sight line HD\,148579 (Eq. \ref{eq:8eta}). \\
		$^k$ Normalised equivalent width in HD\,148579. \\
		$^m$ R$_{\rm av}$(148579), ratio of the DIB equivalent width in the sight line HD\,148579 to the average of its equivalent width in all sight lines (Eq.\ \ref{eq:7Rav}). \\
		$^q$ Membership of $\chi_a$ family (Table \ref{tab:C1chia}).\\}
\end{table*}

\begin{table}
	{\tiny
	\caption{Correlation coefficients of $\chi$ DIBs with strongest DIBs.}
	\label{tab:C3chimain}
	\begin{tabular}{lccccccc} 
		\hline 
				  &  &  &  &  &  &   &\\
	DIB	&5849&5797&6196&6613&6270$^c$&5780&6009 \\
	Type$^b$  &C$_2\zeta$  &$\zeta$  & $\zeta\sigma$ & $\zeta\sigma$&$\sigma$  &  $\sigma$ &$\sigma$ \\
		\hline
				  &  &  &  &  &  &   &\\
	q$>$0.95$^a$&  3  & 7  & 4    & 4  &  2  & 3  &  0  \\
	q$>$0.90   	& 10  & 10 & 14   & 11 & 10  & 8  &  3  \\
	q$>$0.85	& 17  & 16 & 26   & 20 & 21  & 18 &  7 \\
	q$>$0.80	& 27  & 28 & 41   & 31 & 32  & 30 & 14 \\
	q$>$0.75	& 35  & 49 & 48   & 44 & 45  & 42 & 21  \\
	q$>$0.70	& 47  & 60 & 56   & 55 & 52  & 52 & 28 \\
		\hline 
	\end{tabular} 
~~~~~~~~~~~~~~~~~~~~~~~~~~~~~~~~~~~~~~\\
 $^a$ Number of $\chi$ DIBs, among a total of 111, for which, e.g.  q$_i$ $>$\,0.95, where q$_i$ is their regular correlation coefficient with one of the seven major classical DIBs $i$ listed here  (excluding the sight lines of HD\,175156 and HD\,148579).\\
 $^b$ DIB family type from Fan et al.\ (2022). \\
 $^c$ $\lambda$6270 = $\lambda$6289.66 is a relatively narrow $\sigma$ DIB close to the limit of $\zeta\sigma$ DIBs in Fig.\ B2 of Fan et al.\ (2022). It is highly correlated  with $\lambda$6613 (Table \ref{tab:D4vib6613}) and $\lambda$6196 (q\,$>$\,0.97).
}
\end{table}

\subsection{Correlations between $\chi$ DIBs with indeterminate coefficients}
\label{app:C.2indeter}	

Figs. \ref{fig:15C3-examplesnew} \& \ref{fig:16C4-undeternew} display a few representative  examples of correlation diagrams for $\chi$ sources, where the formal calculations yield meaningless high values ($\ga$0.80) for the normalised Pearson coefficients, because the correlation is dominated by one (or rarely two) sight line, where the normalised equivalent widths 
of both DIBs are more than twice the average value of EN observed in all other sight lines. In such cases only lower limits can be estimated for the Pearson coefficients, such as those reported in the inserts of Fig. \ref{fig:16C4-undeternew}.  The most frequent cases are those of Figs. \ref{fig:15C3-examplesnew} \& \ref{fig:16C4-undeternew}a, where all other sight lines than HD\,175156 and HD\,148579 display no significant correlation. 
Nevertheless, the very high values of EN observed for both DIBs in at least one sight line (generally HD\,175156) show that there is a very significant correlation between both DIBs when including all sight lines. Such a correlation is conservatively estimated by the value kn\,$>$\,0.50 (Fig. \ref{fig:16C4-undeternew}a). In other cases, there is a significant correlation between the two DIBs when excluding the sight lines of HD\,175156 and HD\,148579 (Figs. \ref{fig:16C4-undeternew}b,c,d). Depending on the strength of this correlation, a lower limit for the actual correlation, including all sight lines, is conservatively estimated as kn\,$>$\,0.60, kn\,$>$\,0.70 or kn\,$>$\,0.80 (Figs. \ref{fig:16C4-undeternew}b,c,d). Intermediate lower limits of 0.55, 0.65 or 0.75 may also be used for intermediate cases.

Using such an approximate correlation gauge, it is possible to identify the $\chi_a$ and $\chi_b$ DIBs that are the most strongly correlated within their class. In this way, a small group of three DIBs was identified in each class. It is thought that the average correlation coefficient kn of the other DIBs of the same class provides an estimate of their degree of correlation with the whole class. The chosen core DIBs are 
$\lambda\lambda$6087, 6093 \& 6128 for $\chi_a$ DIBs. They are used for defining the average coefficient, 
\begin{equation}	\label{eq:C2kn175}
kn_{175} = (kn_{6087} + kn_{6093} + kn_{6128})/3, 
\end{equation}
that is reported in Table \ref{tab:C1chia}. Similarly, Table \ref{tab:C2chib} reports the value of the average coefficient, 
\begin{equation}	\label{eq:C3kn148}
	kn_{148} = (kn_{5257} + kn_{6412} + kn_{6845})/3. 
\end{equation}
As most individual correlation coefficients have only lower limits, their averages,  kn$_{175}$, kn$_{148}$ and  kn$_{\rm chr}$ (Eq.\ \ref{eq:C1knchr}), are themselves lower limits. 
Tables \ref{tab:C1chia}-\ref{tab:C2chib} show that, for $\chi_a$ DIBs, 73\% (resp. 90\%) of the values of  kn$_{175}$ are $\ge$\,0.5 (resp. $\ge$\,0.45), and 90\% of the values of kn$_{148}$ are $\ge$\,0.5 for $\chi_b$ DIBs. This means that, within each class, $\sim$80-90\% of the DIBs are significantly correlated with the whole class. 

\subsection{Prominent $\chi$ DIBs}
\label{app:C.3prominent}	

Only two $\chi_a$ and one $\chi_b$ DIB candidates have EN$_{av}$\,$\ga$\,100\,m\AA /mag, namely: the well-known peculiar C$_2$ DIB $\lambda$4726 (172\,m\AA /mag), $\lambda$6591 (108\,m\AA /mag) and $\lambda$6412 (102\,m\AA /mag, see below). All three are broad, FWHM\,=\,2.8, 7.8 and 7.6\,\AA , respectively. Only two additional $\chi_a$ DIBs have EN$_{av}$\,$\sim$\,40-60\,m\AA /mag, $\lambda$4780 and  the well-known DIB, $\lambda$5849, that is intermediate between C$_2$ and $\zeta$ DIBs. 

 As  discussed in Section \ref{sec:5.2chiproperties}, the most  extraordinary $\chi$ DIB is $\lambda$6412 (EN$_{148579}$\,=\,267\,m\AA /mag). The DIB $\lambda$6128 (EN$_{175156}$\,=\,42\,m\AA /mag) is significantly weaker and narrower. However, it is stronger than most  other $\chi_a$ DIBs and similarly highly correlated with many of them [80\% (resp. 94\%) with kn$_{6128}$\,>\,0.50 (resp. 0.45)]. As discussed in Section \ref{sec:6.1coincidences}, $\lambda$6128  might coincide with the strong band of the C$_{14}^+$ ring. 
 
 Another prominent $\chi_a$ DIB is $\lambda$5419 (EN$_{175156}$ = 388\,m\AA /mag). As discussed in Section \ref{sec:6.1coincidences}, it is also very strong and very broad, and it might perhaps match a vibronic band of the HC$_{11}$H$^+$ chain.
 
Other remarkable $\chi$ DIBs first include the four other highly correlated DIBs, $\lambda\lambda$6093, 6087, 5257, 6845 (Eqs. (\ref{eq:C2kn175}-\ref{eq:C3kn148}). The DIB $\lambda$6093 is remarkable since it is both a $\chi_a$ and $\chi_b$ DIB, and  may be classified as a C$_2$-related DIB candidate (Table \ref{tab:B4related}) from its strong UV sensitivity (Figs. \ref{fig:10B1-kq_k6009longnew} \& \ref{fig:11B2-R21_knqlongnew}),  despite its longer wavelength than most other C$_2$-like DIBs.

\subsection{Connection with C$_2$ DIBs and main DIBs}
\label{app:C.4chimain}	

Tables \ref{tab:C1chia}-\ref{tab:C2chib} and \ref{tab:B3robust}-\ref{tab:B4related} show that six  $\chi$ DIBs are also reference C$_2$ DIBs, three are solid C$_2$ candidates, and 18 are possible C$_2$-related DIBs. In total, there is a substantial overlap with C$_2$ DIBs, since 27 out of 111 $\chi$ DIBs (24\%, both $\chi_a$ and $\chi_b$) are related to C$_2$ DIBs. Such an overlap is further characterised in Figs. \ref{fig:19C7-6010_R21chinew} \& \ref{fig:20C8-5797_5849chinew}.

On the other hand, $\chi$ DIBs strongly overlap with $\zeta$, $\zeta\sigma$ and $\sigma$ main DIBs (Fig. \ref{fig:20C8-5797_5849chinew}). 27 of them are strongly correlated with the 47 non-C$_2$ main DIBs defined by Fan et al. (2022), with at least one correlation coefficient q $\ge$\,0.88 (Tables \ref{tab:D1mainass} and \ref{tab:C1chia}-\ref{tab:C2chib}). 

Fig. \ref{fig:20C8-5797_5849chinew} displays the distribution of the $\chi$ DIBs with respect to the  C$_2$-like DIBs and the non-C$_2$ main DIBs, in the diagram q$_{5849}$ $vs$ q$_{5797}$. Similarly, Fig. \ref{fig:19C7-6010_R21chinew} shows that the correlation (or anticorrelation) coefficient kn$_{6010}$ of the $\chi$ DIBs 
extends over the whole range -0.9 to +0.9.
\FloatBarrier

\section{Some properties of  main DIBs}
\label{app:DmainDIBs}

\subsection{DIBs that are strongly correlated with main DIBs.}
\label{app:D.1associated}	

Table \ref{tab:D1mainass} lists the 106 additional non-C$_2$ DIBs that are strongly correlated with the set of 47 $\zeta$, $\zeta\sigma$ \& $\sigma$ main DIBs defined by Fan et al.\ (2022), with, at least,  one regular correlation coefficient q\,$\ge$\,0.88, i.e.\ a value at least comparable with the minimum value of the correlation of these 47 main DIBs between themselves (Fan et al.\ 2022; Smith et al.\ 2022). Table \ref{tab:D1mainass} also identifies the  strong DIB, among  $\lambda$5797 ($\zeta$), $\lambda\lambda$6196, 6613 ($\zeta\sigma$), $\lambda$6270 ($\sigma$ close to the $\zeta\sigma$ limit), $\lambda$5780 ($\sigma$) and $\lambda$6009 (extreme $\sigma$), for which the correlation is maximum. This gives an indication of the family to which the DIB belongs, that needs to be confirmed. 

\begin{table*}
	\caption{DIBs that are strongly correlated with main DIBs}
	\label{tab:D1mainass}
	\begin{tabular}{lcccccccccccccc}
		\hline  
		&&&&&&&&&&&&&& \\
DIB$^a$  & q$_m^b$ & DIBm$^c$ & & DIB & q$_m$ & DIBm & & DIB & q$_m$ & DIBm & & DIB & q$_m$ & DIBm   \\
\hline
	&&&&&&&&&&&&&& \\
		4688.84 & 89     &  6196  &  | &  4974.93 & 95     &  5797  &  |  &  5236.27 & 96     &  6196  &  |  &  5362.99 & 96     &  5797  \\
5363.77 & 96     &  5797  &  | &  5404.58 & 96     &  6270  &  |  &  5450.07 & 90     &  5797  &  |  &  5487.64 & 97     &  5797  \\
5494.10$^{\chi a}$  & 96     &  6196  &  | &  5502.84$^{\chi a}$ & 89     &  6196  &  |  &  5508.33 & 96     &  6270  &  |  &  5535.14 & 92     &  5780  \\
5559.91$^{\chi a}$ & 96     &  5797  &  | &  5580.79 & 93     &  6196  &  |  &  5609.80  & 95     &  5780  &  |  &  5785.00  & 88     &  6196  \\
5795.21 & 96     &  6009  &  | &  5801.47 & 97     &  5797  &  |  &  5844.89$^{\chi a}$ & 91     &  5797  &  |  &  5888.74 & 92     &  5797  \\
5900.59 & 96     &  6196  &  | &  5925.91 & 94     &  6196  &  |  &  5953.32 & 91     &  5780  &  |  &  5958.86 & 88     &  6196  \\
5962.87 & 92     &  6196  &  | &  5982.77 & 89     &  6270  &  |  &  6000.70  & 88     &  6009  &  |  &  6004.96 & 92     &  6613  \\
6048.42$^{\chi a}$ & 88     &  6613  &  | &  6051.50  & 94     &  5780  &  |  &  6071.33 & 91     &  6270  &  |  &  6110.77 & 90     &  6613  \\
6113.22 & 96     &  6613  &  | &  6139.95 & 91     &  5797  &  |  &  6142.21$^{\chi b}$ & 99     &  6613  &  |  &  6159.63 & 94     &  6613  \\
6194.73 & 89     &  6613  &  | &  6198.99 & 91     &  5780  &  |  &  6211.69$^{\chi a}$ & 96     &  6196  &  |  &  6245.14 & 97     &  6270  \\
6268.59 & 93     &  5780  &  | &  6271.40  & 91     &  6613  &  |  &  6318.07 & 90     &  6270  &  |  &  6357.37 & 94     &  6613  \\
6369.28$^{\chi a}$ & 99     &  5780  &  | &  6425.68$^{\chi a}$ & 94     &  6270  &  |  &  6455.92 & 91     &  6613  &  |  &  6460.36 & 91     &  6613  \\
6489.42 & 89     &  6196  &  | &  6492.01 & 90     &  6009  &  |  &  6494.13 & 91     &  6196  &  |  &  6498.00$^{\chi a}$  & 95     &  6196  \\
6523.29 & 90     &  5797  &  | &  6548.99$^{\chi a}$ & 91     &  6270  &  |  &  6594.30  & 97     &  6196  &  |  &  6597.34 & 90     &  6196  \\
6645.99 & 94     &  6196  &  | &  6657.34 & 89     &  6613  &  |  &  6672.23 & 89     &  6613  &  |  &  6689.35 & 94     &  6613  \\
6693.57 & 89     &  6196  &  | &  6694.53 & 89     &  6613  &  |  &  6709.49 & 92     &  6196  &  |  &  6713.79$^{\chi a}$ & 88     &  5780  \\
6719.19 & 88     &  5780  &  | &  6737.26 & 90     &  5780  &  |  &  6740.97 & 88     &  6009  &  |  &  6747.82$^{\chi a}$ & 89     &  6270  \\
6750.76 & 94     &  6196  &  | &  6765.29 & 90     &  5797  &  |  &  6770.17 & 91     &  6196  &  |  &  6788.83$^{\chi b}$ & 93     &  6196  \\
6792.51 & 90     &  6196  &  | &  6795.26 & 94     &  5797  &  |  &  6801.47 & 93     &  6009  &  |  &  6803.35 & 92     &  6196  \\
6811.27 & 92     &  6196  &  | &  6827.39 & 90     &  6196  &  |  &  6841.61$^{\chi a}$ & 91     &  5780  &  |  &  6843.76 & 91     &  6196  \\
6852.52 & 91     &  5797  &  | &  6860.02 & 92     &  5797  &  |  &  6862.49$^{\chi ab}$ & 90     &  6613  &  |  &  6887.03 & 95     &  6009  \\
6904.04 & 97     &  6009  &  | &  6919.26 & 94     &  6009  &  |  &  6998.67 & 94     &  6270  &  |  &  7061.09 & 94     &  6196  \\
7077.90  & 91     &  6196  &  | &  7086.77 & 89     &  6613  &  |  &  7124.74$^{\chi a}$ & 94     &  6613  &  |  &  7198.58 & 92     &  6613  \\
7257.46$^{\chi a}$ & 97     &  5780  &  | &  7276.59 & 89     &  6196  &  |  &  7322.07 & 91     &  6196  &  |  &  7334.53 & 92     &  6009  \\
7339.88 & 93     &  6009  &  | &  7360.55 & 88     &  6196  &  |  &  7382.53 & 93     &  6009  &  |  &  7405.73 & 94     &  5780  \\
7406.26 & 92     &  6009  &  | &  7419.11 & 89     &  6270  &  |  &  7581.47 & 89     &  6196  &  |  &  7832.88$^{\chi b}$ & 97     &  6196  \\
8001.17 & 95     &  6613  &  | &  8026.23 & 94     &  6196  &  |  &       &        &     &  |  &      &       &     \\
		\hline
	\end{tabular}
	~~~~~~~~~~~~~~~~~~~~~~~~~~~~~~~~~~~~~~\\
	{\small $^a$ Wavelength of the 106 additional non-C$_2$ DIBs that are strongly correlated (q$_m$\,$\ge$\,88$^b$) with the set of 47 $\zeta$, $\zeta\sigma$ \& $\sigma$ main DIBs defined by Fan et al.\ (2022). \\
		$^b$ Maximum value of the regular correlation coefficient q with the set of strong DIBs, including $\lambda\lambda$5797, 6196, 6613, 6270, 5780 \& 6009 (excluding the sight lines of HD\,175156 and HD\,148579). \\
		$^c$ Wavelength of the strong DIB for which q is maximum and equal to q$_{\rm m}$.
	} \\
\end{table*}

\subsection{Possible coincidences of wavelengths of C$_{2p+1}$ chains with strong DIBs}
\label{app:D.2C2p+1}

\begin{table}[!htbp]
	\caption{Comparison of Ne-matrix wavelengths of C$_{2p+1}$ carbon chains with very strong narrow DIBs, $\lambda\lambda$5797, 6196, 6613.}
	\label{tab:D2C2pp1}
	\begin{tabular}{llcccccl}
		\hline \hline
		n$^a$ &  $\lambda_{n}Ne^b$ & $\lambda_{n}Ne_{\rm ext}^c$ &   $\lambda_{\rm DIB}^d$ & d$\lambda^e$ &  q$_{5797}$ &  q$_{6196}$ &  q$_{6613}$  \\
		& \AA   &   \AA & \AA &  \AA & & &  \\
		\hline
		&&&&&&&\\
		13& 3796& 3789 &      &    & & &  \\
		15& 4196& 4201 &      &    & & &  \\
		17& 4603& 4613 &      &   & & &   \\
		19& 5032& 5024 &      &    & & &  \\
		21& 5437& 5436 &  &  & & &  \\
		23&     & 5848 & 5797 & 51 & & 0.94&0.94 \\
		25&     & 6260 & 6196 & 64 &0.94 & &0.98  \\
		27&     & 6672 & 6613 & 58 &0.94 &0.98 & \\
		\hline  
		\hline
	\end{tabular}
	\\
	\\
	{\small $^a$  Number of carbon atoms in pure carbon chains C$_{\rm n}$.\\
		$^b$  Wavelength of the strong band $^1\Sigma_{\rm u}^+$ - X$^1\Sigma_{\rm g}^+$ of C$_{\rm n}$ in Ne-matrix (Wyss et al.\ 1999; Forney et al.\ 1996).\\
		$^c$  Linear-fit extrapolated wavelength of the strong band $^1\Sigma_{\rm u}^+$ - X$^1\Sigma_{\rm g}^+$ of C$_{\rm n}$ in Ne-matrix: $\lambda_{n}Ne_{\rm ext}$ = 206 x n + 1112.\\
		$^d$  Very strong narrow DIB.\\
		$^e$  [$\lambda_{n}Ne_{\rm ext}$ - $\lambda_{\rm DIB}$]  should be the Ne-matrix shift for a strictly linear relation between $\lambda$ and n.\\ 
	}
\end{table}

Table \ref{tab:D2C2pp1} shows explicit values of the wavelengths of a very strong  band ($^1\Sigma_{\rm u}^+$ - X$^1\Sigma_{\rm g}^+$) of C$_{2p+1}$ chains in Ne-matrix, either measured by Wyss et al.\ (1999) and Forney et al.\ (1996) up to n = 2p + 1 = 21, or extrapolated from their measurements with a slope $\lambda$/n = 206 \AA . These extrapolated wavelengths, $\lambda_{n}Ne_{\rm ext}$, 
 for n\,=\,23 to 27, are compared with the wavelengths of the major narrow DIBs, $\lambda\lambda$5797, 6196 \& 6613.  There is a striking possible agreement, since the DIB wavelengths are slightly blue shifted from $\lambda_{n}Ne_{\rm ext}$ by a few tens of \AA , respectively, i.e.\ a value compatible with the expected red shift from the Ne-matrix. 

But, the conjecture that these highly correlated  DIBs could be carried by C$_{2p+1}$ chains would face various difficulties, including:

i) The uncertainties on the wavelength extrapolations and the matrix shift preclude to check the actual wavelength match.

ii) There is no good match between any DIB and the actual matrix  measurements for C$_{19}$ and C$_{21}$.

iii) The DIB profile of $\lambda$6613 implies the likely presence of a Q band (see e.g. MacIsaac et al.\ 2022), unlike the profiles of $\lambda\lambda$5797 \& 6196, the profiles of C$_{2p+1}$ chains up to C$_{19}$, and the profiles of most C$_2$ DIBs (Elyajoury et al.\ 2018). This seems hardly compatible with the expected $^1\Sigma_{\rm u}^+$ - X$^1\Sigma_{\rm g}^+$ strong band of linear C$_{27}$, unless, perhaps, this chain was bent. But, there is no other evidence of such a change of structure of C$_{27}$ with respect to  shorter chains.

iv) It seems that such very long chains could hardly survive to cyclisation in the diffuse ISM, after the absorption of a $\sim$10\,eV UV photon. However, detailed models for such cyclisations, including delayed fluorescence, are still lacking.

\subsection{Possible vibronic systems of major DIBs}
\label{app:D.3vibronic}

Several DIBs are highly correlated with $\lambda\lambda$5797, 6196 \& 6613 and blue shifted by less than a few hundred cm$^{-1}$. They might be possible vibronic DIBs (Smith et al.\ 2021; Fan et al.\ 2022). 

Such possible vibronic DIBs include those listed in Table \ref{tab:D3vib5797-6196}, for  $\lambda$5797 and $\lambda$6196. Their number is limited, even taking into account the enhanced difficulty for detecting DIBs at short wavelength. This might perhaps be roughly compatible with the extrapolation to C$_{23}$-C$_{25}$ of the vibronic structures seen in the matrix measurement of Wyss et al.\ (1999) for  C$_{19}$ and C$_{21}$. But, the values of these possible vibronic shifts seem hardly compatible with the extrapolation of the theoretical values of the lowest $\sigma_{g}^+$ mode of C$_{19}$-C$_{21}$ (Ghosh, Reddy \& Mahapatra 2019). 

On the other hand, the high number of possible vibronic DIBs associated with $\lambda$6613 (Table \ref{tab:D4vib6613}) is impressive, especially when compared with $\lambda\lambda$5797, 6196, although it might include  fortuitous coincidences. This difference may be partially explained by the larger DIB wavelength range possible for vibronic DIBs of $\lambda$6613, but it probably implies a different carrier structure. 

\begin{table}[!htbp]
	\tiny
	\caption{Possible vibronic DIBs associated with major DIBs, $\lambda$5797 ($\zeta$) and $\lambda$6196 ($\zeta\sigma$).}
	\label{tab:D3vib5797-6196}
	\begin{tabular}{llcclcccl}
		\hline \hline	
		DIB$^a$    & DIBv$^b$ & $\sigma^c$ & $\sigma_{\rm th}^c$ & q$_{DIB}^d$ & EW$^e$ & FW$^f$ & R$_{21}^g$ & T$^h$  \\
		 \AA   &  \AA & cm$^{-1}$  & cm$^{-1}$ &  & m\AA & \AA & &    \\
		\hline
		&&&&&&&& \\
		6196 & 6108$^l$ & 232 & x & 0.96 & 14  & 0.5 & 0.57 &{\tiny $\zeta\sigma$}  \\
		6196&  {\tiny 6113$^l$} & {\tiny 219} & {\tiny x} & {\tiny 0.94} & {\tiny 33} & {\tiny 0.7} & {\tiny 0.79} & \\	
			\hline 
		&&&&& &  &  &   \\
		5797  &	5766$^j$ & 93	& x &	0.89$^j$ &	17  & 0.8 & 1.4 & $\zeta$  \\
		5797  &	5711$^i$ & 	260 & x &	0.93$^i$ &	36  & 1.2 &  &  $\zeta$  \\
		5797 &	5545 &	784 & 780$^k$	&	0.97 &	39  & 0.8 & 0.9 &  $\zeta$  \\
		\hline \hline
	\end{tabular}
\\
{\small $a$ Main DIB.\\
	$b$ Possible vibronic DIB. \\
	$c$ Vibronic shift. \\
	$d$ Regular Pearson correlation coefficient of the possible vibronic DIB$^b$ with the main DIB. \\
	$e$ Equivalent width in the sight line of HD\,183143. \\
	$f$ APO DIB width (FWHM).\\
	$g$ Ratio of equivalent widths in HD\,204827 to HD\,183143, measured by APO. \\
	$h$ DIB family type of DIBv, from Fan et al.\ (2022). Transition DIBs between the $\zeta$ and $\sigma$ families are denoted $\zeta\sigma$. \\
	$i$ A high correlation coefficient is obtained only when the APO blended DIBs, $\lambda$5710 + $\lambda$5711, are merged. \\
	$j$ The profiles of the main DIB, $\lambda$5797, and the possible vibronic DIB, $\lambda$5766,  are reported as highly correlated by Ebenblicher et al.\ (2024).\\
	$k$  3$\times$260\,cm$^{-1}$.\\
	$l$ $\lambda$6108 and $\lambda$6113 might also be part of the vibronic system of $\lambda$6613 (Table \ref{tab:D4vib6613}).}
\end{table}

\begin{table}
	\caption{Possible DIB vibronic system of the major DIB $\lambda$6613 ($\zeta\sigma$)}
	\label{tab:D4vib6613}
	\begin{tabular}{lcccccccl}
		\hline \hline
		{\tiny Mode} & {\tiny DIB$_{\rm v}^m$} & $\sigma$ &   {\tiny $\sigma_{\rm th}^a$} & {\tiny q$_{6613}^b$} & {\tiny EW$^c$} & {\tiny FW$^d$} & {\small R$_{21}^e$} & {\tiny T$^f$} \\
		& \AA   & {\tiny cm$^{-1}$} & {\tiny cm$^{-1}$} & & {\tiny m\AA } &\AA & &  \\
		\hline
		&  &  &  &  &  &  &  &  \\
		0$^0_0$& {\tiny {\bf 6613}} & {\tiny 0}  &  {\tiny 0}  & x & {\tiny 318} & {\tiny 1.1} & {\tiny 0.52} & {\tiny $\zeta\sigma$}\\
		&  &  &  &  &  &  &  &  \\
		V$^1_0$ & {\tiny 6520$^l$} & {\tiny {\bf 216}} & x & {\tiny 0.97} & {\tiny 48} & {\tiny 1.0} & {\tiny 0.54} & {\tiny $\zeta\sigma$} \\
		V$^2_0$ & {\tiny 6425} & {\tiny 442} & {\tiny 432} & {\tiny 0.91} & {\tiny 21} & {\tiny 0.7} & {\tiny 0.70} &  \\
		& {\tiny 6445?$^j$} & {\tiny 395?} & {\tiny 432} & {\tiny 0.93} & {\tiny 40} & {\tiny 0.6} & {\tiny 0.77} & {\tiny $\zeta\sigma$} \\
		V$^3_0$& {\tiny 6330} & {\tiny 678} & {\tiny 658} & {\tiny 0.93} & {\tiny 16} & {\tiny 0.7} & {\tiny 0.63} & {\tiny $\zeta\sigma$} \\
		V$^4_0$ & {\tiny 6245} & {\tiny 892} & {\tiny 845} & {\tiny 0.95} & {\tiny 7.2} & {\tiny 0.5} & {\tiny 0.60} &  \\
  &  &  &  &  &  &  &  &  \\
	  		W$^1_0$ & {\tiny 6494$^g$} & {\bf{\tiny 278}} & {\tiny x}& {\tiny 0.87$^g$} & {\tiny 266} & {\tiny 7.4}  & {\tiny 0.28$^g$} & {\tiny } \\
	   	 & {\tiny 6498} & {\tiny 269} & {\tiny x} & {\tiny 0.93} & {\tiny 9.4} & {\tiny 0.7} & {\tiny 0.47} & {\tiny } \\
		W$^2_0$& {\tiny 6376} & {\tiny 553} & {\tiny 556} & {\tiny 0.95} & {\tiny 50} & {\tiny 0.8} & {\tiny 0.73} & {\tiny $\zeta\sigma$} \\
		& {\tiny 6377?$^j$} & {\tiny 561} & {\tiny 556} & {\tiny 0.90} & {\tiny 15} & {\tiny 0.6} & {\tiny 1.1?} & {\tiny $\zeta\sigma$} \\
		& {\tiny 6367?$^j$} & {\tiny 585?} & {\tiny 556} & {\tiny 0.97} & {\tiny 18} & {\tiny 0.9} & {\tiny 0.89} & {\tiny $\zeta\sigma$} \\
		W$^3_0$& {\tiny 6270$^h$} & {\tiny 829} & {\tiny 834} & {\tiny 0.97} & {\tiny 134} & {\tiny 1.2} & {\tiny 0.41} & {\tiny $\sigma^h$} \\
		& {\tiny 6271} & {\tiny 825} & {\tiny 834} & {\tiny 0.91} & {\tiny 30} & {\tiny 0.5} & {\tiny 0.33} & {\tiny } \\
		W$^4_0$& {\tiny 6159} & {\tiny 1115} & {\tiny 1112} & {\tiny 0.94} & {\tiny 11} & {\tiny 1.0} & {\tiny 0.41} & {\tiny } \\
		&  &  &  &  &  &  &  &  \\
		{\tiny V$^1_0$W$^1_0$}& {\tiny 6397} & {\tiny 512} & {\tiny 494} & {\tiny 0.91} & {\tiny 39} & {\tiny 1.3} & {\tiny 0.51} & {\tiny $\zeta\sigma$} \\
		{\tiny V$^2_0$W$^1_0$}& {\tiny 6318} & {\tiny 707} & {\tiny 720} & {\tiny 0.90} & {\tiny 147} & {\tiny 4.8} & {\tiny 0.33} & {\tiny  } \\
		{\tiny V$^4_0$W$^1_0$}& {\tiny 6498} & {\tiny 1089} & {\tiny 1072} & {\tiny 0.93} & {\tiny 9.4} & {\tiny 0.7} & {\tiny 0.47} & {\tiny  } \\
		{\tiny V$^1_0$W$^3_0$}& {\tiny 6185} &{\tiny 1046} & {\tiny 1045} & {\tiny 0.94} & {\tiny 12} & {\tiny 0.5} & {\tiny 0.58} &  {\tiny $\zeta\sigma$} \\
		{\tiny W$^1_0$W$^3_0$}& {\tiny6159} & {\tiny 1114} & {\tiny 1107} & {\tiny 0.94} & {\tiny 11} & {\tiny 1.0} & {\tiny 0.41} & {\tiny  } \\
		{\tiny V$^2_0$W$^3_0$}&  {\tiny 6113$^k$} & {\tiny 1258} & {\tiny 1260} & {\tiny 0.96} & {\tiny 33} & {\tiny 0.7} & {\tiny 0.79} & {\tiny  } \\
		&{\tiny 6108$^k$} & {\tiny 1252} & {\tiny 1260} & {\tiny 0.93} & {\tiny 14} & {\tiny 0.5} & {\tiny 0.57} & {\tiny  $\zeta\sigma$} \\
		{\tiny V$^1_0$W$^2_0$}& {\tiny 6384?$^i$} & {\tiny 793} & {\tiny 790} & {\tiny 0.88$^i$} & {\tiny 1958} & {\tiny 4.5} & {\tiny 0.29} & {\tiny $\sigma^i$ } \\
		&  &  &  &  &  &  &  &  \\
		\hline  
		\hline
	\end{tabular}
	~~~~~~~~~~~~~~~~~~~~~~~~~~~~~~~~~~~~~~\\
	{\small $m$  Expected vibronic DIBs, with origin at 6613\,\AA , and possible vibration modes V at 216\,cm$^{-1}$ and W at 278\,cm$^{-1}$, inferred from $\lambda$6520$^l$ and $\lambda$6494$^g$, respectively.\\
		$a$  Expected wave number of the vibronic DIB. \\
		$b$ Regular Pearson correlation coefficient with $\lambda$6613.\\ 
		$c$ Equivalent width in HD\,183143 sight line, measured by APO or inferred. \\
		$d$ APO DIB width (FWHM).\\
		$e$ Ratio of equivalent widths in HD\,204827 to HD\,183143, measured by APO or inferred. \\
		$f$ DIB family type of DIB$_{\rm v}$, from Fan et al.\ (2022). Transition DIBs between the $\zeta$ and $\sigma$ families are denoted $\zeta\sigma$. \\
		$g$ Very broad DIB, possibly implying uncertain values for q$_{6613}$ and R$_{21}$.\\
		$h$ Major DIB, $\lambda$6269.66, classified as $\sigma$ by Fan et al.\ (2022), close to the limit of $\zeta\sigma$ DIBs in their Fig.\ B2.\\
		$i$ An additional DIB at about 6285\,\AA\ might be hidden within the very strong and broad DIB $\lambda$6284. \\
		$j$ Tentative identification, that may be wrong because of poor wave-number match, low correlation or poor R$_{21}$ fit.\\
		$k$ $\lambda$6108 and $\lambda$6113 might also be part of the vibronic system of $\lambda$6196 (Table \ref{tab:D3vib5797-6196}).\\
		$l$ The profiles of the main DIB, $\lambda$6613, and the possible vibronic DIB, $\lambda$6520, are reported as highly correlated by Ebenblicher et al.\ (2024).\\
	}
\end{table}
\end{document}